\documentclass[aps,preprintnumbers,nofootinbib,onecolumn, superscriptaddress]{revtex4}

\usepackage{graphicx}
\usepackage{caption}
\usepackage{pdfpages}
\usepackage[shortlabels]{enumitem}
\usepackage{mathtools}
\usepackage{fancyhdr}
\usepackage{nicefrac}
\usepackage{mathrsfs}
\usepackage{subcaption}
\usepackage{bbm}
\makeatletter

\usepackage{graphics}
\usepackage{amsfonts,amssymb,amsmath}            
\usepackage{ifthen}
\usepackage{amsthm, thm-restate}
\usepackage{dsfont}
\usepackage{float}                       
\usepackage{MnSymbol}
\usepackage{cancel}

\PassOptionsToPackage{hyphens}{url}
\usepackage{hyperref}
\usepackage{xcolor}
\definecolor{mylinkcolor}{rgb}{0,0,0.7} 
\hypersetup{unicode=true,%
  bookmarksnumbered=false,bookmarksopen=false,bookmarksopenlevel=1, %
  breaklinks=true,pdfborder={0 0 0},colorlinks=true}%
\hypersetup{%
  anchorcolor=mylinkcolor,citecolor=mylinkcolor, %
  filecolor=mylinkcolor,linkcolor=mylinkcolor, %
  menucolor=mylinkcolor,runcolor=mylinkcolor, %
  urlcolor=mylinkcolor}

\usepackage[capitalise]{cleveref}

\usepackage{tikz}
\usetikzlibrary{arrows}
\usetikzlibrary{arrows.meta}
\usepackage{rotating, xcolor}
\usetikzlibrary{shapes}
\usetikzlibrary{hobby}
\usetikzlibrary{decorations.markings}
\usetikzlibrary{calc,intersections,through,backgrounds}
\usetikzlibrary{decorations.pathreplacing,angles,quotes}
\usetikzlibrary{arrows,decorations.markings}
\usetikzlibrary{arrows.meta}
\usetikzlibrary{snakes} 
\usepackage{tikz-cd}
\tikzset{degil/.style={
            decoration={markings,
            mark= at position 0.5 with {
                  \node[transform shape] (tempnode) {$\setminus$};
                  }
              },
              postaction={decorate}
}
}

\newcommand\longrsquigarrow{
\begin{tikzpicture}
\draw [decorate, decoration={zigzag, segment length=+6pt, amplitude=+.95pt,post length=+2pt}, arrows={-Classical TikZ Rightarrow}]  (0,0.1) -- (0.6,0.1); \draw[draw=none] (0,0)--(0.6,0);
\end{tikzpicture}
}

\theoremstyle{plain}

\theoremstyle{definition}
\newtheorem{definition}{Definition}
\newtheorem{remark}{Remark}
\newtheorem{example}{Example}
\newtheorem{notation}{Notation}

    \newcommand{\ket}[1]{\vert  #1 \rangle}
    \newcommand{\bra}[1]{\langle #1 |}
    \newcommand{\inprod}[2]{\langle #1 | #2 \rangle}

	\newcommand{\id}{\mathbbm{1}}

	\newcommand{\tr}{\operatorname{Tr}  }
	\DeclareMathOperator{\Tr}{Tr}

\usepackage[nodayofweek]{datetime}
\newcommand{\comment}[1]{}

\newcommand{\cE}{\mathcal{E}}
\newcommand{\cF}{\mathcal{F}}
\newcommand{\cG}{\mathcal{G}}
\newcommand{\cH}{\mathscr{H}}
\newcommand{\cI}{\mathcal{I}}
\newcommand{\cL}{\mathcal{L}}
\newcommand{\cM}{\mathcal{M}}
\newcommand{\cN}{\mathcal{N}}
\newcommand{\cP}{\mathcal{P}}
\newcommand{\cQ}{\mathcal{Q}}
\newcommand{\cR}{\mathcal{R}}
\newcommand{\cS}{\mathcal{S}}
\newcommand{\cT}{\mathcal{T}}

\tikzset{
carrow/.style={thick, decorate, decoration={zigzag, segment length=+6pt, amplitude=+.95pt,post length=+4pt}, arrows={-stealth}},
nodeshade/.style={shade,ball color = lightgray, opacity = 0.5},
nodeshade2/.style={fill=blue!80!white, opacity = 0.6},
lightcone/.style={thick, brown!80!yellow, draw opacity=0.6},
dashedarrow/.style={thick,dashed, arrows={-stealth}, blue!50!black},
solidarrow/.style={thick,arrows={-stealth}, blue!50!black},
augnodeshade/.style={shade,ball color = red!60, opacity = 0.4},
}

\makeatletter
\pgfkeys{/pgf/.cd,
  cube offset x/.initial=2mm,
  cube offset y/.initial=2mm
}
\pgfdeclareshape{cube}
{
  \inheritsavedanchors[from=rectangle] 
  \inheritanchorborder[from=rectangle]
  \inheritanchor[from=rectangle]{north}
  \inheritanchor[from=rectangle]{north west}
  \inheritanchor[from=rectangle]{north east}
  \inheritanchor[from=rectangle]{center}
  \inheritanchor[from=rectangle]{west}
  \inheritanchor[from=rectangle]{east}
  \inheritanchor[from=rectangle]{mid}
  \inheritanchor[from=rectangle]{mid west}
  \inheritanchor[from=rectangle]{mid east}
  \inheritanchor[from=rectangle]{base}
  \inheritanchor[from=rectangle]{base west}
  \inheritanchor[from=rectangle]{base east}
  \inheritanchor[from=rectangle]{south}
  \inheritanchor[from=rectangle]{south west}
  \inheritanchor[from=rectangle]{south east}
  \backgroundpath{
    \southwest \pgf@xa=\pgf@x \pgf@ya=\pgf@y
    \northeast \pgf@xb=\pgf@x \pgf@yb=\pgf@y
    \pgfmathsetlength\pgfutil@tempdima{\pgfkeysvalueof{/pgf/cube offset x}}
    \pgfmathsetlength\pgfutil@tempdimb{\pgfkeysvalueof{/pgf/cube offset y}}
    \def\ppd@offset{\pgfpoint{\pgfutil@tempdima}{\pgfutil@tempdimb}}
    \pgfpathmoveto{\pgfqpoint{\pgf@xa}{\pgf@ya}}
    \pgfpathlineto{\pgfqpoint{\pgf@xb}{\pgf@ya}}
    \pgfpathlineto{\pgfqpoint{\pgf@xb}{\pgf@yb}}
    \pgfpathlineto{\pgfqpoint{\pgf@xa}{\pgf@yb}}
    \pgfpathclose
    \pgfpathmoveto{\pgfqpoint{\pgf@xb}{\pgf@ya}}
    \pgfpathlineto{\pgfpointadd{\pgfpoint{\pgf@xb}{\pgf@ya}}{\ppd@offset}}
    \pgfpathlineto{\pgfpointadd{\pgfpoint{\pgf@xb}{\pgf@yb}}{\ppd@offset}}
    \pgfpathlineto{\pgfpointadd{\pgfpoint{\pgf@xa}{\pgf@yb}}{\ppd@offset}}
    \pgfpathlineto{\pgfqpoint{\pgf@xa}{\pgf@yb}}
    \pgfpathmoveto{\pgfqpoint{\pgf@xb}{\pgf@yb}}
    \pgfpathlineto{\pgfpointadd{\pgfpoint{\pgf@xb}{\pgf@yb}}{\ppd@offset}}
  }
}
\makeatother

\DeclareRobustCommand\ground{\begin{tikzpicture}[scale=0.8]
		\draw [thick](-0.4,0)--(0.4,0);\draw [thick](-0.3,0.1)--(0.3,0.1);\draw [thick](-0.2,0.2)--(0.2,0.2);\draw [thick](-0.1,0.3)--(0.1,0.3);
	\end{tikzpicture}}

\begin{document}

\title{Embedding cyclic information-theoretic structures in acyclic spacetimes: no-go results for indefinite causality}

\author{V. Vilasini}
\email{vilasini@inria.fr}
\affiliation{Université Grenoble Alpes, Inria, 38000 Grenoble, France}
\affiliation{Institute for Theoretical Physics, ETH Zurich, 8093 Z\"{u}rich, Switzerland}
\author{Renato Renner}
\email{renner@ethz.ch}
\affiliation{Institute for Theoretical Physics, ETH Zurich, 8093 Z\"{u}rich, Switzerland}

\date{\today}
\begin{abstract}

The notions of causality adopted within the quantum information and spacetime physics communities are distinct. Although experience tells us that these notions play together in a compatible manner in physical experiments, their general interplay is little understood in theory. Therefore, we develop a theoretical framework that connects the two causality notions, while also clearly distinguishing them. The framework describes a composition of quantum operations through feedback loops, and the embedding of the resulting, possibly cyclic information-theoretic structure in an acyclic spacetime structure. Relativistic causality (which forbids superluminal communication) then follows as a graph-theoretic compatibility condition between the two structures. Demonstrating that indefinite causal order (ICO) processes, widely studied in the quantum information community, can be formulated within our framework, we shed light on the links between indefinite and cyclic causality, and on questions regarding their physicality. In particular, there are several experiments that claim to implement ICO processes in Minkowski spacetime, presenting an apparent theoretical paradox: how can an \emph{indefinite} information-theoretic causal structure be consistent with a \emph{definite} spacetime structure? We address this through no-go theorems, showing that as a consequence of relativistic causality, (a) realisations of ICO processes necessarily involve the non-localisation of systems in spacetime and (b) will nevertheless admit an explanation in terms of a definite and acyclic causal order process, at a more fine-grained level. These results are made possible by introducing the concept of fine-graining that allows causal structures to be analysed at different levels of detail. This fully resolves the apparent paradox and bears implications for the physical interpretation of ICO experiments. Our work also sheds light on the limits of quantum information processing in spacetime and offers concrete insights on the operational meaning of indefinite causality, both within and beyond the context of a fixed spacetime.

\end{abstract}

\maketitle

\tableofcontents

\section{Introduction and summary of contributions}
\label{sec: intro}

The notion of \emph{causality}, although fundamental to science, appears in several different forms across scientific disciplines. A defining feature of a \emph{causal structure} is that it specifies an \emph{order relation} between certain \emph{events}, the meaning of the events and the relation varies depending on the notion of causality considered. In relativity theory, causality is a property of the geometry of spacetime, the events are points (or regions) in spacetime, and the causal relation is specified by the metric tensor together with a time direction.\footnote{Considering the fact that the metric tensor can be recovered from the causal structure up to a scaling factor, one can basically identify the causal relation with the spacetime geometry.} We refer to this as \emph{spacetime causality}. In the context of quantum theory, causality is more commonly used to describe how interventions, such as the choices of parameters by an agent (such as an experimenter), influence observations. The events are associated to random variables (or, more generally, quantum Hilbert spaces) and the order relation is specified by information-theoretic channels connecting them, which can generate correlations between them. This enables analyses of the flow of information in quantum communication protocols between multiple agents. We refer to this as \emph{information-theoretic causality}. The information-theoretic approach also aligns with how causality is defined within causal modelling and causal inference approaches originally developed in classical statistics \cite{Pearl2009}, which have found diverse applications across data driven fields such as machine learning, economics and medicine. 

\bigskip

\begin{figure}
    \centering
    \subfloat[\label{fig: compat}]{\includegraphics[scale=1.0]{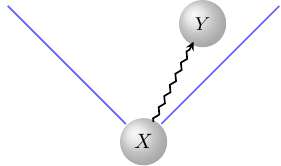}}\qquad\qquad
\subfloat[\label{fig: demand_price_coarse}]{\includegraphics[scale=1.0]{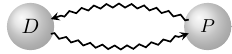}}\qquad\qquad\subfloat[\label{fig: demand_price_fine}]{\includegraphics[scale=1.0]{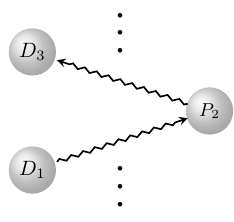}}
 \caption{(a) A simple information-theoretic causal structure, over nodes $X$ and $Y$ embedded in Minkowski spacetime (time along vertical and space along horizontal axis). The 45 degree lines (blue) indicate light like surfaces emanating from the spacetime point where $X$ is embedded, and we see that the information-theoretic causal influence flows within the lightcone and the two causal orders are compatible. (b) A cyclic information-theoretic causal structure over variables $D$ and $P$ depicting the demand and price of a commodity which can mutually influence each other. (c) A fine-grained acyclic causal structure that would physically explain (b), this captures that demand at one time influences price at a later time and so on. Notice that (b) would not be compatible with an acyclic spacetime if we associate individual spacetime locations to $D$ and $P$, but we can obtain the physical fine-grained picture given in (c) if we associate spacetime regions (or sets of spacetime points) to nodes of (b), in which case we can also recover compatibility with the spacetime. This highlights that the physical interpretation of seemingly exotic cyclic information-theoretic causal structures depends on how they are linked to spatio-temporal degrees of freedom.}
 \label{fig: demand_price}
\end{figure}

In physical experiments, information-theoretic structures are \emph{embedded} in spacetime and these two notions of causality play together in a compatible manner. In fact, \emph{relativistic causality principles} such as the impossibility of superluminal signalling impose such compatibility conditions by requiring the direction of information flow to align with the light-cone structure and time direction of the spacetime (see also \cref{fig: compat}), and this has been formalised in general physical theories, in recent works involving one of the authors \cite{VilasiniColbeckPRA, VilasiniColbeckPRL, VilasiniColbeckJamming}. As a consequence of Bell's theorem \cite{Bell1964}, the interplay between spacetime and information-theoretic causality becomes particularly subtle in quantum experiments where freely chosen measurements are performed on entangled quantum systems. The theorem indicates a fundamental incompatibility between classical information-theoretic notions of causality and the spatio-temporal notion of causality \cite{Wiseman2015,Wood2015} in explaining such experiments. 
 In light of this, there has been significant progress in the development of quantum causal modelling frameworks \cite{Tucci_1995,Leifer_2006,Laskey2007,Leifer_2008,Leifer2013,Henson2014,Wood2015,Pienaar2015,Ried_2015,Costa2016,Fritz_2015,Allen2017,Barrett2020A,Pienaar_2020}, which reveal that free choice along with relativistic principles of causality in the background spacetime can be simultaneously preserved if we replace our classical intuitions with a quantum information-theoretic notion of causation. These formalisms provide a faithful and compatible explanation of both notions of causation in quantum experiments where information-theoretic events are localised in an acyclic spacetime.
 
\bigskip
 
However, we can have physical scenarios, even in a classical background spacetime, where quantum systems (associated with the in and outputs of quantum channels) are not localised in space and time (see e.g., \cite{Portmann2017, Chiribella_2019, Rubino_2021}). More generally, in quantum gravitational settings, a fixed background spacetime structure may no longer be available \cite{Hardy2005, Hardy_2007, Zych2019}. This has motivated information-theoretic formalisms for analysing multi-agent information-processing protocols where a definite acyclic ordering of the agents' operations is not assumed \cite{Hardy2005, Chiribella2013, Oreshkov2012}. A prominent example is the process matrix framework \cite{Oreshkov2012}. Process matrices which
are incompatible with a definite acyclic (information-theoretic) causal order between the operations of the agents are regarded as \emph{indefinite causal structures} in this framework (we do not necessarily endorse this terminology, but we will use it in order to remain consistent with the literature. It is to be noted that this refers to an indefinite order of operations in a quantum information protocol, not indefiniteness in the order or geometry relating spatio-temporal events). The theoretical framework encompasses regular quantum circuits, quantum combs \cite{Chiribella2009}, superpositions of direct and common cause scenarios \cite{Feix_2017} as well as more exotic processes such as \emph{causally non-separable processes} and those that violate so-called \emph{causal inequalities} \cite{Oreshkov2012, Araujo2015, Oreshkov2016}\footnote{Analogous to non-separable quantum states and quantum states that violate Bell inequalities.}. These more general classes have been widely studied for the potential advantages that they may provide over regular quantum circuits in various information-theoretic tasks \cite{Guerin2016, Chiribella_2021,Chiribella_2012,Zhao_2020,Araujo2014,Guha_2020, Felce_2020, guha2022}. 

 \bigskip

However, there are several intriguing open questions and debates regarding the physicality of such processes as well as their link to spatio-temporal concepts. A first question is regarding the set of process matrices that can be realised in accordance with standard quantum theory in Minkowski spacetime. Through bottom-up constructions, a general class of quantum circuits realising quantum controlled superpositions of the order (of agents' operations) have been proposed \cite{Wechs2021, Purves2021}, and an important open question is whether these correspond to the largest class of physically realisable processes. On the other hand, numerous table-top experiments have been performed that claim to physically implement an indefinite causal structure (\emph{the quantum switch} \cite{Chiribella2013}) in Minkowski spacetime \cite{Procopio2015, Rubino2017, Goswami2018, Wei_2019, Ho2019, Guo_2020, Goswami_2020, Taddei_2021, Rubino_2021, Felce2021}, and their interpretation has been a subject of much debate in the community \cite{Portmann2017,Vilasini_thesis,Paunkovic2019, Oreshkov2019, Ormrod2022, Kabel2024}. A large class of previous theoretical and experimental results appear to suggest that there is a sense in which certain indefinite causal structure processes can be realised in Minkowski spacetime. A necessary step for clarifying these questions about quantum causality is to address the apparent paradox-- how can an indefinite information-theoretic causal structure be consistent with relativistic causality in a definite and acyclic spacetime causal structure? We develop a general top-down approach for formally addressing such questions and in a manner that accommodates and links different causality notions. Such a general framework must incorporate the following desiderata, we first state them and then justify them. 
\begin{enumerate}
    \item The framework must clearly disentangle the information-theoretic and spacetime notions of causality, and characterise both under minimal but operational assumptions, which means that the former need not in general be acyclic.
    \item To have relevance for physical experiments, the framework must formalise what it entails for a non-acyclic information-theoretic structure to be compatible with relativistic causality in an acyclic spacetime.
    \item The embedding must be general enough to model quantum systems that are not necessarily localised in spacetime.
    \item The framework should permit the analysis of causal structures at different levels of detail, as the acyclicity of a causal structure can depend on the information captured by its nodes.
\end{enumerate}

Both indefinite as well as cyclic causal structures correspond to non-acyclic causal structures. Interestingly, previous works \cite{Oreshkov2012, Chiribella2013,Baumeler_2016, Araujo2017,Barrett2020} have demonstrated mathematical correspondences between the two, which provide very useful insights for achieving the first criterion above. However, these works focus on the details of the information-theoretic notion and do not consider spacetime causality or compatibility as required by the second criterion. 
In recent works involving one of the authors \cite{VilasiniColbeckPRA, VilasiniColbeckPRL}, a framework applicable to quantum as well as post-quantum probabilistic theories, that meets both the first and second criteria has been developed. The present work is inspired by the general approach to information-theoretic and spatio-temporal causal structures introduced there. However, the previous framework mainly focuses on signalling between classical settings and outcomes (even if these may be generated by measuring quantum or post-quantum systems), and the criteria 3 and 4 were not considered there, which are necessary for analysing realisations of indefinite causal order processes. 

\bigskip

There has also been much progress, in particular, coming from the research program relating to \emph{quantum reference frames} (QRF), towards instantiating process matrices with temporal information \cite{Zych2019,Castro_Ruiz_2020, Oreshkov2019,baumann2021}. While these cover parts of points 1, 2 and 3, these approaches typically only describe fully unitarised quantum circuits. Moreover all systems involved, such as the reference frame and the observer whose perspective is described, are modelled as fully quantum systems. QRF approaches can thus also capture theoretical quantum gravitational realisations of process matrices where spacetime geometry is subject to quantum superpositions. However, this is a different physical regime than that of presently realisable quantum experiments, which involve classical observers (our experimental colleagues) who have access to classical reference frames (such as their wall clock) which they can utilise to order and control their actions. Such experiments effectively take place in a well-defined classical background spacetime. This is an important yet often overlooked distinction, bearing implications for the concept of causality and events, and we discuss this further in \cref{sec: discussions}.

\bigskip

To the best of our knowledge, previous approaches do not fully address point 2 specifically in the context of a background spacetime, while accounting for arbitrary measurements along with unitary transformations that can be physically performed by the agents involved.
Point number 4 above, which has not been previously considered, is crucial for this purpose as the following classical example illustrates. Cyclic causal models are often used in classical data sciences for modelling physical scenarios with feedback \cite{Bongers2021}, for instance the demand $D$ and price $P$ for a commodity can causally influence each other. However, we know that the situation is ultimately described by an acyclic causal structure (possibly over a larger number of nodes) where the demand $D_1$ at time $t_1$ influences the price $P_2$ at time $t_2>t_1$ which in turn influences the demand $D_3$ at time $t_3>t_2$ and so on. Indeed agents can verify this by intervening on $D$ or $P$ at some specific time $t$ (such as through appropriate choice of market policies) and observing changes in the other variable at a later time. If we coarse-grain over the time information in this acyclic causal structure, we recover the original cyclic causal structure. This example (illustrated in Figures~\ref{fig: demand_price_coarse} and \ref{fig: demand_price_fine}) highlights that the cyclicity or acyclicity of a causal structure can depend on the level of detailed information encoded in its nodes. 
Therefore, the apparently exotic cyclic (information-theoretic) causal structure has a clear physical interpretation in such scenarios.
Here we develop a framework that meets all the above criteria, which enables a similar clear and operational interpretation for spacetime realisations of so-called indefinite causal order processes.

\bigskip

Within our formalism we derive multiple classes of results: characterisation of purely information-theoretic causal structures and their compatibility with spacetime, no-go theorems for spacetime realisations of indefinite causal structures, examples and results establishing the applicability of our framework to a range of existing approaches to causality. 

\bigskip

{\bf Summary of contributions} We provide a summary of these contributions below, which would enable a reader interested in a subset of the results to navigate to the relevant sections and theorems. For those seeking a concise yet formal introduction to our framework, particularly the methods relevant for understanding our main no-go results for ICOs (\cref{theorem: nogo_main}, \cref{theorem: PMFinegrain} and \cref{corollary: nogoB}), along with an intuitive proof sketch for the theorems, please refer to the associated letter \cite{us_short} and its supplementary materials. In \cref{sec: abstract_causal_str} of this paper, we start with general abstract definitions that apply to both notions of causality and the following paragraphs summarise the contents of the remaining sections. Proofs of all results from the main text can be found in \cref{appendix: proofs}.


\begin{itemize}
    \item {\bf A top-down approach to quantum information-theoretic causation} In \cref{sec: info_th}, we describe causal influences in a purely information-theoretic sense by defining (possibly cyclic) quantum networks that can be formed through compositions of quantum operations by means of feedback loops. We focus on the operationally relevant notion of signalling for such networks, while taking into account that signalling and causation are distinct \cite{Wood2015, Barrett2020, VilasiniColbeckPRA, VilasiniColbeckPRL}. We introduce the concepts of 
    \emph{embedding} and \emph{fine-graining} of quantum networks which enables us to formulate their \emph{compatibility} with abstract graphs (\cref{sec: compat_graph}), and to study different realisations of such networks with the possibility of analysing them at different levels of detail (\cref{sec: finegraining}).

    \item {\bf Compatibility between cyclic quantum networks and acyclic spacetimes} Defining spatio-temporal causal structures in terms of graphs (\cref{sec: spacetime}), we apply the concepts of embedding and compatibility to link the two notions of causality and provide an order-theoretic formalisation of \emph{relativistic causality} (\cref{sec: infoth-spacetime}). In theorem~\ref{theorem: embedding_sig}, we show that any (possibly cyclic) signalling structure can be compatibly embedded in an acyclic spacetime if we allow non-localisation of information in the spacetime.

    \item {\bf Recovering process matrices, linking indefinite and cyclic causality} After reviewing the process matrix formalism (\cref{sec: PM}), we show that process matrix protocols can be recovered as a special case of the general cyclic quantum networks of our framework. In particular, we recover the probability rule of the process formalism through loop composition (\cref{lemma: probabilities}) and derive a tight correspondence between the non-definiteness of causal order in the process formalism and the cyclicity of an underlying causal structure (inferred solely through signalling relations) in our framework (\cref{theorem: indef_cyclic}).

    \item {\bf No-go result 1: non-localisation of information in spacetime}
Our first main no-go theorem (\cref{theorem: nogo_main}) implies that any realisation of an indefinite causal order process which satisfies relativistic causality in a fixed acyclic spacetime must necessarily involve the non-localisation of information in spacetime, while indicating the degree to which it must be non-localised. As a corollary it implies the impossibility of such realisations where the in and output systems of agents' operations are time localised in a global reference frame (\cref{corollary: time}). These results shed light on the physical resources in spacetime which are necessary for realising such quantum processes.
    
    \item {\bf No-go result 2: recovering definite causal structure under fine-graining} Our second set of no-go results (\cref{theorem: PMFinegrain} and \cref{corollary: nogoB}) show that any realisation of an indefinite causal order process which satisfies relativistic causality in a fixed acyclic spacetime will ultimately admit, at a fine-grained level, an explanation in terms of fixed causal order process over a larger number of agents, where the (information-theoretic) causal order is definite and acyclic. This holds even for realisations where agents exchange quantum messages at superpositions of spacetime locations. This resolves the apparent paradox regarding the two causality notions in a fixed spacetime.

    \item {\bf Implications for quantum switch experiments} We consider the special case of the quantum switch (QS, an indefinite causal order process) and derive stronger versions of our two main no-go results for this case (\cref{lemma: nogo_QS} and \cref{corollary: QSExpt}), which imply that existing experimental realisations of this process in Minkowski spacetime can be explained in terms of a definite and acyclic causal order at a fine-grained level. In \cref{sec: QS_Minkowski}, we identify a new realisation of QS in Minkowski spacetime where the in/output systems of each agent are time localised in their own reference frame, a property that was previously observed only in quantum gravitational realisations of QS. However, the conclusions of our no-go theorems also apply to such realisations, and we discuss the physical interpretation of such experiments in detail (\cref{sec:qs_expts}). 

\item {\bf  Operational meaning of indefinite causation in and beyond fixed spacetimes} In \cref{sec: discussions}, we connect the assumptions of our no-go theorem to the set-up assumptions of the process matrix framework which sheds light on the operational meaning of indefinite causation and its witnesses such as causal inequality violations. We also discuss the outlook that our work provides for studying different notions of events, and understanding causality and information-processing beyond the context of a fixed background spacetime.
    
    \item {\bf Relation to other approaches to causality} In the appendices, we show that other distinct approaches to causality can also be studied in our general framework. The results of \cref{appendix: framework_recovers_CB} show that the most general realisations of (cyclic) quantum networks satisfying relativistic causality in a fixed acyclic spacetime correspond to so-called \emph{causal boxes} that were previously proposed for studying relativistic quantum cryptography \cite{Portmann2017, Vilasini_crypto}. \cref{appendix: examples_fg} describes how classical \emph{functional causal models on cyclic graphs} (studied in classical statistics \cite{Bongers2021}) as well as \emph{quantum Bayesian networks} \cite{Henson2014} can be studied in our formalism, while illustrating how the concept of fine-graining introduced here applies to such causal models. Finally, \cref{appendix: relation_to_VC} discusses the relation to recent research \cite{VilasiniColbeckPRA, VilasiniColbeckPRL}, where it was shown that it is theoretically possible to have genuine causal loops between events in Minkowski spacetime, without superluminal signalling. 
    
\end{itemize}

Finally, we refer to \cref{appendix: notation} for an overview of the notations
used in the paper.

\section{Abstract causal structures}
\label{sec: abstract_causal_str}

As mentioned in the introduction, a basic feature  that is common across different notions of causality is that a causal structure specifies an order relation between certain events. We will therefore use the following minimal definition of a causal structure in this paper. This abstract notion will be specified further depending on the flavour of causality (information-theoretic, spatio-temporal etc.) being considered. 

\begin{definition}[Causal structures]
    \label{def: abstract_causal_str}
    A causal structure is any directed graph $\cG$ where $\text{Nodes}(\cG)$ denotes the set of all nodes of $\cG$ or ``events'' of the causal structure $\cG$, and $\text{Edges}(\cG)$ denotes the set of all edges of $\cG$ or ``causal relations'' of $\cG$. We say that $N_1$ is a cause of $N_2$ in $\cG$ if there exists a directed path from the node $N_1$ to the node $N_2$ in $\cG$, and in particular that $N_1$ is a direct cause of $N_2$ if there is a directed edge from $N_1$ to $N_2$ in $\cG$.
\end{definition}

\bigskip

To achieve the main goals laid out in the introduction, in particular to capture the fourth desideratum of being able to analyse causal structures at different levels of detail, we introduce the concept of \emph{fine-graining of causal structures}. The following minimal definition defines fine-graining for directed graphs and applies to both information-theoretic and spacetime causal structures. For the information-theoretic case, our example with demand and price illustrated in Figures~\ref{fig: demand_price_coarse} and \ref{fig: demand_price_fine} paints an intuitive picture of fine-graining, the acyclic causal structure over $D_1$, $P_2$, $D_3$,... in the latter figure is a fine-graining of the cyclic causal structure over $D$ and $P$ in the former figure. For the spatio-temporal case, considering spacetime regions as the nodes of the causal structure, we can obtain a more fine-grained description by splitting the regions into sub-regions or individual spacetime events (cf. \cref{sec: spacetime}). Throughout the paper, we will use $\Sigma(.)$ to denote the powerset of the set in its argument and we say that there is a directed path from a subset $S_1$ to a subset $S_2$ of nodes in a directed graph $\cG$ whenever there exists $N_1\in S_1$ and $N_2\in S_2$, and some nodes $M_1,...,M_k$ such that $\cG$ contains directed edges from $N_1$ to $M_1$, from $M_l$ to $M_{l+1}$ for all $l\in \{1,...,k-1\}$, and from $M_k$ to $N_2$.\footnote{Here, the sets $S_1$ and $S_2$ can have overlap and $N_1$ and $N_2$ can be the same, in which case a directed path from $N_1$ to itself would correspond to a directed cycle.}

\begin{figure}[t!]
    \centering
  \subfloat[\label{figure: FinegrainingA}]{\includegraphics[scale=1.1]{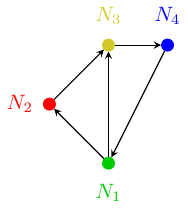}}\qquad \subfloat[\label{figure: FinegrainingB}]{\includegraphics[scale=0.9]{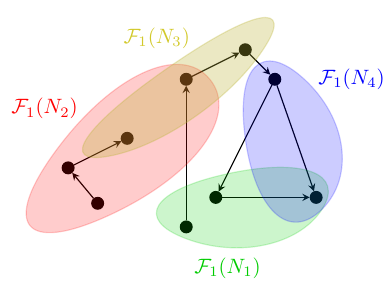}}\qquad  \subfloat[\label{figure: FinegrainingC}]{\includegraphics[scale=0.9]{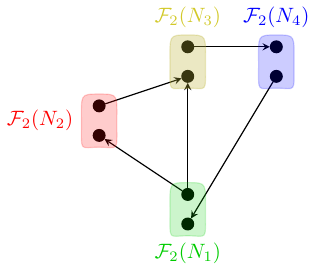}}
    \caption{ Fine-grainings of a directed graph (color online):  A directed graph $\mathcal{G}$ (a) and two possible fine-grainings $\mathcal{G}^{f}_1$ (b) and $\mathcal{G}^{f}_2$ (c) of the same graph, both of which satisfy \cref{definition: fg_graph} but relative to different maps $\cF^{\text{graph}}_1$ and $\cF^{\text{graph}}_2$ (which are denoted as $\cF_1$ and $\cF_2$ to avoid clutter). Each node $N_i$ in the original graph $\mathcal{G}$, with each $i\in\{1,2,3,4\}$ depicted in a distinct color, maps to a set of nodes $\mathcal{F}_j^{\mathrm{graph}}(N_i)$ ($j\in\{1,2\}$) in the fine-grained graph $\mathcal{G}^{f}_j$ lying within a blob of the same color. Notice that (a) is a cyclic graph while its fine-grainings are acyclic. Definition~\ref{definition: fg_graph} is general and allows for overlaps in the image of the fine-graining map, as seen in (b). This will be relevant the spacetime causal structure, as it will allow us to consider overlapping spacetime regions. On the other hand, we will see that the more specific definition of fine-graining for information-theoretic structures will forbid such overlaps as it will map each quantum system in the coarse-grained description to a distinct set of systems in the fine-grained description. }
    \label{fig:finegraining_graph}
\end{figure}

\begin{definition}[Fine-graining of a directed graph]
\label{definition: fg_graph}
A directed graph $\cG^f$ is called a fine-graining of a directed graph $\cG$ if there exists a map $\cF^{\text{graph}}: \text{Nodes}(\cG)\mapsto \Sigma(\text{Nodes}(\cG^f))$ such that whenever there is a directed edge from $N_1$ to $N_2$ in $\cG$ (for $N_1,N_2\in \text{Nodes}(\cG)$), there exists a directed path from $\cF^{\text{graph}}(N_1)$ to $\cF^{\text{graph}}(N_2)$ in $\cG^f$.
\end{definition}

The definition is illustrated in \cref{fig:finegraining_graph}. A further property that one would intuitively expect from a concept of fine-graining is that the number of nodes in the image of $\cF^{\text{graph}}$ is larger than the number of nodes of $\cG$. While this is not required for our main results, it will be the case in all examples we illustrate in this paper (including those of \cref{fig:finegraining_graph}). In the next sections, we will describe how we model general information-theoretic and spatio-temporal causal structures, and introduce more detailed definitions of fine-graining for these.

\section{Information-theoretic causal structures}
\label{sec: info_th}

The operational formulation of quantum theory describes quantum networks (such as circuit diagrams) formed by the composition of quantum operations. This enables information flow from an in/output system of one operation to that of another in the network, allowing for a purely information-theoretic notion of causal influence to be defined. Typically, only combinations of operations resulting in acyclic causal structures are considered such that every network singles out a direction, which we can regard as ``time''. 
More generally, we need not restrict to acyclic information-theoretic causal structures. We often have physical scenarios with feedback where the output of a physical device is looped back and fed in to its input, and in classical statistics, the mathematics of cyclic causal structures has therefore been developed \cite{Bongers2021}. 

\bigskip

Here, we develop a formalism for composing quantum operations through feedback loops that can result in cyclic quantum networks. We will make statements about the underlying causal explanations of a network by focusing on the operational notion of signalling which can be detected through physical interventions. Finally, we introduce the concept of \emph{fine-graining of quantum networks}, which will enable us to relate cyclic and acyclic networks in a manner that preserves the relevant information and to study different realisations of such networks.

\subsection{Cyclic quantum networks}
\begin{notation}
   We consider completely positive linear maps (henceforth, simply referred to as CP maps or CPMs) associated with specified sets of in and output subsystems. For a CPM $\cM$, we denote the sets of in and output subsystems as In$(\cM)$ and Out$(\cM)$ respectively, and only consider scenarios where these are finite sets. Whenever the map $\cM$ is evident from context, we simply refer to these sets as $\text{In}$ and $\text{Out}$. Each element $S\in \text{In}(\cM)\cup\text{Out}(\cM)$ is associated with a Hilbert space $\cH^S$ and a corresponding set $\cL(S)$ of quantum states (or density matrices) on $\cH^S$. More generally, if $\cS$ is a subset of in or output systems, $\cL(\cS):=\bigotimes_{S\in \cS}\cL(S)$ denotes the joint state space of all systems in $\cS$. Moreover, for any set $\{\cM_i\}_i$ of CPMs we denote the union of input systems as $\text{In}(\{\cM_i\}_i):=\bigcup_i \text{In}(\cM_i)$, and similarly for the output systems. We will assume that each Hilbert space $\cH$ of a $d$-dimensional quantum system has a well defined \emph{computational basis} $\{\ket{i}\}_{i\in\{0,...,d-1\}}$ consisting of orthonormal vectors $\ket{i}$, where $d$ is finite. 
\end{notation}


We consider two types of composition operations on CPMs, parallel and loop composition \cite{Portmann2017}, as defined below. 

\begin{definition}[Parallel composition]
Parallel composition of two CPMs $\cM$ and $\cN$ is a CPM $\cM||\cN:=\cM\otimes \cN$, with $ \text{In}(\cM||\cN)=  \text{In}(\cM)\cup\text{In}(\cN)$ and $\text{Out}(\cM||\cN)=\text{Out}(\cM)\cup\text{Out}(\cN)$.
\end{definition}

\begin{definition}[Loop composition \cite{Portmann2017}]
Consider a CPM $\cM$ where $S\in \text{In}(\cM)$ and $S'\in \text{Out}(\cM)$ have isomorphic Hilbert spaces $\cH^{S}\cong \cH^{S'}$. 
Then we can compose the output subsystem $S$ with the input subsystem $S'$ through loop composition to obtain a CPM $\cM^{S \hookrightarrow S'}$, with $ \text{In}(\cM^{S \hookrightarrow S'})=  \text{In}(\cM)\backslash \{S'\}$ and $ \text{Out}(\cM^{S \hookrightarrow S'})=\text{Out}(\cM)\backslash \{S\}$.
The action of $\cM^{S \hookrightarrow S'}$ is given as follows, where we take $\rho^{\text{In}\backslash \{S'\}}$ to be an arbitrary state in $\cL(\text{In}\backslash \{S'\})$,
and treat, for simplicity (and without loss of generality), $S'$ as the last input subsystem.

\begin{equation}
\label{eq: loops}
    \cM^{S \hookrightarrow S'}(\rho^{\text{In}\backslash \{S'\}}):= \sum_{i,j} \bra{i}^{S}\cM\Big(\rho^{\text{In}\backslash \{S'\}}\otimes \ket{i}\bra{j}^{S'}\Big)\ket{j}^{S},
\end{equation}
where the bases appearing in the above equation are taken by convention to be the computational basis of the relevant system. 
\end{definition}

\begin{figure}[t!]
\centering
\includegraphics[scale=1.0]{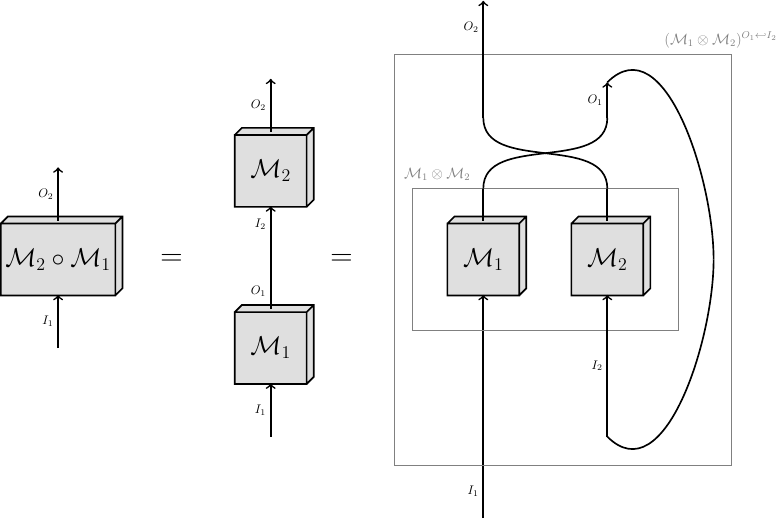}

\caption{Illustration of \cref{equation: sequential}. Sequential composition of CPMs can be equivalently described as a parallel composition followed by loop composition. 
}
\label{fig: sequential}
\end{figure}

In \cref{appendix: framework_loop} we show that the set of CPMs is indeed closed under parallel and loop composition as suggested by the above definitions. 
Now, consider two CPMs $\cM_1$ and $\cM_2$ where $\text{In}(\cM_j)=\{I_j\}$, $\text{Out}(\cM_j)=\{O_j\}$ for $j\in \{1,2\}$ and
 $\cH^{O_1}\cong \cH^{I_2}$. The sequential composition of $\cM_1$ followed by $\cM_2$ is a CPM with inputs $\text{In}(\cM_1)$ and outputs $\text{Out}(\cM_2)$ that acts as $\cM_2(\cM_1(\rho))$. This sequential composition can be equivalently obtained by first composing the two CPMs in parallel and then performing a loop composition on the single CPM obtained through the parallel composition \cite{Portmann2017} i.e.,
 \begin{equation}
     \label{equation: sequential}
     (\cM_1\otimes \cM_2)^{O_1\hookrightarrow I_2}(\rho)= \cM_2(\cM_1(\rho))\quad \forall \rho\in \cL(\text{In}(\cM_1)).
 \end{equation}

This is illustrated in \cref{fig: sequential}. More generally, the maps may have an arbitrary number of in and output subsystems and the sequential composition may take place by connecting one of the output subsystems of $\cM_1$ to one of the input subsystems of $\cM_2$, and this can also be equivalently described as a loop composition on those subsystems in the parallel composition of the two maps. We are now ready to define what we mean by a network of CPMs.


\begin{definition}[Network of CPMs ] A network $\mathfrak{N}:=(\mathfrak{N}^{\text{maps}},\mathfrak{N}^{\text{comp}})$ of CPMs is specified by two elements: a set $\mathfrak{N}^{\text{maps}}$ of CPMs and a set $\mathfrak{N}^{\text{comp}}$ of connections between the CPMs in $\mathfrak{N}^{\text{maps}}$ given through loop compositions. More specifically, $\mathfrak{N}^{\text{comp}}$ is a set of the form 

\begin{equation}
  \mathfrak{N}^{\text{comp}}  =\{S\hookrightarrow S'|S\in \text{Out}(\mathfrak{N}^{\text{maps}}), S'\in \text{In}(\mathfrak{N}^{\text{maps}}), \cH^{S}\cong \cH^{S'}\}, 
\end{equation}
where no output $S$ is paired with more than one input $S'$ and vice versa. 
\end{definition}

\begin{definition}[Induced map of a network]
    The induced map $\cN$ of a network $\mathfrak{N}=(\mathfrak{N}^{\text{maps}},\mathfrak{N}^{\text{comp}})$ is defined by composition of the maps in $\mathfrak{N}^{\text{maps}}$ according to  $\mathfrak{N}^{\text{comp}}$ i.e.,

\begin{equation}
\label{eq: network}
    \cN:=\Big(\bigotimes_{\cM\in \mathfrak{N}^{\text{maps}}} \cM\Big)^{\{S\hookrightarrow S'\in \mathfrak{N}^{\text{comp}}\}}
\end{equation}

\end{definition}
This definition allows for possibly cyclic networks where there can be a directed path from an output of $\cM_1$ to an input of $\cM_2$ and a directed path from an output of $\cM_2$ to an input of $\cM_1$ (see \cref{fig: eg1_map} for an example). Moreover, in \cref{appendix: framework_loop}, we show that given a CPM and multiple loop compositions to be performed, the loop composed CPM is well-defined independently of the order in which the compositions are performed. This implies that the induced map $\cN$ is itself a CPM  and it is uniquely defined by the set of maps $\mathfrak{N}^{\text{maps}}$ and the set of loop compositions $\mathfrak{N}^{\text{comp}}$. The relevant set of systems in a network is then defined as follows.

\begin{definition}[Systems in a network]
The set of systems in a network  $\mathfrak{N}$ is defined as
\begin{equation}
\mathfrak{N}^{\text{sys}}:=\text{In}(\cN)\cup\text{Out}(\cN)\cup \Big(\bigcup\limits_{S\hookrightarrow S'\in \mathfrak{N}^{\text{comp}}} S\Big).
\end{equation}

\end{definition}
That is, $\mathfrak{N}^{\text{sys}}$ consists of all systems in the network which are not involved in a composition (which are precisely the systems in $\text{In}(\cN)\cup\text{Out}(\cN)$), along with a single system $S$ representing each pair $S$ and $S'$ of systems composed through loop composition.


\begin{definition}[Sub-networks]
A sub-network $\mathfrak{N}_{\text{sub}}=(\mathfrak{N}^{\text{maps}}_{\text{sub}},\mathfrak{N}^{\text{comp}}_{\text{sub}})$ of a network $\mathfrak{N}=(\mathfrak{N}^{\text{maps}},\mathfrak{N}^{\text{comp}})$, is itself a network where $\mathfrak{N}_{\text{sub}}^{\text{maps}}\subseteq \mathfrak{N}^{\text{maps}}$ and  $\mathfrak{N}_{\text{sub}}^{\text{comp}}\subseteq \mathfrak{N}^{\text{comp}}$. 
\end{definition}
The induced map $\cN_{\text{sub}}$ of any sub-network is obtained by applying \cref{eq: network} to $\mathfrak{N}_{\text{sub}}$. Moreover, note that every network $\mathfrak{N}$ is a sub-network of itself, and for every map $\cM\in \mathfrak{N}^{\text{maps}}$, $(\{\cM\}, \emptyset)$ is a sub-network of $\mathfrak{N}$ with induced map $\cM$. Of particular relevance will be networks of completely positive and trace preserving maps (CPTPMs), defined below.


\begin{definition}[Network of CPTPMs]
    A network $\mathfrak{N}$ of CPMs is called a network of CPTPMs if the induced map $\cN_{\text{sub}}$ of every sub-network $\mathfrak{N}_{\text{sub}}$ of $\mathfrak{N}$ is a CPTPM.
\end{definition}

\begin{remark}
Simply requiring each $\cM\in \mathfrak{N}^{maps}$ to be a CPTPM does not guarantee that $\mathfrak{N}$ is a network of CPTPMs. This is because the TP property is not always preserved by loop compositions. As we expect physical dynamics of any system under consideration, as well as of any subsystem thereof to be CP and TP, networks of CPTPMs will model examples of physical interest. As we will show later in the paper, indefinite causal order processes \cite{Oreshkov2012, Chiribella2013} can be modelled using such networks of CPTPMs (\cref{sec: PM_network}). An interesting question for future work is to characterise the conditions on the maps and compositions of a network that would ensure that it is a network of CPTPMs. Closely related open questions are: which is the largest subset of CPTPMs that is closed under arbitrary loop compositions, and given a set of CPTPMs, which is the largest set of loop compositions which preserve the TP property?  
\end{remark}

\begin{remark}
 A related framework is the category-theoretic approach to quantum information \cite{abramsky2007, abramsky2008}, which can also describe
cyclic information-theoretic causal structures 
    from a general class of operational theories. The formalism is more abstract than ours, which focuses on quantum theory to derive more specific results for cyclic quantum causal structures. The two are however related, in particular, our formalism of cyclic quantum networks would correspond to an instance of a compact closed category. This is because loop composition as defined here can be equivalently described through pre and post-selection on maximally entangled states (i.e., in terms of post-selected teleportation, see \cref{appendix: framework_loop}), which would be instances of the ``cups'' and ``caps'' of such a category. 
\end{remark}

\subsection{Signalling and causal structures of a network}
\label{sec:signalling}

\begin{figure}[t!]
\centering
\includegraphics[scale=1.0]{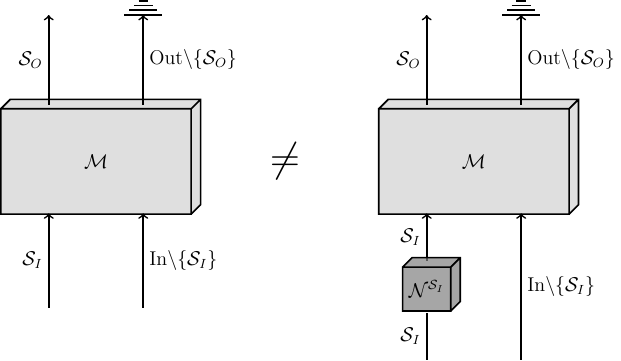}
\caption{Diagrammatic representation of \cref{definition: signalling} of signalling. Here $\cS_I$ and $\cS_O$ represent a subset of input and out subsystems of the CPM $\cM$, and the ground symbol \ground\ denotes trace.
}
\label{figure: signalling}
\end{figure}

\begin{definition}[Signalling structure of a CPM]
\label{definition: signalling}
We say that there is a \emph{signalling relation} from a subset $\cS_I\subseteq \text{In}(\cM)$ of input systems to a subset $\cS_O\subseteq \text{Out}(\cM)$ of output systems of the map $\cM$ and denote it as $\cS_I\longrightarrow \cS_O$ if there exists a CPTPM $\cN^{\cS_I}:\cL(S_I)\mapsto \cL(\cS_I)$ acting locally on $\cS_I$ such that the following holds.
\begin{equation}
    \label{eq: signalling}
 \tr_{\text{Out}\backslash \cS_O} \circ\cM\neq  \tr_{\text{Out}\backslash \cS_O} \circ\cM\circ \cN^{\cS_I}.
\end{equation}
 The set of all signalling relations of $\cM$ forms a directed graph $\cG_{\cM}^{\text{sig}}$ whose
nodes are elements of the powerset (or sigma algebra) $\Sigma(\text{In}\cup\text{Out})$ of $\text{In}\cup\text{Out}$, with a directed edge $\longrightarrow$ between two nodes whenever there is a signalling relation between those subsets. We refer to this graph as the signalling structure of $\cM$. 
\end{definition}

Signalling from $\cS_I$ to $\cS_O$ captures the idea that an agent with access to the inputs can communicate information to an agent with access to the outputs $\cS_O$ by a suitable choice of local intervention on $\cS_I$ (as illustrated in \cref{figure: signalling}). Notice that the definition applies to CPMs $\cM$ but the local operation $\cN^{\cS_I}$ must be a CPTPM, otherwise, we could choose $\cN^{\cS_I}$ to be a constant factor times a CPTPM which would make two sides of \cref{eq: signalling} trivially differ, even in maps $\cM$ where no physical operation could enable information flow or signalling from the inputs to the outputs. Thus the definition only holds operational meaning when the local operation $\cN^{\cS_I}$ are also TP. For our main results, we will only need to apply the definition to cases where $\cM$ is also a CPTPM, in which case, the above definition of signalling is equivalent to other natural ways in which one might formalise the ability for $\cS_I$ to transmit information to $\cS_O$, as shown in \cite{Schumacher2005, Ormrod2022}. We use this formulation as our primary definition of signalling, as it encodes more explicitly how agents can communicate to each other through local interventions on nodes of the network.


\begin{definition}[Signalling structure of a network of CPMs]
The signalling structure $\cG^{\text{sig}}_{\mathfrak{N}}$ of a network $\mathfrak{N}$ is a directed graph whose nodes are elements of the powerset $\Sigma(\mathfrak{N}^{\text{sys}})$ of $\mathfrak{N}^{\text{sys}}$.
We have $N_1 \longrightarrow N_2$ in
$\cG^{\text{sig}}_{\mathfrak{N}}$ (where $N_1$ and $N_2$ are disjoint subsets of $\mathfrak{N}^{\text{sys}}$) if and only if there exists a sub-network $\mathfrak{N}_{\text{sub}}$ of $\mathfrak{N}$ whose induced map $\cN_{\text{sub}}$ enables signalling from $N_1'$ to $N_2$. Here, $N_1'$ is equivalent to $N_1$ up to replacing every $S\in N_1$ with $S'$  whenever $S\hookrightarrow S'\in \mathfrak{N}^{\text{comp}}$.
\end{definition}

The above substitution of $N_1$ with $N_1'$ when checking for signalling relations is to account for the identification of loop composed systems $S\hookrightarrow S'\in \mathfrak{N}^{\text{comp}}$ as a single system $S$.
For example, consider $\cM_1:\cL(I)\mapsto \cL(S)$ and $\cM_2: \cL(S')\mapsto \cL(O)$ which are sequentially composed by connecting the output $S$ of $\cM_1$ to the input $S'$ of $\cM_2$ resulting in a network $\mathfrak{N}$ with an induced map $\cN:\cL(I)\mapsto \cL(O)$. Then $\mathfrak{N}^{\text{sys}}:=\{I,O,S\}$ and we have a signalling relation $\{S\}\longrightarrow \{O\}$ in $\cG_{\mathfrak{N}}^{\text{sig}}$, whenever $\{S'\}\longrightarrow \{O\}$ in the sub-network whose induced map is $\cM_2$.

\begin{notation}
    Stochastic maps and deterministic functions on random variables (RV) can be viewed as special cases as CPTPMs, by encoding a RV in the computational basis of a Hilbert space. Whenever we consider purely classical examples, we will use the same labels to refer to RV as well as the quantum state space which encodes the RV.
\end{notation}

\paragraph{Causal structure(s) of a network} So far, we have focused on the concept of signalling, which can be operationally determined in any quantum network and we have treated causal structures only as abstract graphs. We have not referred to the concept of information-theoretic causal structures that can be associated with a quantum network, although many such definitions exist in the literature. Most of our main results including the general no go theorems of this paper will only rely on the concept of signalling. This allows our framework and results to be generally applicable to different information-theoretic notions of causation that have been proposed in the literature, all of which agree on the definition of signalling. 
We will therefore not provide a full definition of information-theoretic causality here. Rather, 
we provide some intuition on how causality for a quantum network is typically modelled in information-theoretic approaches, through concrete physical examples. This will also help the reader appreciate the distinction between signalling and causation, even without a fully general definition of the latter.

\bigskip

As we have seen, the nodes of the signalling structure of a network are generally subsets of systems in the network. On the other hand, a network of maps also induces possible directed graphs on individual systems of the network, through its connectivity, which captures how information flows through systems in the network. A quantum circuit is an example of a network, and the wires in the circuit diagram tell us about the information flow. For example, in simple network involving sequential composition of quantum channels $\cM_1: A \mapsto B$ and $\cM_2: B \mapsto C$ through the $B$ system, suggests that information can flow from $A$ to $B$ to $C$\footnote{Whether it does flow depends on the maps, e.g., if $\cM_1$ is a trace and replace operation, information cannot flow from $A$ to $B$}, but information cannot flow from $C$ to $A$ in this network. One would say that $C$ is not a cause of $A$ in any information-theoretic causal structure that explains this network (but $A$ can be a cause of $B$ for instance). In this simple example, it is clear that even without fixing the graph, we know that it must be acyclic: this network admits an explanation in terms of an acyclic information-theoretic causal structure. Notice that here, knowing that $A$ signals to $B$ in $\cM_1$ would tell us that information does flow from $A$ to $B$, hence the signalling structure tells us something about the causal structure. However, generally the signalling structure of a network (even a single CPTP map) need not fix the causal structure uniquely: we can have wires in an internal decomposition of a network which may nevertheless not lead to observable signalling at the level of the input/output behaviour of the network. The following example illustrates this fact.

\bigskip

\begin{figure}[t!]
    \centering
    \subfloat[]{\includegraphics[scale=1.0]{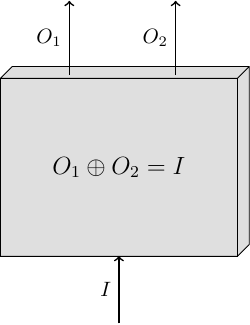}}\qquad\qquad\subfloat[ \label{fig:jamming_example1}]{\includegraphics[scale=1.0]{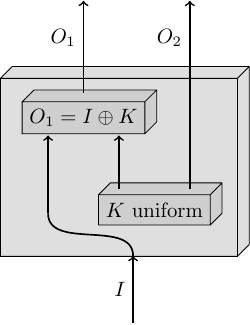}}\qquad\qquad\subfloat[ \label{fig:jamming_example2}]{\includegraphics[scale=1.0]{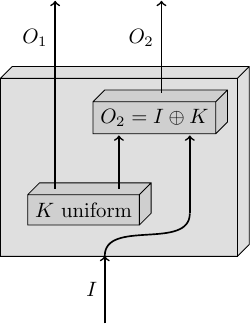}}
    \caption{(a) A classical stochastic map acting on bits (a special case of a CPTP map) where the XOR of the two output bits $O_1$ and $O_2$ equals the input bit $I$ (b) One possible decomposition of this map where $I$ is a cause of $O_1$ but not of $O_2$ (c) Another possible decomposition of the same map where $I$ is not a cause of $O_1$ but is a cause of $O_2$.}
    \label{fig:jamming_example}
\end{figure}

Consider a CPTPM $\cM$ with input $I$ and two outputs $O_1$ and $O_2$, all of which correspond to classical bits such that $\cM$ is effectively a stochastic map fully described by a conditional distribution $P(O_1O_2|I)$ (see \cref{fig:jamming_example}). Suppose that $P(O_1O_2|I)$ is such that $O_1\oplus O_2=I$ always holds and $P(O_1|I)$ and $P(O_2|I)$ are both uniform distributions, independently of $I$. This would imply that $\{I\}$ does not signal to $\{O_1\}$ and $\{I\}$ does not signal to $\{O_2\}$ in $\cM$, however $\{I\}$ signals to $\{O_1,O_2\}$ in $\cM$. Now consider the following two possible ways of implementing this map. One way is the following: $\cM$ generates a uniform bit $K$ internally and sets $O_2=K$ and $O_1=I\oplus K$. We can then easily verify that this realises the behaviour $P(O_1O_2|I)$ of the channel as described above \cite{VilasiniColbeckPRA, VilasiniColbeckPRL}. Another way is to exchange the roles of $O_1$ and $O_2$ and set $O_1=K$ and $O_2=I\oplus K$ which also implements the same channel. If we analyse the internal connections/flow of information in the two implementations, we would say that $I$ causally influences $O_1$ (since $O_1$ is a non-trivial function of $I$) but not of $O_2$ (which is not a function of $I$) in the first while $I$ causally influences $O_2$ but not $O_1$ in the second. This example also highlights that causation does not imply signalling \cite{Barrett2020A, VilasiniColbeckPRA, VilasiniColbeckPRL}, in both implementations, we have that $I$ is a cause of either $O_1$ or $O_2$ but in both cases $\{I\}$ signals to neither $\{O_1\}$ nor to $\{O_2\}$.


\bigskip

In summary, signalling is a property of an information-theoretic network that can be uniquely determined by an operational procedure, even when given black-box access to the network. Causal structure on the other hand, is related to particular internal decompositions of the maps in a network, and is not unique to a network. Given a particular decomposition of the maps in a network, we can infer an absence of causation from one system $S$ to another system $S'$ whenever there is an absence of a directed path of wires from $S$ to $S'$ in the decomposition. Therefore, in some situations, knowing that there exists a decomposition of a given network with the absence of directed paths between certain systems would be sufficient to conclude that the network admits an explanation in terms of an acyclic information-theoretic causal structure (even when that causal structure is not uniquely determined). As stated before, for the results of this paper, we will only require the signalling structure of the network for statements about general unknown networks. The signalling structure can nevertheless allow us to infer something about the underlying causal structure(s) of the network, and we will use this when discussing specific examples of known networks/decompositions. For further details on how information-theoretic causation can be defined, we refer the reader to the literature on quantum causality (e.g., \cite{Henson2014, Costa2016, Allen2017, Barrett2020A, Lorenz2021, Perinotti2021}, section 2.5 of the thesis \cite{Vilasini_thesis} and references therein).

\begin{notation}[Signalling vs causation]
We will use different types of arrows to distinguish between signalling and information-theoretic causation. In consistency with notation from a previous work \cite{VilasiniColbeckPRA}, $\mathcal{S}$ signals to $\mathcal{S}'$ is denoted $\cS \longrightarrow \cS'$ (where $\cS$ and $\cS'$ are sets of systems) and $S$ is a direct (information-theoretic) cause of $S'$ will be denoted as $S \longrsquigarrow S'$ (where $S$ and $S'$ are individual systems). 
For instance, in \cref{fig:jamming_example1}, we have $I \longrsquigarrow O_1$ but $\{I\}\not\longrightarrow \{O_1\}$, which highlights the distinctions in the edges and the nodes, explicitly in the notation. 
\end{notation}

\subsection{Compatibility with an abstract causal structure}
\label{sec: compat_graph}

The following general definitions will allow us to consider whether a signalling structure (operationally obtained from a possibly unknown network of maps) is compatible with different notions of causality. 

\begin{definition}[Embedding systems in a causal structure]
\label{definition: embedding_general}
An embedding $\cE$ of a set $\cS$ of systems in a causal structure $\cG$ is a mapping $\cE: \cS\mapsto \text{Nodes}(\cG)$ from elements of $\cS$ to nodes of $\cG$. For each system $S\in\cS$, we will use $S^{\cE(S)}$ to denote the system embedded on the node $\cE(S)$ of $\cG$, and refer to $S^{\cE(S)}$ as the $\cG$-embedded system. 

\end{definition}

\begin{definition}[Embedding networks of CPMs in a causal structure]
\label{definition: network_embedding_graph}
An embedding of a network $\mathfrak{N}$ of CPMs in a causal structure $\cG$ with respect to an embedding $\cE$ of the systems $\cS:=\mathfrak{N}^{\text{sys}}$ in $\cG$ is a network $\mathfrak{N}_{\cG,\cE}$ that is equivalent to $\mathfrak{N}$ up to a relabelling of each system $S\in \cS$ to the corresponding $\cG$-embedded system $S^{\cE(S)}$. Further, for every $S\hookrightarrow S'\in \mathfrak{N}^{\text{comp}}$, we require that the embedding satisfies $\cE(S)=\cE(S')$.
\end{definition}

\begin{definition}[Compatibility of a signalling structure with a causal structure]
\label{definition: graph_compatibility}

We say that a signalling structure $\cG^{\text{sig}}$ over a set $\cS$ of systems is compatible with an embedding $\cE$ in a causal structure $\cG$ if the following holds.

 \begin{center}
    $\cS_1\longrightarrow \cS_2$ in $\cG^{\text{sig}}$
 \end{center}
 \begin{center}
     $\Downarrow$
 \end{center}
 \begin{center}
     $\exists S_1\in \cS_1, S_2\in\cS_2$ such that there is a directed path from $\cE(S_1)$ to $\cE(S_2)$ in $\cG$.
 \end{center}
\end{definition}

Once we embed a network in a causal structure by associating systems in the network to nodes in the causal structure, then compatibility allows us to infer a causal influence from a node associated with some $S_1\in\cS_1$ to a node associated with some $S_2\in\cS_2$ whenever $\cS_1$ signals to $\cS_2$, or equivalently, compatibility ensures that in the absence of any causal influences from the nodes associated with $\cS_1$ to those associated with $\cS_2$, we cannot have signalling from $\cS_1$ to $\cS_2$. For example, in the case of information-theoretic causal structures of CPTP maps, a decomposition such as \cref{fig:jamming_example1} where there is no causal influence (in this context, no internal wires) from an input system $I$ to an output system $O_2$ implies (as we can easily check) that $I$ does not signal to $O_1$ in that decomposition. In the spatio-temporal case, if we have two systems $S_I$ and $S_O$ of a network embedded at space-like separated locations in a spacetime, there are no directed paths between the corresponding locations in the spatio-temporal causal structure and compatibility would require that $S_I$ does not signal to $S_O$ in the network. In \cref{sec: spacetime}, we will apply these concepts to generally define relativistic causality in a spacetime.

\begin{remark}
    A compatibility condition between cyclic information-theoretic causal models and acyclic graphs, for the purpose of providing a general formulation of relativistic causality principles, was also proposed in recent works \cite{VilasiniColbeckPRA, VilasiniColbeckPRL} (in the context of general theories, not necessarily restricted to quantum theory). The relationships and differences between the current formalism and that of \cite{VilasiniColbeckPRA, VilasiniColbeckPRL} are discussed in \cref{appendix: relation_to_VC}. 
\end{remark}

\subsection{Fine-graining of quantum networks}
\label{sec: finegraining}

\begin{figure}[t!]
    \centering
 \subfloat[\label{fig: eg1_map}]{\includegraphics[scale=1.0]{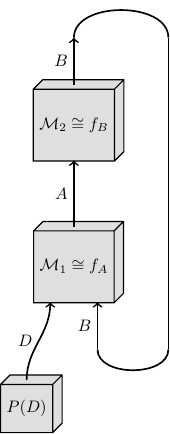}}\qquad\qquad\qquad\qquad \subfloat[\label{fig: eg1_map_fg}]{\includegraphics[scale=1.0]{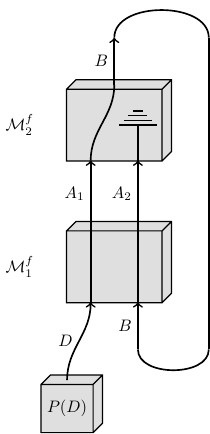}}

\vspace{7mm}
    
\subfloat[\label{fig: eg1}]{\includegraphics[scale=1.0]{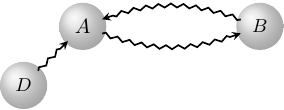}}\qquad\qquad\qquad\subfloat[\label{fig: eg1_fg}]{\includegraphics[scale=1.0]{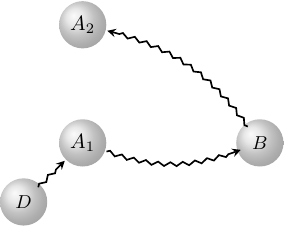}}

    \caption{(a) is a cyclic network which is described in \cref{example: fine-graining1} (in \cref{appendix: examples_fg}). The network can be uniquely associated with a cyclic causal structure (c). The network (a) can be fine-grained to an acyclic network shown in (b) which can be uniquely associated with the acyclic causal structure of (d). The causal structure (d) is then a fine-graining of the causal structure of (c). In this example, all systems are classical and the maps $\cM_1$ and $\cM_2$ implement the deterministic functions $f_A$ and $f_B$ which are described in the text of \cref{example: fine-graining1}. More generally, quantum networks with cyclic causal structures, and the associated fine-grained networks can also be modelled in our formalism.}
    \label{fig:fg_example1}
\end{figure}

The general formalism laid out thus far allows for networks which can be associated with cyclic information-theoretic causal structures. For instance, imagine a network consisting of two maps $\cM_1: A\mapsto B$ and $\cM_2: B \mapsto A$ with $\cH^A\cong \cH^B$, such that we compose the in/output systems with the same label to obtain a causal structure with $A\longrsquigarrow B$ and $B\longrsquigarrow A$ as illustrated in \cref{fig: eg1_map} and \cref{fig: eg1}\footnote{We would have this causal structure whenever both maps allow signalling from their single input to single output system, which is taken to be the case in this example}. Upon a closer inspection, it may turn out that we can encode the information about this network in a new network where the information about $A$ is now encoded in two systems $A_1$ and $A_2$ such that the causal structure over these ``fine-grained'' systems is acyclic, with $A_1\longrsquigarrow B$ and $B\longrsquigarrow A_2$ as shown in \cref{fig: eg1_fg}. 

\bigskip

In \cref{appendix: examples_fg}, we provide illustrative examples of the concept of fine-graining of cyclic networks and their causal structures. The examples highlight (at least) two distinct physical ways in which fine-graining becomes relevant: namely \emph{fine-graining by splitting into subsystems} and \emph{fine-graining through uncertainty in location (relative to an abstract causal structure)}. The former case corresponds to a scenario where a cyclic network involving a 4-dimensional system $A$, when equivalently representing $A$ in terms of two 2-dimensional subsystems $A_1$ and $A_2$, turns out to be an acyclic network over the subsystems and can therefore be explained by an acyclic fine-grained causal structure. 
The latter models a situation where the information content of $A$ is encoded either into a system $A_1$ or a system $A_2$, where $A_1$ and $A_2$ are some systems (of any dimension) embedded in an abstract causal structure.
This would correspond to a scenario where the information content of $A$ is no longer localised in the abstract causal structure, this kind of non-localisation of information can also be due to purely classical uncertainty as illustrated by \cref{example: fine-graining2} in \cref{appendix: examples_fg}, or it may be quantum, as will be the case in spacetime realisations of certain processes such as the \emph{quantum switch} \cite{Chiribella2013}, which we will discuss later in the paper (\cref{sec: QS}). In the following, we define the concept of fine-graining more rigorously and generally for arbitrary cyclic networks.

\begin{definition}[Fine-graining of systems]
We say that a set $\cS^f$ of systems is a fine-graining of a set $\cS$ of systems if there exists an injective map $\cF:\cS\mapsto \Sigma(\cS^f)$ that maps each system $S\in \cS$ to a subset of systems $\cF(S)\subseteq \cS^f$ such that $\cF(S)\cap \cF(S')=\emptyset$ whenever $S\neq S'$. We refer to such a map $\cF$ as a systems fine-graining. 
\end{definition}

  \begin{sloppypar} 
\begin{definition}[Fine-graining of a CPM]
\label{definition: fg_map}
Consider two CPMs $\cM$ and $\cM^f$ whose in and output systems are related by a systems fine-graining $\cF:\text{In}(\cM)\cup\text{Out}(\cM)\mapsto \Sigma(\text{In}(\cM^f)\cup\text{Out}(\cM^f))$ such that $\cF(S')\subseteq \text{In}(\cM^f)$ for all $S'\in \text{In}(\cM)$ and $\cF(S)\subseteq \text{Out}(\cM^f)$ for all $S\in \text{Out}(\cM)$. Then the CPM $\cM^f$ is called a fine-graining of the CPM $\cM$ with respect to the system fine-graining $\cF$ if there exist a pair of CPTPMs $\mathrm{Enc}:\cL(\text{In}(\cM))\mapsto \cL(\text{In}(\cM^f))$ and $\mathrm{Dec}: \cL(\text{Out}(\cM^f))^*\mapsto \cL(\text{Out}(\cM))$, which we refer to as the \emph{encoder} and the \emph{decoder} respectively, such that the following properties are satisfied. Here, $\cL(\text{Out}(\cM^f))^*$ is the subspace of $\cL(\text{Out}(\cM^f))$ defined by the image of $\cM^f\circ\mathrm{Enc}$. 
\begin{enumerate}

 \item {\bf Recovery of $\cM$ from $\cM^f$:} The CPM $\cM$ can be recovered from the CPM $\cM^f$ through the encoding and decoding procedure i.e.,   \begin{equation}
 \label{eq:fg_cond1}
        \cM=\mathrm{Dec}\circ\cM^f\circ \mathrm{Enc}
    \end{equation}
    
    \item {\bf Preservation of signalling relations: } For every directed edge $\cS_1\longrightarrow \cS_2$ in the signalling structure $\cG^{\text{sig}}_{\cM}$ of $\cM$, there exists a corresponding directed edge $\cF(\cS_1)\longrightarrow \cF(\cS_2)$ in the signalling structure $\cG^{\text{sig}}_{\cM^f}$ of $\cM^f$, where $\cF(\cS_i):=\bigcup_{S\in \cS_i}\cF(S)$, for $i\in \{1,2\}$. 

\end{enumerate}
\end{definition}
\end{sloppypar}

  \begin{figure}[t!]
        \centering
\includegraphics[scale=1.0]{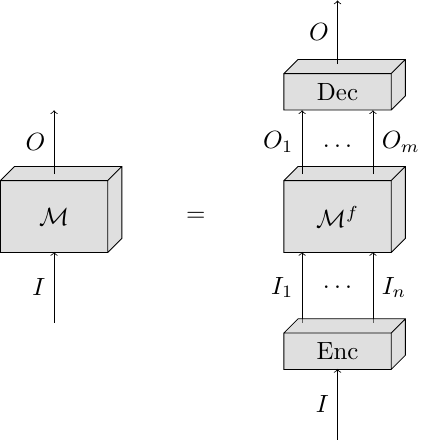}
\caption{Illustration of the first condition of \cref{definition: fg_map}, which requires that $\cM$ can be recovered from $\cM^f$ through an encoding-decoding scheme. We have illustrated the simplified case where $\cM$ has a single input system $I$ and a single output system $O$, but \cref{definition: fg_map} also applies to $\cM$ with arbitrary number of in and output subsystems. We can regard $\cM^f$ as a fine-graining of $\cM$ if this condition is satisfied, and the signalling relations of $\cM$ are preserved in $\cM^f$, for the given map $\cM$, the latter condition translates to he requirement that $\{I\}\longrightarrow \{O\}$ in $\cM$ implies $\{I_1,...,I_n\}\longrightarrow \{O_1,...,O_m\}$ in $\cM^f$. }
\label{fig: fine-graining_CPM}
    \end{figure}

See \cref{fig: fine-graining_CPM} for an illustration of \cref{eq:fg_cond1}.
The above definition only requires that any signalling properties of the coarse-grained (original) map $\cM$ must be replicated in the fine-grained map $\cM^f$, but not the converse. Indeed, the fine-grained description can have additional structure that can reveal itself at the level of the individual fine-grained systems, that may not be reproduced when grouping them together as a single coarse-grained system. This is illustrated in the following simple example. 

\begin{example}[``Coarse-graining'' does not preserve signalling relations]
\label{example: fine-graining_sig}
Let $\cM$ correspond to a classical stochastic map from an input bit $I$ to an output bit $O$ that discards the input $I$ and deterministically prepares $O=0$ as the output. Then we immediately have $\{I\}\not\longrightarrow \{O\}$ in $\cM$. We now construct a fine-graining of $\cM$ where $\cF(I)\longrightarrow \cF(O)$ holds. Consider the map $\cM^f$ with the input systems $I_1,I_2$ and output systems $O_1,O_2$ (all of which correspond to classical bits) acting as $\cM^f: (I_1,I_2)\mapsto (O_1=I_1\oplus I_2,O_2=I_1\oplus I_2)$ (where $\oplus$ denotes modulo-2 addition). Consider the systems fine-graining $\mathcal{F}(I)=\{I_1,I_2\}$ and $\mathcal{F}(O)=\{O_1,O_2\}$, an encoder that maps
 the input value $I=0$ randomly to one of the values $(I_1,I_2)\in\{(0,0),(1,1)\}$ and $I=1$ to one of the values in $(I_1,I_2)\in\{(0,1),(1,0)\}$\footnote{Note that the encoder has a single input $I$ and two outputs $I_1$, $I_2$ such that $I_1\oplus I_2=I$ i.e., it is precisely the map depicted in \cref{fig:jamming_example} with $I_1$ and $I_2$ playing the role of $O_1$ and $O_2$.}, and a decoder that
 maps the values $(O_1,O_2)\in\{(0,0),(1,1)\}$ to $O=0$ and $(O_1,O_2)\in\{(0,1),(1,0)\}$ to $O=1$ i.e., $O=O_1\oplus O_2$. In other words, the coarse grained variables encode the parity of the corresponding two fine-grained variables. We can see that $\cM^f$ is indeed a fine-graining of $\cM$ relative to these encoders and decoders. It satisfies \cref{eq:fg_cond1} since $\cM$ maps every $I\in\{0,1\}$ to $O=0$ while $\cM^f$ maps every input to outputs that satisfy $O_1=O_2$ which will always be  decoded to $O=0$. It trivially satisfies the second condition of \cref{definition: fg_map} since $\cM$ has no non-trivial signalling relations to be preserved. However, we can see that $\{I_1,I_2\}\longrightarrow \{O_1,O_2\}$ in $\cM^f$ since $\cM^f(I_1=0,I_2=0)\neq \cM^f(I_1=0,I_2=1)$.
\end{example}
\begin{definition}[Fine-graining of a network of CPMs]
\label{definition: fg_network}
Let $\mathfrak{N}$ and $\mathfrak{N}^f$ be two networks of CPMs such that there exists a systems fine-graining $\cF: \mathfrak{N}^{\text{sys}}\mapsto \Sigma(\mathfrak{N}^{f,\text{sys}})$. Then the network $\mathfrak{N}^f$ is called a fine-graining of the network $\mathfrak{N}$ with respect to $\cF$ if for every sub-network $\mathfrak{N}_{\text{sub}}$ of $\mathfrak{N}$, there exists a unique sub-network $\mathfrak{N}_{\text{sub}}^f$ of $\mathfrak{N}^f$ such that the induced CPM $\cN_{\text{sub}}^f$ is a fine-graining of the induced CPM $\cN_{\text{sub}}$, with respect to the systems fine-graining $\cF$. 


\end{definition}

With these definitions, the following lemma immediately follows.

\begin{restatable}{lemma}[Fine-graining a network preserves its signalling relations]
\label{lemma: fg_network_sig}
Let the network $\mathfrak{N}^f$ be a fine-graining of the network $\mathfrak{N}$ with respect to a systems fine-graining $\cF$. Then for every directed edge $\cS_1\longrightarrow \cS_2$ in the signalling structure $\cG^{\text{sig}}_{\mathfrak{N}}$ of $\mathfrak{N}$, there exists a corresponding directed edge $\cF(\cS_1)\longrightarrow \cF(\cS_2)$ in the signalling structure $\cG^{\text{sig}}_{\mathfrak{N}^f}$ of $\mathfrak{N}^f$, where $\cF(\cS_i):=\bigcup_{S\in \cS_i}\cF(S)$ for $i\in \{1,2\}$. 
\end{restatable}
This lemma follows from the above definitions because the signalling structure of a network $\mathfrak{N}$ is obtained from the signalling relations present in the CPMs induced by each sub-network $\mathfrak{N}_{\text{sub}}$ of $\mathfrak{N}$. Since the CPM induced by each sub-network $\mathfrak{N}_{\text{sub}}^f$ of the fine-grained network $\mathfrak{N}^f$ must itself be a fine-graining of the corresponding induced CPM of the original network $\mathfrak{N}$ (by \cref{definition: fg_network}), it follows from \cref{definition: fg_network} that the signalling relations of each sub-network in $\mathfrak{N}$ is preserved in a corresponding sub-network of $\mathfrak{N}^f$ and therefore that the signalling relations of $\mathfrak{N}$ are preserved in $\mathfrak{N}^f$.

\bigskip





\section{Spatio-temporal causal structures}
\label{sec: spacetime}

We model a spacetime in general and order-theoretic terms, without assuming any manifold topology or symmetries, but focusing on its causal structure.

\begin{definition}[Fixed acyclic spacetime]
\label{def: spacetime}
A fixed acyclic spacetime corresponds to a partially ordered set $\cT$ associated with the order relation $\preceq$. $P\preceq Q$ is denoted as $P\prec Q$ whenever $P,Q\in\cT$ are distinct elements. $P\prec Q$, $P\succ Q$ and $P\nprec\nsucc Q$ represent $P$ being in the past of, future of and neither in the past nor future of $Q$ respectively, with respect to this order relation. 
\end{definition}

In the remainder of the paper, we will often refer to a fixed acyclic spacetime simply as a spacetime. The partial order is an abstraction of the light-cone structure. In particular, any globally hyperbolic spacetime with a well-defined arrow of time can be described this way. The partial order on spacetime events induces and order relation on spacetime regions as follows.

\begin{figure}
    \centering
\subfloat[]{\includegraphics[scale=1.0]{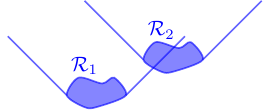}}\qquad\subfloat[]{\includegraphics[scale=1.0]{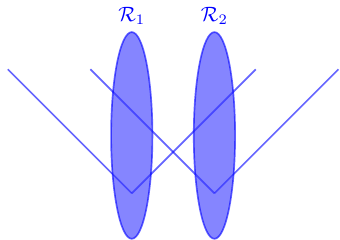}}\qquad\subfloat[]{\includegraphics[scale=1.0]{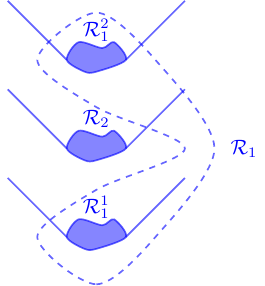}}
    \caption{Examples of region causal structures in Minkowski spacetime, where the 45 degree lines denote forward light-like surfaces. The region causal structure depicted in (a) is acyclic with $\cR_1\xrightarrow[]{R} \cR_2$ (but $\cR_2\cancel{\xrightarrow{R}} \cR_1$) while the region causal structures depicted in (b) and (c) are cyclic with $\cR_1\xrightarrow[]{R} \cR_2$ and $\cR_2\xrightarrow[]{R} \cR_1$. In (c), it is evident that the region $\cR_1$ is composed of two disjoint regions $\cR_1^1$ and $\cR_1^2$ such that the region causal structure over $\cR_1^1$, $\cR_1^2$ and $\cR_2$ is acyclic with $\cR_1^1\xrightarrow[]{R} \cR_2 \xrightarrow[]{R} \cR_1^2$. Since the spacetime is acyclic, it is easy to see that (b) can also be partitioned into a set of sub-regions such that the resulting region causal structure is acyclic. }
    \label{fig: regions}
\end{figure}

 \begin{definition}[Order relation  on spacetime regions]
  \label{def: region_order}
Let $\cR_1, \cR_2\subseteq \cT$ be two distinct subsets of locations (or ``regions'') in a spacetime $\cT$. We say that $\cR_1\xrightarrow[]{R} \cR_2$ if there exists at least one pair of distinct locations $P_1\in \cR_1$ and $P_2\in \cR_2$ such that $P_1\prec P_2$. More generally, we will refer to a directed graph $\cG^{\text{reg}}_{\cT}$ as a region causal structure of $\cT$ if $\mathrm{Nodes}(\cG^{\text{reg}}_{\cT})\subseteq \Sigma(\cT)$ and its edges are given by the order relation $\xrightarrow[]{R}$ induced by the partial order of $\cT$. 
  \end{definition}
Note that $\xrightarrow[]{R}$ is neither a pre-order nor a partial order relation as it is non-transitive and we can in general have $\cR\xrightarrow[]{R} \cR'$ as well as $\cR\xrightarrow[]{R} \cR'$ for two spacetime regions (see \cref{fig: regions} for illustrative examples of region causal structures). 

\bigskip

\paragraph{Region partitions}
Choosing a partition of each spacetime region $\cR\in \text{Nodes}(\cG^{\text{reg}}_{\cT})$ into mutually disjoint sub-regions $\cR=\bigcup_i\cR_i$ ($\cR_i\cap \cR_{i'}=\emptyset$ for all $i\neq i'$) defines a fine-graining $\cG^{\text{reg},f}_{\cT}$ of the graph $\cG^{\text{reg}}_{\cT}$ (Definition~\ref{definition: fg_graph}), where $\cG^{\text{reg},f}_{\cT}$ is generated by considering each node $\cR\in \text{Nodes}(\cG^{\text{reg}}_{\cT})$ and including every sub-region $\cR_i$ appearing in the partition of $\cR$ as a node of $\cG^{\text{reg},f}_{\cT}$. The order relation (defining the edges of  $\cG^{\text{reg},f}_{\cT}$) is then induced by the spacetime order relation $\prec$ as per \cref{def: region_order}.\footnote{That $\cG^{\text{reg},f}_{\cT}$ is a fine-graining of the graph $\cG^{\text{reg}}_{\cT}$ follows from Definition~\ref{def: region_order} which implies that whenever $\cR=\cup_i\cR_i\xrightarrow[]{R} \cR'=\cup_j\cR'_j$, there exist $\cR_i$ and $\cR'_j$ such that $\cR_i\xrightarrow[]{R} \cR'_j$.} While the spacetime $\cT$ and regions $\cR\subseteq \cT$ can in general be infinite sets, we only consider finite partitions where a given region is split into finitely many sub-regions. This is because for any physical protocol, agents' actions occur only within a finite set of well-defined spacetime regions. In fact, it is often useful to take the spacetime $\cT$ to be a finite poset, which represents the finite set of spacetime events which are relevant to the physical protocol that one wishes to model on the spacetime. In such cases, we can partition each region $\cR\in \mathrm{Nodes}(\cG^{\text{reg}}_{\cT})$ in a region causal structure of $\cT$ in terms of the individual spacetime locations in $\cT$ that comprise it $\cR=\bigcup_{P\in\cR}P$. Consequently, the spacetime $\cT$ itself, when represented as a directed acyclic graph corresponds to a fine-graining of any region causal structure $\cG^{\text{reg}}_{\cT}$ of the spacetime, which we will refer to as the \emph{maximal fine-graining} of $\cG^{\text{reg}}_{\cT}$. Note that while $\cG^{\text{reg}}_{\cT}$ can in general be cyclic, its maximal fine-graining, with nodes corresponding to elements of $\cT$, would always be acyclic since $\cT$ is a partially ordered set. Our results do not rely on the existence of such a maximal fine-graining, but it will often be useful to refer to it in examples and explanations.

\section{Linking information-theoretic and spatio-temporal causal structures}
\label{sec: infoth-spacetime}
Until this point, the spacetime structure is simply an abstract (acyclic) causal structure, which is devoid of any operational meaning before we embed physical systems in it. An agent may assign physical meaning to a spacetime point $P$ when it can be associated with some operational event such as ``I received message from Bob at the spacetime location $P$''. In our formalism, the in and output systems of our networks of maps model such operational events. Physically realising a network of maps in a spacetime, involves associating spacetime regions with the in and output systems of the network. We formalise these ideas below, by applying the general concepts that we have already defined in the previous sections.

\subsection{Realisations of cyclic quantum networks in fixed acyclic spacetimes}

We have already defined what it means to embed systems and networks of CPMs in an abstract causal structure, in \cref{definition: embedding_general} and \cref{definition: network_embedding_graph}. The following two definitions of spacetime embeddings naturally follow from applying these general definitions to region causal structures of a spacetime.

\begin{definition}[Spacetime embedding of systems] 
A spacetime embedding $\cE: \cS\mapsto \Sigma(\cT)$ of a set $\cS$ of systems in a fixed acyclic spacetime $\cT$ assigns a spacetime region $\cE(S)\in \Sigma(\cT)$ to each $S\in \cS$. The image of $\cE$ defines a region causal structure $\cG_{\cT}^{\text{reg}}$ of $\cT$ where $\cE(S)$ for every $S\in \cS$ corresponds to a distinct node of $\cG_{\cT}^{\text{reg}}$. Then $\cE$ corresponds to an embedding of $\cS$ in the directed graph $\cG_{\cT}^{\text{reg}}$ (according to \cref{definition: embedding_general}), and can equivalently be viewed as a mapping $\cE:\cS\mapsto \mathrm{Nodes}(\cG_{\cT}^{\text{reg}})$.
\end{definition}

It is possible to have $\cE(S)=\cE(S')$ for distinct systems $S,S'\in \cS$, however, these will correspond to distinct nodes in the region causal structure implied by the image of $\cE$. This is helpful for modelling situations where distinct agents act within the same spacetime region on distinct set of systems e.g., $S$ may be Alice's system and $S'$ may be Bob's system and Alice and Bob may be situated within the same lab.

\begin{definition}[Spacetime embedding of a network of CPMs]
\label{definition: network_embedding_spacetime}
A spacetime embedding $\cE:\cS\mapsto \Sigma(\cT)$ of the systems $\cS:=\mathfrak{N}^{\text{sys}}$ of a network $\mathfrak{N}$ induces an embedding $\mathfrak{N}_{\cG^{\text{reg}}_{\cT},\cE}$ of $\mathfrak{N}$ (cf.\ Definition~\ref{definition: network_embedding_graph}) in the region causal structure $\cG^{\text{reg}}_{\cT}$ defined by the image of $\cE$. We will denote $\mathfrak{N}_{\cG^{\text{reg}}_{\cT},\cE}$ as $\mathfrak{N}_{\cT,\cE}$ for short.
\end{definition}

\paragraph{Spacetime embedded networks and fine-graining} 
Let $\mathfrak{N}_{\cT,\cE}$ be a spacetime embedding of a network $\mathfrak{N}$ and $\cG^{\text{reg}}_{\cT}$ be the region causal structure defined by the image of the embedding $\cE$. 
Recall from \cref{sec: spacetime} that
by choosing a partition for each region $\cR\in \mathrm{Nodes}(\cG^{\text{reg}}_{\cT})$ into disjoint sub-regions $\cR=\bigcup_i \cR_i$, we can define a fine-graining  $\cG^{\text{reg},f}_{\cT}$ of $\cG^{\text{reg}}_{\cT}$.
This induces a systems fine-graining $\cF(S)=\{S^{\cR_i}\}_i$, of the systems $S\in \mathfrak{N}^{\text{sys}}$ (or equivalently, the corresponding $\cG^{\text{reg}}_{\cT}$-embedded systems) into $\cG^{\text{reg},f}_{\cT}$-embedded systems, where $S^{\cR_i}$ labels the system embedded in the region $\cR_i\in \text{Nodes}(\cG^{\text{reg},f}_{\cT})$. We can then consider a fine-graining $\mathfrak{N}^f_{\cT,\cE}$ of the network $\mathfrak{N}$ (or equivalently, the spacetime embedded network $\mathfrak{N}_{\cT,\cE}$ cf. \cref{definition: network_embedding_spacetime}) with respect to each such system fine-graining $\cF$ induced by a choice of partition on regions in $\cG^{\text{reg}}_{\cT}$).
In particular, when we maximally fine-grain each region in the image of $\cE$ in terms of the individual spacetime points, $\cE(S)=\{P\in\cT\}_{P\in\cE(S)}$, each system $S$ is correspondingly fine-grained to a set of systems $\cF_{max}(S)=\{S^{P}\}_{P\in\cE(S)}$, such that each spacetime location $P\in \cE(S)$ becomes associated with a Hilbert space $\cH^{S^{P}}$. 

\begin{definition}[Fixed spacetime realisation of a network of CPMs]
\label{definition: spacetime_realisation}
A realisation of a network $\mathfrak{N}$ in a fixed acyclic spacetime $\cT$ with respect to an embedding $\cE: \mathfrak{N}^{\text{sys}}\mapsto \Sigma(\cT)$ is specified by the spacetime embedded network $\mathfrak{N}_{\cT,\cE}$ along with at least one fine-graining $\mathfrak{N}^f_{\cT,\cE}$ of $\mathfrak{N}_{\cT,\cE}$ which is associated with an acyclic region causal structure $\cG^{\text{reg},f}_{\cT}$.
\end{definition}

If the image of the spacetime embedding $\cE$ itself corresponds to an acyclic region causal structure, then the spacetime realisation is fully specified by the spacetime embedded network $\mathfrak{N}_{\cT,\cE}$ (which is equivalent to the original network $\mathfrak{N}$, up to the labelling of each system $S$ with the spacetime region it is embedded in). If this is not the case, we can partition the regions in the image of $\cE$ until we obtain an acyclic region causal structure and specify a fine-graining $\mathfrak{N}^f_{\cT,\cE}$ of $\mathfrak{N}_{\cT,\cE}$ relative to that partitioning of regions, in order to fully specify the spacetime realisation. In particular, any fine-graining $\mathfrak{N}^f_{\cT,\cE}$ of the network defined relative to the maximal systems fine-graining $\cF_{\text{max}}$ will always have this property. However, it is also  possible to have other non-maximal partitions of the regions that still lead to an acyclic region causal structure, and it would be sufficient to only consider these in the specification of a spacetime realisation.

\begin{remark}
Notice that the above definition only refers to the acyclicity of the fine-grained region causal structure and does not place any constraints on the causal or signalling structure of the fine-grained network $\mathfrak{N}^f_{\cT,\cE}$. This is a minimal requirement which captures the essence of regarding a background spacetime as being acyclic: we can in-principle describe any protocol immersed in an acyclic spacetime in terms of what happens (or does not happen) within a set of spacetime regions which have a well-defined ordering.
\end{remark}

\subsection{Relativistic causality}
\label{sec: rel_caus}
For networks embedded/realised in spacetime as discussed in the previous sub-section, both information-theoretic and spatio-temporal causal orders come into play, and both are defined over the same set of nodes: the spacetime embedded systems which carry both system labels and spacetime region labels. The information-theoretic order is given by the edges $\longrsquigarrow$, the properties of which we can infer from the signalling order $\longrightarrow$ defined over subsets of spacetime embedded systems. The latter is given by the order relation $\xrightarrow{R}$ on regions. Relativistic causality principles impose constraints on the relations between these informational and spatio-temporal types of orders, which have so far been treated independently in our framework. In particular, we consider the relativistic causality principle of no signalling outside the future light-cone.  The following definitions provide a natural sufficient condition for capturing this principle in our framework, as a compatibility condition between $\longrightarrow$ and $\xrightarrow{R}$. In general there are distinct relativistic causality principles which can be considered, for instance, depending on whether they are formulated using the causal order $\longrsquigarrow$ or signal order $\longrightarrow$ and whether we consider necessary or sufficient conditions \cite{VilasiniColbeckPRA, VilasiniColbeckPRL, VilasiniColbeck2024} (see \cref{appendix: relation_to_VC} for further details). In this paper, we will focus on the following definitions for capturing relativistic causality, which follows as a special case of \cref{definition: graph_compatibility} of compatibility between  a signalling structure and an abstract causal structure, when the latter is a region causal structure of a spacetime.

\begin{definition}[Relativistic causality for signalling structures]
\label{def: rel_causality_sig}
Let $\cG^{\text{sig}}$ be a signalling structure over a set $\cS$ of systems and $\cE$ be an embedding of the systems $S\in\cS$ in a region causal structure $\cG^{\text{reg}}_{\cT}$ associated with a spacetime $\cT$. Then, we say that the signalling structure $\cG^{\text{sig}}$ satisfies relativistic causality with respect to the embedding $\cE$ in the spacetime $\cT$ if $\cG^{\text{sig}}$ is compatible with the causal structure $\cG^{\text{reg}}_{\cT}$ (cf. Definition~\ref{definition: graph_compatibility}).
\end{definition}

Then we have the following theorem.

\begin{restatable}{theorem}{EmbeddingSig}[Embedding arbitrary, cyclic signalling relations in spacetime] 
\label{theorem: embedding_sig}
For every signalling structure $\cG^{\text{sig}}$, there exists a fixed acyclic spacetime $\cT$ and an embedding $\cE$ of $\cG^{\text{sig}}$ in a region causal structure $\cG^{\text{reg}}_{\cT}$ of $\cT$ that respects relativistic causality.
\end{restatable}

\begin{definition}[Relativistic causality for spacetime realisations of networks]
\label{def: rel_causality_network}
A spacetime realisation of a network $\mathfrak{N}$ of CPMs is said to satisfy relativistic causality relative to an embedding $\cE$ in a spacetime $\cT$ only if the signalling structure of each network in the specification of the spacetime realisation satisfies the relativistic causality condition of Definition~\ref{def: rel_causality_sig}.
We will then refer to such a spacetime realisation as a \emph{physical spacetime realisation}.
\end{definition}


The above is a minimal definition of what constitutes a physical realisation of a network in a spacetime, it ensures that no interventions that an agent could perform could allow them to signal outside the future lightcone of the spacetime. In the next sections, we apply our general framework and results to address questions relating to the physical spacetime realisations of indefinite causal order quantum processes.

\section{Indefinite causal structures in the process matrix framework}
\label{sec: PM}

The concept of indefinite causal structures was independently proposed within different information-theoretic frameworks, to model scenarios where agents can perform quantum operations that are not applied in a definite and acyclic causal order. The causaloid framework of Lucien Hardy \cite{Hardy2005}, the formalism of higher-order computation and maps \cite{Chiribella2013}, and the process matrix framework \cite{Oreshkov2012} are prominent examples. Here, we consider the process matrix framework (which is known to be mathematically equivalent to higher-order maps), since the question of physicality of indefinite causal structures has most commonly been posed and discussed in this formalism. The framework describes multi-agent information-processing scenarios without assuming a global, acyclic causal order between the operations of the agents, but while assuming the local validity of quantum theory within the lab of each agent. 

\bigskip

\paragraph{Local behaviour: quantum instruments.}
Each agent $A$ acts within their respective local laboratory, which is associated an input Hilbert space $\mathcal{H}^{A^I}$ of dimension $d_{A^I}$ and an output Hilbert space $\mathcal{H}^{A^O}$ of dimension $d_{A^O}$. $A^I$ and $A^O$ will be used as shorthand to represent the set of all linear operators over $\mathcal{H}^{A^I}$ and $\mathcal{H}^{A^O}$ respectively. The operations that can be performed within the lab of each agent $A$ are described by a set of \emph{quantum instruments}, one for each setting $a$ which is a classical variable specifying the choice of operation performed on $A^I$. For each setting $a$, a quantum instrument is a set of CP maps, $\mathcal{J}^A_a=\{\mathcal{M}_{x|a}^A\}_{x=1}^m$ with $\mathcal{M}_{x|a}^A: A^I \mapsto A^O$ and $\sum_{x=1}^m\mathcal{M}_{x|a}^A$ being a CPTP map \cite{Oreshkov2012, Araujo2015}, where $x$ parametrizes the possible outcomes of the operation (for some set $\{1,...,m\}$ of possible outcome values).  Quantum instruments can be equivalently represented by the set of Choi-Jamio\l{}kowski states $\{M_{x|a}^{A^IA^O}=\Big[\mathcal{I}\otimes\mathcal{M}^A_{x|a}\Big(|\mathds{1}\rrangle\llangle\mathds{1}|\Big)\Big]^T\}_{x=1}^m$, where $|\mathds{1}\rrangle:=\sum_j\ket{j}^{A^I}\ket{j}^{A^I}$ and $T$ denotes matrix transposition with respect to the chosen orthonormal basis $\ket{j}^{A^I}$ of $\mathcal{H}^{A^I}$. 

\bigskip

\paragraph{Global behaviour: process matrix.}
If the scenario consists of $N$ agents $\{A_1,...,A_N\}$,
the probability $P(x_1,...,x_N|a_1,...,a_N)$ that the $N$ agents observe the outcomes $(x_1,...,x_N)$ for a choice of settings $(a_1,...,a_N)$ can be expressed using the Choi-Jamio\l{}kowski  representation of the local operations as follows \cite{Oreshkov2012, Araujo2015},
\begin{equation}
\label{eq: prob}
P(x_1,...,x_N|a_1,...,a_N)=P\left(\mathcal{M}^{A_1}_{x_1|a_1},...,\mathcal{M}^{A_N}_{x_N|a_N}\right)
=\tr\left[\left(M_{x_1|a_1}^{A_1^IA_1^O}\otimes...\otimes M_{x_N|a_N}^{A_N^IA_N^O}\right)W\right],
\end{equation}
for a Hermitian operator $W \in A^I_1\otimes A^O_1\otimes...\otimes A^I_N\otimes A^O_N$,  known as the \emph{process matrix}. 
The set of valid process matrices is characterised by the set of all such Hermitian operators $W$ that yield positive normalised probabilities for all possible CP maps $\{\mathcal{M}^{A_k}_{x_k|a_k}\}_{k=1}^N$. This is required to hold also for local operations that can additionally act on ancillary quantum systems, where the ancillas between multiple labs may be entangled \cite{Oreshkov2012}.

\bigskip

The global behaviour can equivalently be described as a \emph{higher-order map} which maps each set of local operations to a probability distribution \cite{Chiribella2013, Araujo2016}. This higher-order map can itself be represented as a CPTP map $\mathcal{W}$ from the input systems $A^O_1$,...,$A^O_N$ (corresponding the outputs of the agents) to the output systems $A^I_1$,...,$A^I_N$ (corresponding to the inputs of the agents), as illustrated in \cref{fig: PM_Composition}. The process matrix $W=\mathcal{I}\otimes\mathcal{W}\Big(|\mathds{1}\rrangle\llangle\mathds{1}|\Big)$ is then the Choi matrix of the CPTPM $\mathcal{W}$, where $|\mathds{1}\rrangle$ corresponds to the unnormalised maximally entangled state over two copies of the input Hilbert space of $\mathcal{W}$ \cite{Araujo2015}.\footnote{Note that the Choi matrix of $\mathcal{W}$ and that of the local operations defined in the previous paragraphs differs by a transpose, this is a choice of convention made in the process matrix framework, that makes the notation and calculations more convenient.} In the rest of this paper, we will refer to the CPTPM $\mathcal{W}$ as the process map, and this will be our object of interest.



\bigskip

\paragraph{Different classes of processes} 
Several different classes of processes have been proposed and studied in the process matrix framework. Here, we review these concepts for the bipartite case and refer the reader to \cite{Araujo2016, Oreshkov2016} for the general definitions. Firstly, we have the class of \emph{fixed order processes}, which are processes that are compatible with a definite acyclic causal order. In particular, this includes standard causally ordered quantum circuits. For instance a process $W^{A\prec B}$ is said to be compatible with the fixed order Alice before Bob if Alice's outcome $x$ is independent of Bob's setting $b$ i.e., $P(x|ab)=P(x|a)$. 
Similarly, $W^{B\prec A}$ is a fixed order process that is compatible with the order $B\prec A$ whenever $P(y|ab)=P(y|b)$. Note that a process where neither agent can communicate to the other is compatible with both orders according to this definition. The general definition of a fixed order process is reviewed in \cref{appendix: PM}.


\bigskip

More generally, we can have probabilistic mixtures of fixed order processes, which leads to the set of causally separable processes as well as causally non-separable processes that cannot be cast in this form. These concepts are analogous to separable and non-separable quantum states. Moreover, in analogy to Bell local and Bell non-local quantum states and correlations, we also have the concepts of causal and non-causal processes and correlations. These are reviewed below for the bipartite case, the general definitions can be found in \cite{Araujo2015, Oreshkov2016}.

\begin{definition}[Causally separable and non-separable processes \cite{Araujo2015}]
\label{definition:causal_sep}
A bipartite process matrix $W$ is said to be \emph{causally separable} iff it decomposes as 
 \begin{equation}
     \label{eq: causalsep}
     W=qW^{A\prec B}+(1-q)W^{B\prec A},
 \end{equation}
 for some $q\in[0,1]$, where $W^{A\prec B}$ and $W^{B\prec A}$ are process matrices compatible with the fixed ordering between agents indicated in the respective superscripts. $W$ is said to be \emph{causally non-separable} otherwise. 
\end{definition}

 \begin{definition}[Causal and non-causal processes and correlations \cite{Oreshkov2012}]
\label{definition:causal}
A bipartite process matrix $W$ is said to be \emph{causal} iff for all choices of local operations, the joint probability $P(xy|ab)$ generated by $W$ (for outcomes $x$ and $y$ and settings $a$ and $b$ of agents $A$ and $B$ respectively) decomposes as follows for some $q\in[0,1]$
 \begin{equation}
    \label{eq: PMcausal} 
    P(xy|ab)=qP^{A\prec B}(xy|ab)+(1-q)P^{B\prec A}(xy|ab),
 \end{equation}
 where $P^{A\prec B}$ is a probability distribution compatible with the causal order $A\prec B$ by disallowing signalling from $B$ to $A$ i.e., $P^{A\prec B}(x|ab)=P^{A\prec B}(x|a)$, and similarly for $P^{B\prec A}$. The process $W$ is called \emph{non-causal} otherwise.  Similarly, distributions $P(xy|ab)$ are said to be causal/non-causal depending on whether they can be decomposed as above. 
\end{definition}

Both causally non-separable processes and non-causal processes are regarded as having indefinite causal orders in this framework, as they are incompatible with an explanation in terms of a fixed acyclic order of the operations, or any convex mixture of such fixed orders. Moreover, non-fixed order but causally separable processes can be regarded as representing classically indefinite orders.

\bigskip

\paragraph{Set-up assumptions}
The process matrix framework rests on the assumption that each agent acts within a closed laboratory, that interacts with the environment only to let in one input system $A^I$, apply an operation on it and then let out one output system $A^O$, with the local events unfolding in that order (even though no global order between the operational events of different agents is assumed)\footnote{Note that no spacetime information is explicitly considered but even in the absence of information about the absolute time of occurrence of these operational events, it is in principle possible to ensure that they are ordered in a certain way. } and the lab being closed to in/output otherwise. These correspond to the closed labs (CL) and local order (LO) assumptions, and they are meant to ensure that each agent acts once and only once.
Moreover, free choice (FC) is also assumed: that the agents can freely chose the local operation performed in their lab, that is the setting modelling the choice of operation in each lab has no other causes relevant to the scenario.

\bigskip

\paragraph{Causal inequalities}
\cite{Oreshkov2012} derives a linear inequality constraint on the joint probabilities $P(xy|ab)$ under four assumptions, this is referred to as a \emph{causal inequality} and is shown to be a necessary constraint on causal distributions. The first three assumptions are the set-up assumptions LO, CL and FC of the process matrix framework. The fourth is an additional assumption referred to as \emph{causal structure} (CS) which states that the input and output events of the agents are localised in a fixed partial order such as a background spacetime that prohibits signalling outside the future. 
Therefore, it is argued that if the correlations violate a causal inequality in a scenario where CL, LO and FC are satisfied, this would be certify a violation of the assumption CS which relates to the existence of a definite causal structure. Hence causal inequalities are commonly regarded as device-independent witnesses of the indefiniteness of causal order. In \cref{ssec: causal_ineq}, we will analyse the CS assumption in more detail in light of our results to draw more fine-grained insights on the interpretation of causal inequalities and their violation.

\section{Indefinite causal structures as instances of cyclic quantum networks}
\label{sec: PM_network}

In this section, we show that indefinite causal structures as described in the process matrix formalism can be modelled as instances of the cyclic quantum networks allowed in our framework, and derive some connecting results. 

\bigskip

\paragraph{Extended local maps} 
To capture the set of quantum instruments associated with an agent in the process framework as a single CPTP map, we introduce additional in and output systems corresponding to the classical setting and outcome of the agent.
Observe that the local operation $\mathcal{M^A}$ of an agent $A$ can be equivalently modelled as a CPTP map from the input systems $A^I$ and $A^s$ to the output systems $A^O$ and $A^o$, where $A^I$ and $A^O$ are the quantum in/output systems we have seen before, $A^s$ and $A^o$ model the local in/output systems carrying the classical setting $a$ (that specifies the choice of operation to be applied on the input system $A^I$) and a possible measurement outcome $x$ of the agent. We can encode the classical setting and outcome, $a$ and $x$ as quantum states $\ket{a}$ and $\ket{x}$ in the computational basis.
If the map does not implement a measurement but implements a transformation from $A^I$ to $A^O$ depending on a setting choice $a$ on $A^s$, the output state on $A^o$ will correspond to the deterministic outcome represented by a fixed state $\ket{\perp}$, and the output system $A^o$ can simply be ignored in this case. An outcome is non-trivial if it does not equal $\perp$. 
We refer to the map $\mathcal{M}^{A_k}: A^I_k\otimes A^s_k\mapsto A^O_k\otimes A^o_k$ for each agent $A_k$ as the \emph{extended local map} or \emph{extended local operation} of that agent. From the extended map, we can then recover the action the CP maps $\mathcal{M}_{x|a}^A$ (defined in the process framework) on an input state $\rho^{A^I}$ as follows.
\begin{equation}
\label{eq: projection}
    \mathcal{M}_{x|a}^A(\rho^{A^I}):=\tr_{A^o}\Bigg[\Big(\ket{x}\bra{x}^{A^o}\otimes\mathcal{I}^{A^O}\Big)\Big[\mathcal{M^A}(\ket{a}\bra{a}^{A^s}\otimes\rho^{A^I})\Big]\Bigg].
\end{equation} 

Notice that $\mathcal{M}_{a}^A(.):=\sum_x \mathcal{M}_{x|a}^A(.)=\tr_{A^o}\Big[\mathcal{M^A}(\ket{a}\bra{a}^{A^s}\otimes(.))\Big]$ is then a CPTP map from $A^I$ to $A^O$, since we sum over a complete set of outcomes, $\sum_x\ket{x}\bra{x}=\id$. We will refer to $\mathcal{M}_{a}^A$ as a \emph{fixed local map} as opposed to the extended local map $\cM^A$, since it corresponds to the local map of agent $A$ associated with plugging in a fixed choice of setting $a$.

\bigskip

\begin{figure}[t!]
    \centering
    \includegraphics[scale=0.7]{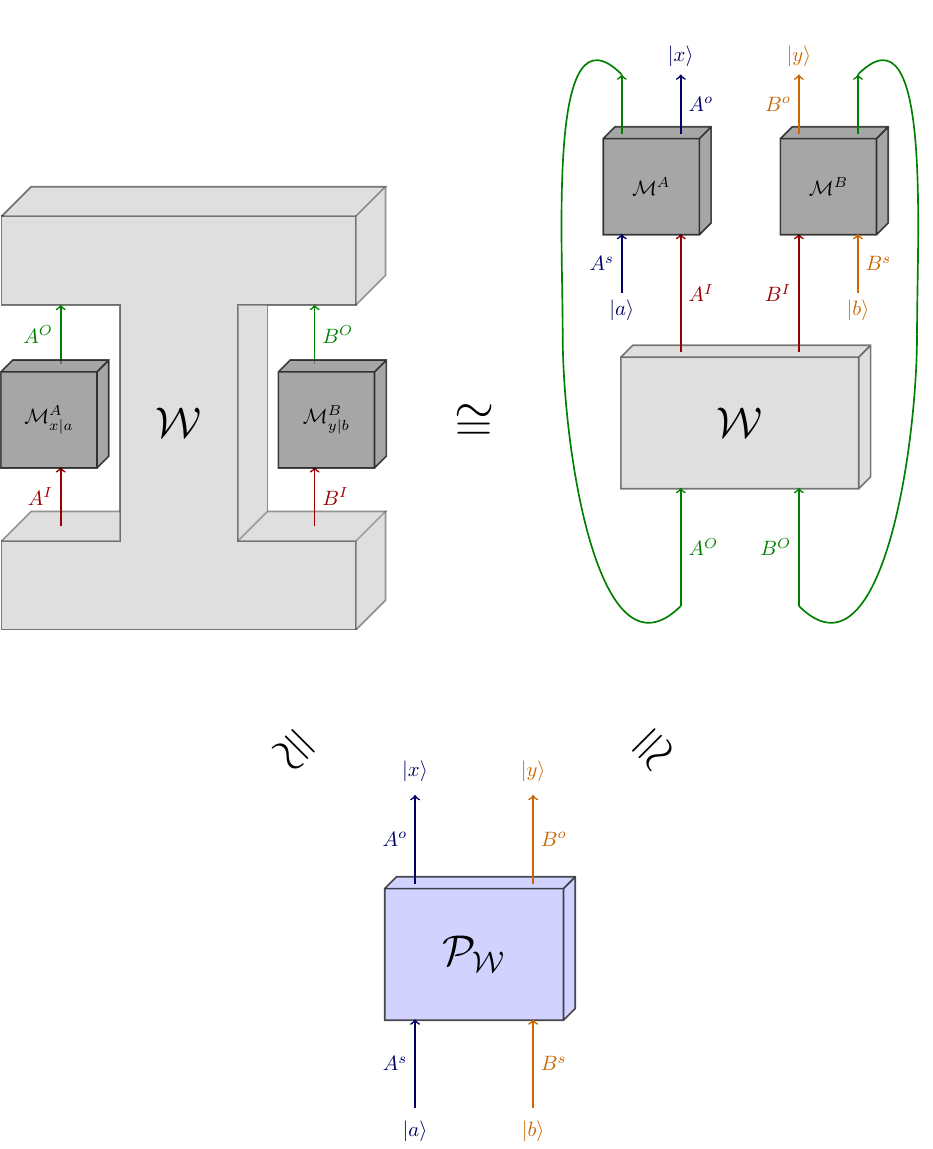}
    \caption{The process matrix $W$ can be modelled as a higher-order transformation which, given a set of CP (but possibly trace decreasing) maps $\mathcal{M}^A_{x|a}$ and $\mathcal{M}^B_{y|b}$ (one for each agent), maps them to a probability distribution $P(xy|ab)$ given by Equation~\eqref{eq: prob}. This is illustrated in the top left. This can be equivalently viewed in terms of a network formed by the loop composition of the process map $\mathcal{W}$ with the extended local operations of the agents as shown on the top right, which models their classical setting and outcome as additional in and output wires. The induced map of the network formed in this way is denoted as $\mathcal{P}_{\mathcal{W}}$ (bottom middle), it is a map with classical in and outputs that encodes the joint probability distribution $P(xy|ab)$ as shown in Lemma~\ref{lemma: probabilities}. }
    \label{fig: PM_Composition}
\end{figure}

\paragraph{The process network and joint probabilities} The action of the process matrix on the local operations can be equivalently understood as a loop composition between the process map $\mathcal{W}$ and 
the extended local maps $\{\mathcal{M}^{A_k}\}_{k=1}^N$, as shown in Figure~\ref{fig: PM_Composition}. Let us denote the in/outputs of the local maps as $A^{'I}_{k}$, $A^O_k$  and corresponding systems of $\mathcal{W}$ as $A^I_k$, $A^{'O}_{k}$ (note that the prime is associated with systems that are inputs of the corresponding map). Then for an $N$ partite process matrix $W$, this action generates the network $\mathfrak{N}_{\mathcal{W},N}$ where 
\begin{align}
\label{eq: process_network}
    \begin{split}
    \mathfrak{N}_{\mathcal{W},N}^{maps}&:=\mathcal{W} \cup \{\mathcal{M}^{A_k}\}_{k=1}^N,\\
    \mathfrak{N}_{\mathcal{W},N}^{comp}&:=\{A^I_k \hookrightarrow A^{'I}_{k}, A^O_k \hookrightarrow A^{'O}_{k}\}_{k=1}^N.
    \end{split}
\end{align}

We refer to the network $\mathfrak{N}_{\mathcal{W},N}$ as the \emph{process network} associated with the process matrix $W$ (or process map $\mathcal{W}$). The induced map of this network will be denoted as $\cN_{\mathcal{W},N}:=\mathcal{P}_{\mathcal{W}}$ as it encodes the probabilities $P(x_1,...x_N|a_1,...,a_N)$ and the number of parties $N$ is usually clear from context. Notice that the induced map $\mathcal{P}_{\mathcal{W}}$ obtained in this way has only classical input systems ($\{A^s_k\}_k$ carrying the setting choices of all agents) and classical output systems ($\{A^o_k\}_k$ carrying the outcomes of the agents).

\bigskip

If we input a choice $\{a_k\}_k$ for settings to the map $\mathcal{P}_{\mathcal{W}}$ and post-select on a set $\{x_k\}_k$ of outcomes on the corresponding output, we get the joint probability of obtaining those outcomes given those setting choices as the success probability of that post-selection. Explicitly, given an $N$-partite process map $\mathcal{W}$ and a set $\{\mathcal{M}^{A_k}\}_{k=1}^N$ of extended local maps, the joint probability of obtaining a set $\{x_k\}_k$ of outcomes given a choice $\{a_k\}_k$ of the settings is obtained as follows.
\begin{equation}
\label{eq: prob_composition1}
   P(x_1,...,x_N|a_1,...,x_N)=\frac{\tr\left[\Pi_{x_1}\otimes...\otimes\Pi_{x_N}\Big( \mathcal{P}_{\mathcal{W}}(\ket{a_1}\bra{a_1}\otimes...\otimes\ket{a_N}\bra{a_N})\Big)\right]}{\sum_{x_1,...,x_N} \tr\left[\Pi_{x_1}\otimes...\otimes\Pi_{x_N}\Big( \mathcal{P}_{\mathcal{W}}(\ket{a_1}\bra{a_1}\otimes...\otimes\ket{a_N}\bra{a_N})\Big)\right]},
\end{equation}
where the projector $\Pi_{x_k}=\ket{x_k}\bra{x_k}$ projects the state on the system $A^{o}_k$ to $\ket{x_k}$. The following result shows that this reproduces the probability rule of the process matrix framework.

\begin{restatable}{lemma}{ProbRule}[Probabilities from loop composition]
\label{lemma: probabilities}
For every process map $\mathcal{W}$, the joint probability distribution obtained through the induced map $\mathcal{P}_{\mathcal{W}}$ of the network $\mathfrak{N}_{\mathcal{W},N}$ as in Equation~\eqref{eq: prob_composition1} is equivalent to that obtained in the process matrix framework through Equation~\eqref{eq: prob}. Moreover, $W$ being a valid process matrix guarantees that the denominator of Equation~\eqref{eq: prob_composition1} equals unity.
\end{restatable}

Notice that for a general map $\mathcal{W}$, the denominator is needed to ensure that we get normalised probabilities, since a map such as $\mathcal{P}_{\mathcal{W}}$ formed by loop composition of CPTP maps could in general be trace decreasing and considering the numerator alone may not result in a valid normalised distribution. The above result shows that in the special case where the Choi matrix of $\mathcal{W}$ is a valid process matrix, the numerator alone defines a normalised probability distribution.

\bigskip

{\bf Sub-networks and reduced processes} The network $\mathfrak{N}_{\mathcal{W},N}$ involves composing the $N$-partite process map $\mathcal{W}$ together with the local operations of all $N$ agents, the sub-networks of $\mathfrak{N}_{\mathcal{W},N}$ include networks $\mathfrak{N}_{\mathcal{W},l<N}$ where $\mathcal{W}$ is composed with the local operations of $l$ out of $N$ agents, with $l<N$. Without loss of generality, we take these to be the first $l$ agents, $\{A_1,...,A_l\}$, then the maps and compositions in the network $\mathfrak{N}_{\mathcal{W},l<N}$ are defined exactly as in \cref{eq: process_network} but with the index $k$ going from $1$ to $l$ (instead of $N$).


\bigskip

In the following lemma, we relate the sub-networks $\mathfrak{N}_{\mathcal{W},l<N}$ with the notion of a reduced process matrix \cite{Araujo2015}. Given an $N$-partite process matrix $W$ and a CPTP map $\mathcal{M}^{A_j}_{a_j}$ (associated with a specific setting choice $a_j$) for the $j^{th}$ agent with Choi matrix $M^{A^I_jA^O_j}_{a_j}$, the reduced process matrix \cite{Araujo2015} for the remaining $N-1$ agents is given as
\begin{equation}
\label{eq: ReducedPM}
    \overline{W}(M^{A^I_jA^O_j}_{a_j}):=\tr_{A^I_jA^O_j}\Bigg[\Big(\mathds{1}^{A^I_1A^O_1}\otimes...\otimes M^{A^I_jA^O_j}_{a_j}\otimes...\otimes \mathds{1}^{A^I_NA^O_N}\Big).W\Bigg],
\end{equation}

where $\mathds{1}^{A^I_iA^O_i}$ is an identity matrix on $A^I_i\otimes A^O_i$.
Analogously, given a fixed local operation (which is a CPTPM) for each of the first $l<N$ agents, the reduced process $\overline{W}(M^{A^I_1A^O_1}_{a_1},...,M^{A^I_lA^O_l}_{a_l})$ associated with the remaining $N-l$ can be defined.

\begin{restatable}{lemma}{ReducedPM}[Sub-networks and reduced processes]
\label{lemma: reducedPM}
For any $N$-partite process map $\mathcal{W}$, the sub-network  $\mathfrak{N}_{\mathcal{W},l<N}$ of the process network formed by composing $\mathcal{W}$ with some set of fixed local operations $\{\mathcal{M}_{a_k}^{A_k}\}_{k=1}^l$ for the first $l<N$ agents is a CPTP map whose Choi matrix is the reduced process matrix $\overline{W}(M^{A^I_1A^O_1}_{a_1},...,M^{A^I_lA^O_l}_{a_l})$. 
\end{restatable}


These results lead to the following useful theorem, 
that every valid process network is a network of CPTP maps. 

\begin{restatable}{theorem}{PMNetworkCPTP}
The network $\mathfrak{N}_{\mathcal{W},N}$  formed the composition of any valid process map $\mathcal{W}$ with a set of extended local operations $\{\cM^{A_k}\}_{k=1}^N$ is a network of CPTP maps. 
\end{restatable}

These results enable us to apply our framework to process matrices and infer their properties using the observable signalling relations. Recall that the signalling relations between sets of quantum systems in a network is defined through the induced CPTPMs of its sub-networks. On the other hand, we can also consider signalling at the level of the probabilities, by considering whether the outcome of a set of agents is correlated with a freely chosen setting of another agent. Say $N=2$ with the two agents being $A$ and $B$. One notion of signalling from $A$ to $B$ is at the level of the probabilities, which would say $P(y|ab)\neq P(y|b)$. This corresponds to signalling at between the classical systems $A^s$ and $B^o$ of the network. On the other hand, we also can define signalling from $A$ to $B$ at the level of the quantum systems of
the process map $\mathcal{W}$ (which corresponds to a sub-network of the process network) by considering whether $\{A^O\}$ signals to $\{B^I\}$ in $\mathcal{W}$. Does $P(y|ab)\neq P(y|b)$ imply $\{A^O\}$ signals to $\{B^I\}$ in $\mathcal{W}$ and vice-versa? In \cref{appendix: equiv_signalling}, we show that this is indeed the case whenever $W$ is a valid process matrix. More generally we prove that for any $N$-partite process, there is an equivalence between signalling at the level of the probabilities over classical settings and outcomes and signalling at the level of the corresponding quantum systems in the network. An upcoming work \cite{Ferradini} shows with a counter example, that when $\mathcal{W}$ is a CPTP map whose Choi matrix is not a valid process matrix, $P(y|ab)\neq P(y|b)$ does not in general imply $\{A^O\}$ signals to $\{B^I\}$ in $\mathcal{W}$. Therefore $\mathcal{W}$ being associated with a valid process matrix is sufficient for this equivalence to hold in the corresponding (possibly cyclic) network. Whether it is also a necessary condition, is left as an interesting open question for future work.

\bigskip

{\bf Indefinite causality and cyclicity of compatible causal structures} To conclude this section, we present the following theorem that connects the non-fixed order processes to cyclicity of causal structures compatible with the corresponding network, whose proof interestingly relies on the correspondence between the two notions of signalling mentioned above. 

\begin{restatable}{theorem}{IndefCyclic}
\label{theorem: indef_cyclic}
    A process matrix $W$ is not a fixed order process if and only if the signalling structure of the corresponding network $\mathfrak{N}_{\mathcal{W},N}$ certifies the cyclicity of any causal structure $\mathcal{G}$ that it is compatible with. 
\end{restatable}

Here, by certifying the cyclicity of $\mathcal{G}$, we mean that compatibility of the signalling structure with 
$\mathcal{G}$ will imply the presence of at least one directed cycle in $\cG$. This theorem has implications both for causal inference in cyclic quantum causal models, as well as for physical processes that can be implemented, consistently with relativistic causality in a spacetime. These different implications can be derived based on what $\mathcal{G}$ represents, for causal inference it would be the information-theoretic causal structure of the network and for relativistic causality, it would be the causal structure defined through the light-cones of a spacetime. The next section focuses on the implications of this result for spacetime realisations of processes.

\section{No-go results for spacetime realisations of quantum processes}
\label{sec: nogo}
Applying our general framework, we now establish and discuss a number of no-go results for spacetime realisations of indefinite causal order processes. 

\subsection{Localisation of information in spacetime}

The following no-go theorem follows from the general result we have shown in \cref{theorem: indef_cyclic}. It tells us that three, assumptions of physical interest and relevance cannot be simultaneously satisfied. The first assumption requires that the process that we wish to realise in a spacetime is not of fixed order, which is necessary for it to be considered an indefinite causal structure in the process matrix framework. The second assumption ensures that the realisation is physically meaningful, i.e., satisfies relativistic causality in the given spacetime embedding of the process network. The third assumption concerns the embedding, and captures that the systems of the process network are sufficiently localised in the spacetime, an assumption that is usually satisfied when embedding typical abstract quantum information protocols in spacetime. 

\begin{restatable}{theorem}{NogoA}[No-go theorem for spacetime realisations of processes]
\label{theorem: nogo_main}
No realisation  (\cref{definition: spacetime_realisation}) of an $N$-partite process network $\mathfrak{N}_{\mathcal{W},N}$ in a fixed spacetime structure $\mathcal{T}$ (Definition~\ref{def: spacetime}) with respect to a spacetime embedding $\mathcal{E}$ can simultaneously satisfy the following three assumptions.

\begin{enumerate}
\item $W$ is not a fixed order process (Definition~\ref{definition: causallyordered}).
    \item  The spacetime realisation satisfies relativistic causality (cf. \cref{def: rel_causality_network}).
    \item The region causal structure $\mathcal{G}^{\text{reg}}_{\mathcal{T}}$ induced by the embedding $\mathcal{E}$ (\cref{definition: network_embedding_spacetime}) is acyclic.
\end{enumerate}
\end{restatable}

A special case of an embedding that respects condition 3 is one which assigns to each system $S$ of the network $\mathfrak{N}_{\mathcal{W},N}$ a single spacetime location $\cE(S)\in \cT$, since the region causal in this case is a subgraph of $\cT$ which will be acyclic since $\cT$ is a poset. This gives the following corrollary regarding the impossibility of perfectly localising systems in the spacetime.

\begin{restatable}{corollary}{STLocalisation}[Spacetime localisation]
\label{corollary: localisation}
For every non-fixed order process $\mathcal{W}$, it is impossible to implement the corresponding network $\mathfrak{N}_{\mathcal{W},N}$ in a fixed spacetime consistently with relativistic causality, through an embedding that localises all the in/output systems in the spacetime. 
\end{restatable}

All the above statements hold irrespective of the choice of reference frame used to describe the spacetime, since they only depend on the order relation between spacetime points which is an agent/frame independent notion according to Definition~\ref{def: spacetime} (as is the case also in special relativity theory). We can also obtain a frame-dependent statement by considering the spatial and temporal coordinates of all the spacetime locations from the perspective of a single agent. For this, we must first add some more structure to our definition of spacetime to include details about spacetime co-ordinates. 

\begin{definition}[spacetime co-ordinates and time localisation]
\label{definition: coordinates}
Given a spacetime structure $\mathcal{T}$ (cf. \cref{def: spacetime}), each agent $A$ can express every location $P\in\mathcal{T}$ in terms of a spatial co-ordinate $\vec{r}_P^A\in \mathbb{R}^n$ and temporal co-ordinate $t_P^A\in \mathbb{R}$ as $(\vec{r}_P^A,t_P^A)$. Moreover, we require that whenever $P\prec Q$ for some $P,Q\in \mathcal{T}$, $t_P^A< t_Q^A$ for all agents $A$ i.e., agents always agree on the order relations even if they may disagree on the co-ordinate assignment. Then, we say that a spacetime region $\cR\subseteq \mathcal{T}$ is time localised with respect to an agent $A$ if $t^A_P=t^A_{P'}$ for any $P,P'\in\cR$.
\end{definition}

We then obtain the following corollary.

\begin{restatable}{corollary}{Time}[Time localisation in a global frame]
\label{corollary: time}
No realisation of an $N$-partite process network $\mathfrak{N}_{\mathcal{W},N}$ in a fixed spacetime structure $\mathcal{T}$ with respect to a spacetime embedding $\mathcal{E}$ can simultaneously satisfy the following three assumptions.
\begin{enumerate}
\item $W$ is not a fixed order process (cf. \cref{definition: causallyordered}).
    \item  The spacetime realisation satisfies relativistic causality (cf. \cref{def: rel_causality_network}).
    \item The region causal structure $\mathcal{G}^{\text{reg}}_{\mathcal{T}}$ induced by $\mathcal{E}$ is such that there exists an agent $A$ from whose perspective every region $\cR\in \mathrm{Nodes}(\mathcal{G}^{\text{reg}}_{\mathcal{T}})$ is time-localised (cf. \cref{definition: coordinates}).

\end{enumerate}
\end{restatable}

The above theorem requires all regions to be time-localised when described in the single global frame of $A$. A weaker requirement would be to require each spacetime region associated to an in/output system in the process network is time-localised from the perspective of the agent associated with that in/output system. However, the theorem no longer holds when replacing the third condition with this weaker requirement. To illustrate this, in Section~\ref{sec: QS_Minkowski}, we propose a protocol for realising the quantum switch (a causally non-separable process) in Minkowski spacetime where the two agents in relative inertial motion perceive their respective in/output systems to be time-localised in their own frame.


\begin{remark}
A purely information-theoretic notion of \emph{time-delocalised subsystems} was previously introduced \cite{Oreshkov2019}, and the realisation of processes in terms of such subsystems has been studied. While the broad message of \cref{corollary: time} accords with the intuition suggested by these works, that physical realisations of indefinite causal structure processes are linked to the non-localisation of information in time, the definitions and assumptions backing the mathematical model of this information non-localisation are different in our framework as compared to this previous work. In particular, as we show in the next section, even though information contained in the systems of the original process are not necessarily localised in (space)time (in the sense defined here, see \cref{theorem: nogo_main} and \cref{corollary: localisation}), spacetime realisations of any process satisfying relativistic causality in a fixed spacetime will ultimately be compatible with a definite acyclic fine-grained causal structure. This leads to different conclusions and implications for experimental realisations of the quantum switch, than those drawn in the framework of \cite{Oreshkov2019}, as we discuss in \cref{sec:qs_expts}.
\end{remark}

\subsection{Existence of a fine-grained explanation in terms of a fixed order process}
Theorem~\ref{theorem: nogo_main} permits the realisation of networks associated with non-fixed order processes consistently with relativistic causality, as long as the associated region causal structure is cyclic. The results of this section concern the fine-grained description of such realisations, showing that this description will ultimately be associated with that of a process with a fixed acyclic causal order over a larger number of operations. This generalises the intuition from the the classical demand and price example (\cref{fig: demand_price}) to cyclic quantum causal structures described by process matrices.

\bigskip

To simplify the presentation of this theorem, we incorporate a natural and physically motivated property, which is satisfied in experiments where the devices implementing the operations have a fixed (non-zero) processing time for converting in to outputs. More generally, this can be formalised as a correspondence between the spacetime regions assigned to each agent's in and output quantum systems. We say that two regions $\cR_1$ and $\cR_2$ which are the nodes of some region causal structure $\cG^{\text{reg}}_{\cT}$ have a correspondence from $\cR_1$ to $\cR_2$ if for any fine-graining $\cG^{\text{reg},f}_{\cT}$ of $\cG^{\text{reg}}_{\cT}$, through which $\cR_1$ maps to a set of sub-regions $\{\cP_1,...,\cP_n\}\subseteq \mathrm{Nodes}(\cG^{\text{reg},f}_{\cT})$, $\cR_2$ maps to a corresponding set of sub-regions 
 $\{\cQ_1,...,\cQ_n\}\subseteq \mathrm{Nodes}(\cG^{\text{reg},f}_{\cT})$ such that $\cP^i\xrightarrow[]{R} \cQ^i$ for all $i\in \{1,...,n\}$. Then we say that the regions $\cR_1$ and $\cR_2$ have cardinality $n$ relative to this fine-graining, and denote it as $|\cR_1|=|\cR_2|=n$, when the fine-graining is clear from context.  For any given spacetime realisation of the process network $\mathfrak{N}_{\mathcal{W},N}$, $|\cR|$ will denote the cardinality of $\cR$ relative to the fine-graining $\mathfrak{N}_{\mathcal{W},N, \cT,\cE}^{f}$ in the specification of the realisation (\cref{definition: spacetime_realisation}). We will require that the inputs $A^s_k$ and $A^I_k$ of each agent $A_k$ are assigned a spacetime region $\cR^{A^I_k}$ and the outputs $A^O_k$ and $A^o_k$ are assigned a spacetime region $\cR^{A^O_k}$ such that there is a correspondence from the input to the output region.  A physical example of such a correspondence would be a situation where Alice may receive inputs in a region $\cR^{A^I}$ consisting of the spacetime locations $(\vec{x},t_1), ...,(\vec{x},t_n)$ and produce outputs in a region $\cR^{A^O}$ consisting of the locations $(\vec{x},t_1+\delta), ...,(\vec{x},t_n+\delta)$ for some fixed $\delta>0$. For the purpose of the following results, we consider spacetime realisations of processes where such an input-output correspondence is satisfied. Note that this still permits general scenarios where agents' receive/send quantum messages at a superposition of different times (as we will see, for the quantum switch in \cref{sec:qs_expts}).

 \begin{restatable}{theorem}{PMFinegrain}
 \label{theorem: PMFinegrain}
Any spacetime realisation of a process network $\mathfrak{N}_{\mathcal{W},N}$ in a spacetime $\cT$ relative to an embedding $\cE$ that satisfies relativistic causality will admit an explanation in terms of a fine-grained network $\mathfrak{N}_{\mathcal{W}^f,M}$ which will be described by a fixed order and therefore causally separable process $W^f$ associated with $M\geq N$ (possibly communicating) agents. Here $M$ is given in terms of the cardinality of the associated input spacetime regions $M=\sum_{k=1}^N |\cR^{A^I_k}|$.
 \end{restatable}

We clarify what we mean by ``possibly communicating agents'' in the above theorem. In a usual process matrix involving at least two agents $A_1$ and $A_2$, the operations performed by $A_1$ and $A_2$ are taken to be in product form $\cM^{A_1}\otimes \cM^{A_2}$ with any communication between the agents going through the process. In general, plugging in operations on $A^I_1\otimes A^I_2$ which are not of this form into a process matrix can result in non-normalised probabilities.
In the above fine-grained description, we allow for situations where non-product operations can be plugged into the fine-grained process. This models situations where $A_1$ and $A_2$ can be communicating, as long as the communication respects relativistic causality. We show in the proof of the theorem that the fine-grained process $W^f$ will necessarily be such that it will lead to valid probabilities even when acting on these more general set of operations. Allowing for such operations at the level of the information-theoretic description is similar in spirit to multi-round processes introduced in \cite{Hoffreumon2021}, a formalism that generalises the process matrix framework to scenarios where agents can act in multiple rounds while carrying along a local memory from previous rounds.  With these results, we immediately obtain the following corollary.

\begin{restatable}{corollary}{NogoB}
\label{corollary: nogoB}
    No fixed spacetime realisation of a process network $\mathfrak{N}_{\mathcal{W},N}$ in a spacetime $\cT$ relative to an embedding $\cE$ can satisfy the following conditions simultaneously
    \begin{enumerate}
        \item The spacetime realisation satisfies relativistic causality (cf. \cref{def: rel_causality_network})
        \item The spacetime realisation of the network does not admit any explanation in terms of a causally separable process that is consistent with the perspectives of all the agents involved.
    \end{enumerate}
\end{restatable}

In the above, perspectives of agents refer to their description of the spacetime realisation relative to their classical reference frames which they use for specifying their co-ordinate system in the fixed spacetime, which satisfies the natural properties detailed in \cref{definition: coordinates}. Further discussions on the physical implications of these results and the operational meaning of the fine-grained process can be found in \cref{sec:qs_expts}. There, we apply these results to the special case of the quantum switch process and its experimental realisations and also discuss related arguments in the previous literature which were made for this case.

\section{Causality in the quantum switch}
\label{sec: QS}

\begin{figure}[t]
\centering
\begin{subfigure}[b]{0.45\textwidth}
\centering
\includegraphics[scale=0.7]{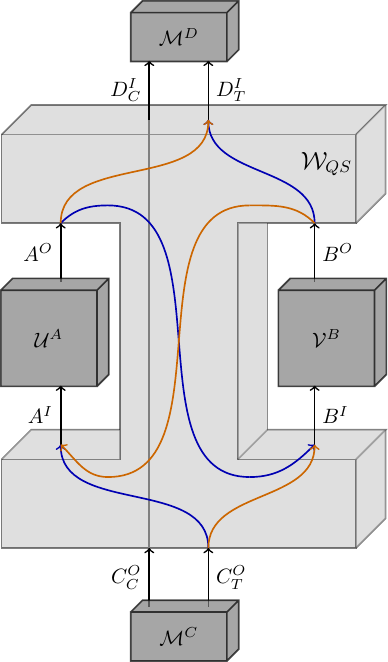} 
   \caption{}
\end{subfigure}\qquad\quad
\begin{subfigure}[b]{0.45\textwidth}
\centering
\includegraphics[scale=1.0]{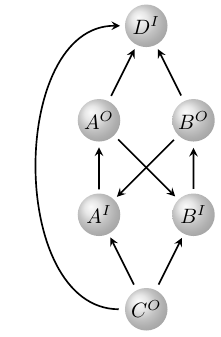} \caption{} 
\end{subfigure}
 
	\caption{(a) The quantum switch can be described as a 4-partite process matrix, where agent $C$ is in the past of all others and their local operation prepares the control and target systems in the initial state given on the right hand side of \cref{eq: QS}, with $C^O_C$ and $C^O_T$ playing the role of $C$ and $T$. The agents $A$ and $B$ apply their local operations $\mathcal{U}^A: A^I\mapsto A^O$ and $\mathcal{V}^B: B^I\mapsto B^O$ respectively, where the process map $\mathcal{W}_{QS}$ sends the target system to $A$ or $B$'s lab in an order that depends on the control. And finally, an agent $D$ in the future of all others receives the target from $A$ or $B$ and control directly from $C$ (gray path) which will be in the final state given in the left hand side of \cref{eq: QS}, with $D^I_C$ and $D^I_T$ playing the role of $C$ and $T$. $D$ can then perform operations/measurements on this state. The process matrix, $W$ for the quantum switch represents a controlled superposition of the orders $C\prec A\prec B\prec D$ (blue path) and $C\prec B\prec A\prec D$ (orange path) of operations on the target system. (b) The causal structure associated with the network formed by the composition of the process map $\mathcal{W}_{QS}$ with non-trivial local operations $\mathcal{U}^A$ and $\mathcal{V}^B$ is cyclic. $\mathcal{W}_{QS}$ is known to be a unitary, thus the causal structure depicted here coincides with the signalling structure over these systems \cite{Barrett2020}. Here we have grouped together $C$'s outputs as $C_O$ and $D$'s inputs as $D^I$.}
	\label{fig: extendedpm}
\end{figure}

To illustrate our general formalism and results, we apply it to the example of the quantum switch (QS), which is a particularly popular indefinite causal order process that has been claimed to be experimentally realised in Minkowski spacetime \cite{Procopio2015, Rubino19, Goswami2018}. QS corresponds to a unitary, causally non-separable process, and we can therefore make stronger statements for this case than Theorem~\ref{theorem: nogo_main} which applies to general processes. This clarifies the physical interpretation of such experiments which have been long debated, while consistently unifying the information-theoretic and spatio-temporal pictures on the causal structures involved. Further, we propose a new realisation of QS in 
Minkowski spacetime, with agents in relative inertial motion, where the agents cannot distinguish the order in which they act even if they were to measure the local time
at which they perform their operation. We explicitly show that this too admits the same physical interpretation, in terms of a definite and acyclic fine-grained causal structure, once we account for both spatial and temporal degrees of freedom.


\subsection{The QS process}
The quantum switch (QS) \cite{Chiribella2013} was originally defined as a higher-order transformation which takes as input two qudit quantum channels (for simplicity, take them to be unitaries $\mathcal{U}^A$ and $\mathcal{V}^B$), and outputs a quantum channel that applies $\mathcal{U}^A$ and $\mathcal{V}^B$ on a target system in a quantum controlled superposition of orders. Explicitly, the final channel obtained from the action of the higher-order transformation on $\mathcal{U}^A$ and $\mathcal{V}^B$ implements the following transformation on an initial state of a control qubit $C$ and a target qudit $T$ given on the left (where $\alpha$ and $\beta$ are arbitrary normalised amplitudes and $\ket{\psi}^T$ is an arbitrary qudit state).
\begin{equation}
\label{eq: QS}
    (\alpha\ket{0}^C+\beta\ket{1}^C)\otimes \ket{\psi}^T \mapsto \alpha \ket{0}^C\otimes \mathcal{V}^B\mathcal{U}^A\ket{\psi}^T+\beta \ket{1}^C\otimes \mathcal{U}^A\mathcal{V}^B\ket{\psi}^T
\end{equation}
It is important that each of $\mathcal{U}^A$ and $\mathcal{V}^B$ are applied no more than once, else, the above transformation can be easily simulated using causally ordered quantum circuits, using multiple queries to the operations \cite{Chiribella2013}. In the process matrix framework, QS can be modelled as a 4-partite process $W$ associated with a cyclic causal structure, as shown in \cref{fig: extendedpm}. The closed labs assumption ensures that each agent in the theoretical process matrix description acts exactly once. Further details on the process matrix description of QS are reviewed in \cref{appendix: PM}. 

\begin{figure}[t]
    \centering
\hspace{-15mm}\begin{subfigure}[b]{0.4\textwidth}
\includegraphics[scale=0.4]{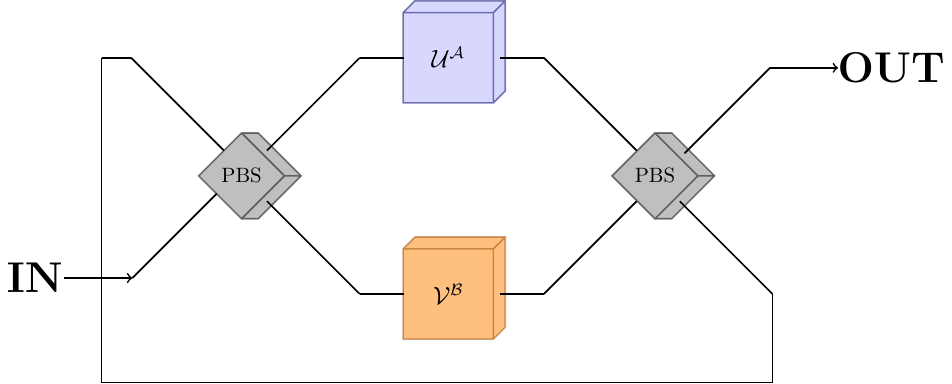}
    \caption{Optical quantum switch}
\end{subfigure}\hspace{2mm}
\begin{subfigure}[b]{0.5\textwidth}
\includegraphics[scale=0.6]{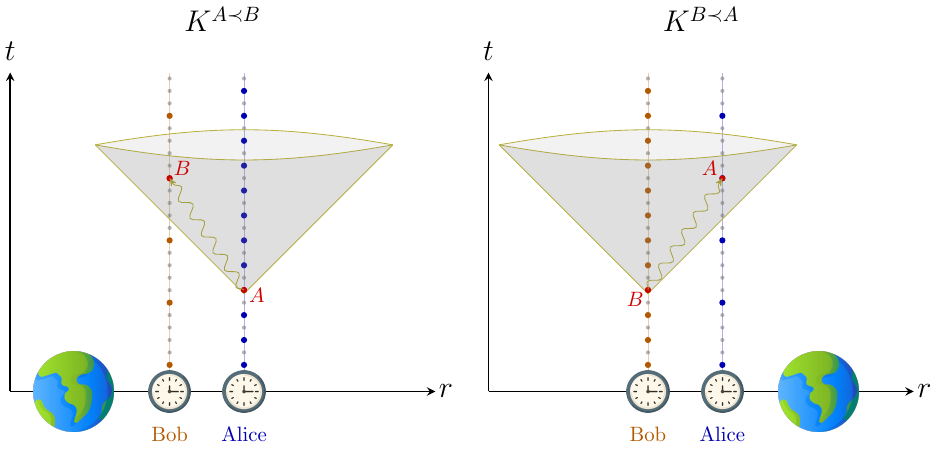}
       \caption{Gravitational quantum switch}
\end{subfigure}
    \caption{Classical and quantum spacetime realisations of the quantum switch:  (a) This is a schematic of a linear optical experimental implementation of QS that can be realised in Minkowski spacetime, reproduced from \cite{Araujo2014}. Here, the control qubit is encoded in the polarisation of a photon and the target qubit is encoded in a different degree of freedom of the same photon e.g., angular momentum modes. A horizontally polarized photon is transmitted by the all the polarizing beam splitters (PBSs) and takes the path where the unitary $\mathcal{U}^A$ is applied first and then $\mathcal{V}^B$ while a vertically polarized photon is reflected by all PBSs and follows the path where $\mathcal{V}^B$ is applied before $\mathcal{U}^A$. The unitaries $\mathcal{U}^A$ and $\mathcal{V}^B$ act only on the target degrees of freedom. This was experimentally realised in several works, such as \cite{Procopio2015, Rubino19}. (b) This is a theoretical proposal \cite{Zych2019} for implementing the QS process using a quantum superposition of gravitating masses, which results in a superposition of spacetime structures. Here, Alice and Bob are each in possession of their own clock $C_A$ and $C_B$ which are initially synchronised. A gravitating mass is prepared in a quantum superposition of macroscopically distinguishable spatial configurations depending on the state of a control qubit. If the control is in the state $\ket{0}$, the mass is placed closer to Bob such that the clock $C_B$ ticks slower that $C_A$ due to gravitational time dilation enabling Alice to send the target system to Bob at a proper time $t_A=3$, such that it is received by Bob at his proper time $t_B=3$. This mass configuration is labelled as $\kappa_{A\prec B}$. If the control is in the state $\ket{1}$, the mass is placed closer to Alice, enabling Bob to send the target system to Alice at $t_B=3$ such that it is received by Alice at $t_A=3$. This mass configuration is labelled as $\kappa_{B\prec A}$. Here, irrespective of the control, both agents receive the target system in their lab at the same local time. The co-ordinate axes are in the frame of a distant observer, whose time intervals (small gray dots) are unaffected by the mass configurations.
    }
    \label{fig: QSexperiments}
\end{figure}

\bigskip

\paragraph{Realisations of QS in different physical regimes}
Figure~\ref{fig: QSexperiments} provides an overview of two proposed realisations of the quantum switch transformation, where it has been argued that there is a sense in which the operations are queried no more than once each. The former corresponds to table-top optical setups in Minkowski spacetime. The latter corresponds to a theoretical proposal for realising the quantum switch through a quantum superposition of gravitating masses \cite{Zych2019}, and involves a quantum indefinite spacetime structure.

\subsection{No go result for QS}

We now derive a stronger version of our general no-go result, \cref{theorem: nogo_main} for the particular case of QS. In the following, we will call a fixed local operation $\mathcal{M}^{A_k}_{a_k}: A^I_k\mapsto A^O_k$ of some agent $A_k$ non-trivial if $\{A^I_k\}\longrightarrow \{A^O_k\}$ in $\mathcal{M}^{A_k}_{a_k}$. For this result, we will consider the network $\mathfrak{N}_{\mathcal{W}_{QS}, U, V}$ which is formed by the composition of the process map $\mathcal{W}_{QS}$ of the quantum switch (which has In $=\{C^O_C,C^O_T,A^O,B^O\}$ and Out $=\{D^I_C,D^I_T,A^I,B^I\}$) with the non-trivial and fixed local operations $\mathcal{U}^A: A^I\mapsto A^O$ and $\mathcal{V}^B: B^I\mapsto B^O$ of Alice and Bob, obtained by loop composing the systems with the same labels. This is distinct from the process network considered in the previous sections, as the local maps $\mathcal{U}^A$ and $\mathcal{V}^B$ here correspond to a fixed choice of setting and are not the same as extended local maps which include all possible setting choices. In the rest of the paper, whenever we refer to the process network of QS, we mean $\mathfrak{N}_{\mathcal{W}_{QS}, U, V}$ and not the network with the extended maps.

\begin{restatable}{lemma}{QS}[No-go result for the quantum switch]
\label{lemma: nogo_QS}
Let $\mathfrak{N}_{\mathcal{W}_{QS}, U, V}$ be the network describing the action of $\mathcal{W}_{QS}$ on any two non-trivial local operations $\mathcal{U}^A: A^I\mapsto A^O$ and $\mathcal{V}^B: B^I\mapsto B^O$, where $\mathcal{W}_{QS}$ is the process map associated with the process matrix $W_{QS}$ of the quantum switch. Then any realisation of this network in a fixed spacetime $\cT$ with respect to an embedding $\cE$ cannot simultaneously satisfy both of the following assumptions
\begin{enumerate}
    \item The spacetime realisation satisfies relativistic causality (cf. \cref{def: rel_causality_network}).
    \item The region causal structure over the nodes $\cE(S)$ for $S\in \{A^I,A^O,B^I,B^O\}$ is acyclic.
\end{enumerate}
\end{restatable}

The first assumption of \cref{theorem: nogo_main} is automatically satisfied here as $W_{QS}$ is a causally non-separable and hence, is not a fixed order process. Moreover, this is a stronger statement than that of \cref{theorem: nogo_main} applied to this process because \cref{theorem: nogo_main} deals with the extended local maps (which include all possible setting choices) while the above statement follows for any fixed choice of settings for $A$ and $B$ associated with non-trivial local operations. For a general process, the analogous statement with fixed operations may not hold.\footnote{For general processes, whether or not agent $A_i$ can signal to an agent $A_j$ can depend on the choice of local operations of $A_j$ as well as those of the remaining agents. In the quantum switch, signalling from $A$ to $B$ (whenever possible, i.e., the initial state $\alpha\ket{0}^C+\beta\ket{1}^C$ of the control has $\alpha\neq 0$) can be verified by $A$ by suitable local choices of operations alone, independently of $B$'s operation, and the symmetric statement holds for $B$.}

\begin{figure}[t]
\centering
\subfloat[]{%
\includegraphics[scale=0.6]{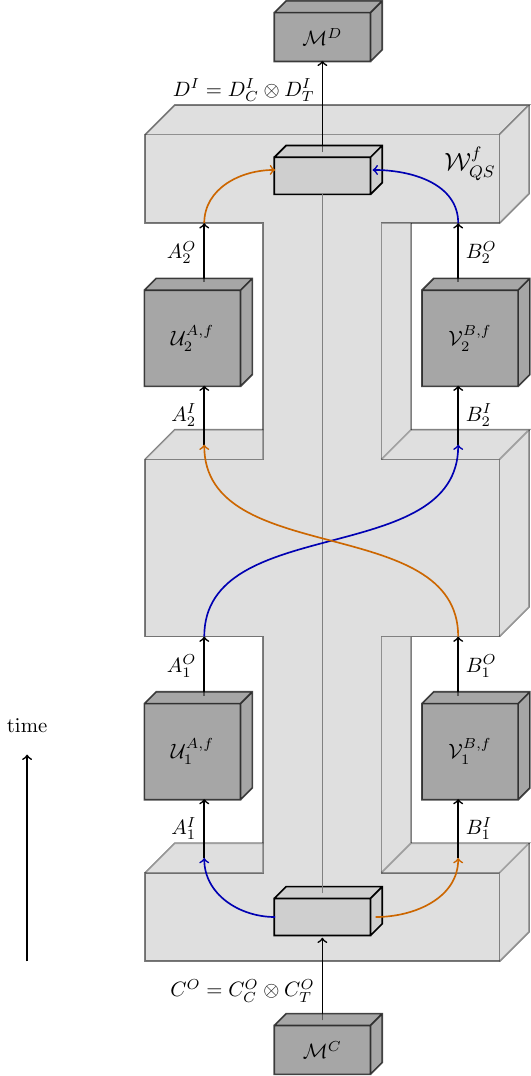}}%
\qquad\qquad\qquad\subfloat[]{%
\includegraphics[scale=1.0]{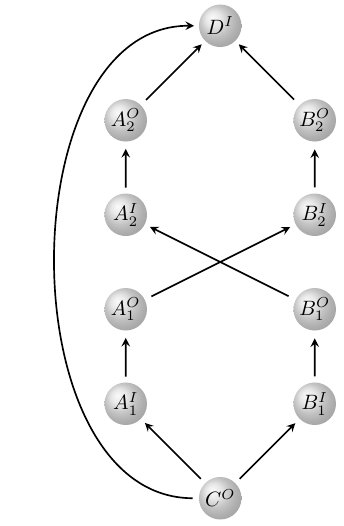}}%
\caption{ Fine-graining of QS associated with its spacetime realisation: (a) A schematic of the fine-grained network associated with typical spacetime realisations of QS. It is associated with a 6-partite process $\mathcal{W}_{QS}^f$ which is a fixed order process but implements, for arbitrary unitaries $\mathcal{U}^A$ and $\mathcal{V}^B$, the QS transformation of \cref{eq: QS} from the control and target received on $C_O$ to corresponding systems in $D^I$ when Alice and Bob (who now act at two distinct times $t_1$ and $t_2>t_1$) apply the same unitary at both times. More detailed descriptions and diagrams of this network, which describe the maps in the internal decomposition of $\mathcal{W}_{QS}^f$ can be found in \cref{appendix: CBQS}. (b) The causal structure of the fine-grained network, which is definite and acyclic. }
	\label{fig: PMQS_fg}
\end{figure}

\subsection{Implications for experimental realisations}
\label{sec:qs_expts}

We first state a corollary of \cref{lemma: nogo_QS} and \cref{corollary: nogoB} which applies to experimental realisations of QS in a fixed spacetime. We then describe a fine-grained network associated with such a realisation, and discuss in detail, the physical interpretation of QS experiments in Minkowski spacetime offered by our results.

\begin{restatable}{corollary}{QSExpt}
\label{corollary: QSExpt}
Any experiment realising the action of the 4-partite process map $\mathcal{W}_{QS}$ on non-trivial local operations $\mathcal{U}^A$ and $\mathcal{V}^B$ without violating relativistic causality in a fixed spacetime will admit an explanation in terms of a fine-grained network with the following properties
\begin{enumerate}
    \item It is a network formed by an $M$-partite process map $\mathcal{W}_{QS}^f$ with $M>4$ together with operations of the $M$ parties. 
    \item $\mathcal{W}_{QS}^f$ is a fixed order process compatible with an acyclic causal structure over the $M$ parties.
     \item The fine-grained description is consistent with the perspectives of all agents involved.
\end{enumerate}

\end{restatable}

These results prove the existence of a fine-grained description with the desired properties. For illustration, we present in detail a particular fine-graining of QS that describes its spacetime realisations while satisfying all these properties. In this case, we can also analyse the causal structure (and not just the signalling structure). The fine-grained network and causal structure are illustrated in \cref{fig: PMQS_fg}. This fine-grained network in fact coincides with a description of QS originally proposed in the causal box framework \cite{Portmann2017}. By connecting both process matrices and causal boxes to our more general framework, our work provides a mapping between the process matrix and causal box descriptions of QS (even though one admits an indefinite and other a definite information-theoretic causal structure) and reveals that these can be understood as coarse and fine-grained descriptions that are related through appropriate encoders and decoders. We describe the encoding and decoding in \cref{appendix: CBQS}. Here we outline the main features of this fine-grained network and associated fixed-order process $\mathcal{W}_{QS}^f$.

\bigskip

\begin{sloppypar}
    The network $\mathfrak{N}_{\mathcal{W}_{QS},U,V}$ associated with the original process, involves the set  $\{C^O_C,C^O_T,A^I,A^O,B^I,B^O,D^I_C,D^I_T\}$ of systems. Suppose it is realised in Minkowski spacetime according to the following embedding.\footnote{In this example, the embedding is specified relative to the maximal fine-graining, i.e., in terms of individual spacetime locations comprising the region.}
\end{sloppypar}

\begin{equation}
\label{eq: embedding_QS}
\begin{split}
       \mathcal{E} (C^O_C)= \mathcal{E} (C^O_T)&:=\{L^C\}\\
        \mathcal{E}(A^I)&:=\{P_1,P_2\},\\
       \mathcal{E}(A^O)&:=\{Q_1,Q_2\},\\
        \mathcal{E}(B^I)&:=\{R_1,R_2\},\\
       \mathcal{E}(B^O)&:=\{T_1,T_2\},\\
        \mathcal{E} (D^I_C)= \mathcal{E} (D^I_T)&:=\{L^D\}.
\end{split}
\end{equation}

Where the spacetime points are ordered as follows
\begin{align}
\label{eq: QS_Minkowski}
 \begin{split}
 &L^C\prec P_1\prec Q_1\prec R_2\prec T_2 \prec L^D,\\
&L^C\prec R_1\prec T_1\prec P_2\prec Q_2 \prec L^D. 
 \end{split}   
\end{align}

In the fine-grained process, $C$ and $D$'s systems have isomorphic states spaces to the original ones while each of $A$ and $B$'s systems split into two systems, one for each spacetime location. For instance, $A^I$ in the coarse-grained process maps to $\{A^{I,P_1},A^{I,P_2}\}$ under the associated systems fine-graining map. To avoid clutter, we simply label these as $\{A^I_1,A^I_2\}$ and similarly for the remaining systems of $A$ and $B$. Then the fine-grained process $\mathcal{W}_{QS}^f$ thus obtained, has In $=\{C^O_C,C^O_T,A^O_1,A^O_2,B^O_1,B^O_2\}$ and Out $=\{D^I_C,D^I_T,A^I_1,A^I_2,B^I_1,B^I_2\}$. The local operation $\mathcal{U}^A$ (taking it to be a unitary for simplicity) of Alice gets fine-grained to $\mathcal{U}^{A,f}_{1,2}=\mathcal{U}^{A,f}_1\otimes \mathcal{U}^{A,f}_2: A^I_1\otimes A^I_2\mapsto A^O_1\otimes A^O_2$ and similarly for Bob. In the fine-grained network, the presence of a $d$-dimensional target system at a spacetime location is modelled as a single $d$-dimensional message at that location and the absence of the target at a location as a vacuum state $\ket{\Omega}$ (zero message state) at the location. The fine-grained systems of Alice and Bob are therefore each $(d+1)$-dimensional when the original target is $d$-dimensional, as they additionally include the 1-dimensional vacuum state $\ket{\Omega}$.

\bigskip

For the sake of illustration, suppose that the spacetime locations are such that the inputs $A^I_1$ and $B^I_1$ are associated with a time co-ordinate $t_1$ while $A^I_2$ and $B^I_2$ with a time co-ordinate $t_2>t_1$ in some global classical reference frame.  Then, 
when the control is $\ket{0}$, the target qudit is sent to Alice at time $t_1$ and then Bob at time $t_2$ (and the vacuum state to Bob at time $t_1$ and Alice at time $t_2$) and when the control is $\ket{1}$, the target is sent to Bob at time $t_1$ and Alice at time $t_2$ (and the vacuum state to Alice at time $t_1$ and Bob at time $t_2$). The fine-grained local operations at each time are such that they leave the vacuum state invariant and apply a qudit unitary $\mathcal{U}_i^A$ whenever the input is non-vacuum, where $i\in \{1,2\}$. That is,
\begin{align}
\label{eq: vacuum}
    \begin{split}
        \mathcal{U}^{A,f}_i\ket{\Omega}^{A^I_i}&=\ket{\Omega}^{A^O_i}\\
        \mathcal{U}^{A,f}_i\ket{\psi}^{A^I_i}&=\mathcal{U}_i^A\ket{\psi}^{A^O_i}
    \end{split}
\end{align}

 In particular, composing the fine-grained process map $\mathcal{W}^f_{QS}$ with the corresponding local operations $\mathcal{U}^{A,f}_{1,2}$ and $\mathcal{V}^{B,f}_{1,2}$ implements the following transformation (see \cref{appendix: CBQS} for further details) on the control and target (where $C$ and $T$ on the left are associated with the corresponding output systems of Charlie, and on the right, to corresponding inputs of Danny).

\begin{equation}
\label{eq: qs_fg}
  (\alpha\ket{0}^C+\beta\ket{1}^C)\otimes \ket{\psi}^T \mapsto \alpha \ket{0}^C\otimes \mathcal{V}^B_2\mathcal{U}^A_1\ket{\psi}^T+\beta \ket{1}^C\otimes \mathcal{U}^A_2\mathcal{V}^B_1\ket{\psi}^T
\end{equation}

\bigskip

Setting $\mathcal{U}_1^A=\mathcal{U}_2^A$ and $\mathcal{V}_1^B=\mathcal{V}_2^B$, this recovers the QS transformation of \cref{eq: QS}. However, accounting for all possible interventions that the agents can physically perform in the spacetime (which includes time-dependent unitaries), we see the fine-grained, 6-partite process of \cref{fig: PMQS_fg} which has a definite acyclic causal structure accurately captures these possibilities. Indeed, in the coarse grained description, the process matrix $W_{QS}$ there is no such degree of freedom available to Alice and Bob through which they can implement a different unitary depending on the order in which they act but this is possible in the spacetime realisation. In this fine-graining, the spacetime location at which Alice/Bob receive a non-vacuum state is perfectly correlated with the control state and therefore the order in which the two agents act. There are QS realisations in Minkowski spacetime \cite{Goswami2018, Goswami_2020} where such perfect correlations don not exist. Our results nevertheless apply here, as discussed in \cref{remark: order_distinguis_qs} in the appendix.

\bigskip
\paragraph{ Role of vacuum states} Similar constructions using vacuum states, for describing experimental realisations of QS have been proposed in the literature \cite{Portmann2017, Paunkovic2019, Vilasini_thesis, Chiribella_2019, Ormrod2022}. The associated fine-graining of QS has the property that each agent acts exactly once on a $d$-dimensional non-vacuum state. One can construct other valid fine-grainings where agents act on multiple $d$-dimensional non-vacuum messages. Arguably however, we could not regard such fine-grainings as corresponding to ``faithful realisations'' of the process as they would not keep with the spirit of the closed labs assumption of the process matrix framework (\cref{sec: PM}) which requires that each agent acts exactly once on the target qudit.  The general results of this paper do not refer to or depend on the existence of such vacuum states, they prove the existence of an acyclic fine-grained causal structure for physical spacetime realisations of any process irrespective of whether the realisation can be considered ``faithful'' in this sense, therefore generalising previous arguments for QS to all processes and also to a larger class of spacetime realisations. Modelling the set-up assumptions of the process framework at the fine-grained level is a subject of an upcoming work, based on \cite{us_QPL}. This enables further no-go results regarding the set of physically and faithfully realisable processes in a fixed spacetime (see also \cite{Salzger}).

\bigskip

\paragraph{Physical interpretation of the experiments}
Whether QS experiments in Minkowski spacetime implement or simulate an indefinite causal structure has been a subject of long-standing debate. Instead of directly focusing on this question, we addressed a fundamental and overlooked question that underpins the discussion: how can an indefinite information-theoretic causal structure (such as that of QS) be compatible with relativistic causality in a definite spatio-temporal causal structure (such as Minkowski spacetime)? We have found that in such experiments, at the fine-grained level, both causal structures can be regarded as being definite and acyclic. This supports the conclusion that these experiments do not implement a fundamentally indefinite causal structure, but they are useful for simulating these exotic causal structures. 

\bigskip

Our work also highlights that such experiments are nevertheless interesting and worthy of further investigation, as they involve the non-localisation of information in spacetime. A better characterisation of the underlying experimental resources, which can distinguish between quantum vs classical forms of non-localisation, would enhance our understanding of spatio-temporal quantum correlations which can play an important role in relativistic quantum information processing. 
Being explainable by an definite information-theoretic causal structure would only make it easier to analyse potentially new applications of such experiments in a fixed spacetime while carrying forth many intuitions from standard quantum information formalism. To this effect, in \cref{appendix: framework_recovers_CB} we show that all physical spacetime realisations of networks in our framework can equivalently be viewed as causal boxes \cite{Portmann2017}, a previously known formalism for modelling quantum messages exchanged at superpositions of spacetime locations, and which has applications in relativistic quantum cryptography \cite{Vilasini_crypto}.



\bigskip

That being said, there are other previous works which present arguments for the opposite conclusion, of regarding these experiments as genuine implementations of indefinite information-theoretic causal structures. A central piece in our results is our formalisation of relativistic causality, which we believe also highlights a core difference between these two opposite sides of the debate. We discuss this aspect further in the next paragraphs.

\bigskip
\paragraph{Relativistic causality and agents' interventions} 

Relativistic causality in spacetime implies that no physical operation an agent can physically perform in one region can transmit information to another region beyond the first region's future light-cone.
It is therefore important to take into account the set of all possible physical operations that an agent can in-principle perform in a given scenario. Let us focus on this aspect, and compare the original process matrix description and the experimental realisations of QS. In process matrix, agents can intervene in their closed laboratories, while the process models the uncontrollable external environment. In contrast, experimental QS setups occur within a single laboratory where experimenters can, in principle, intervene and control any part of the experiment. 
Our fine-grained description of the spacetime realization ensures that even with these interventions, superluminal signaling remains impossible. It is worth noting that the counterfactual possibility of different interventions is also central to information-theoretic approaches to causality based on causal inference \cite{Pearl2009}.

\bigskip

This distinction between considering \emph{all physically possible interventions} vs \emph{actually performed interventions} highlights an important difference between the two sides of the debate. In the fine-grained description of QS, if the two ``Alices'' and ``Bobs'' are constrained to perform identical operations at their respective spacetime locations, the action of the fine-grained network representing a fixed order, causally separable process can become indistinguishable from that of the coarse-grained network which represents a causally non-separable process (also see the paragraph following \cref{eq: qs_fg}). More broadly, when starting with an acyclic information-theoretic causal structure and limiting interventions so they can not individually act on certain nodes, it can appear as if the nodes where agents' operations take place no longer have a definite order. A similar effect also arises in the classical examples of \cref{fig:fg_example1} and \cref{fig:fg_example2}, where by restricting interventions to always act jointly on $A_1$ and $A_2$ without distinguishing them as separate nodes, the locus of agents' operations would define the nodes $\{A_1,A_2\}$ and $B$ between which there is bidirectional causation and signalling, and hence no definite ordering. Analogously, in the spacetime perspective, we could have $P_1\prec Q\prec P_2$ but when focusing on the region $\cR=\{P_1,P_2\}$ as a whole, we cannot assign a definite ordering to $\cR$ and $Q$. In different physical regimes beyond fixed spacetimes, the fundamental structure of allowed physical interventions might enable realizations of indefinite causal structures, where both fine and coarse-grained descriptions agree on the lack of a definite causal order. Thus, it would be intriguing to establish mathematical connections between our framework and other formalisms contributing to this debate,  as this would provide insights also for physical regimes beyond classical spacetime structures and classical reference frames (see \cref{ssec:QRF} for further discussion).

\subsection{A realisation of QS in Minkowski spacetime with systems localised in time}
\label{sec: QS_Minkowski}

\begin{figure}[t!]
    \centering
\includegraphics[scale=1.0]{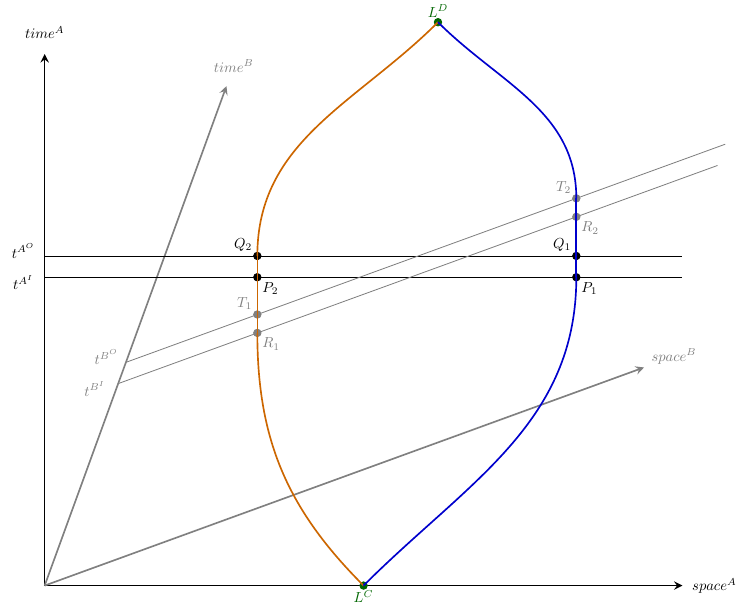}
    \caption{Quantum switch in Minkowski spacetime with time localised systems: Alice labels her spacetime locations with respect to the co-ordinate system associated with the black axes (orthogonal) and Bob with respect to the gray axes (non-orthogonal), which is related to Alice's frame by a Lorenz transformation. \cref{eq: embedding_QS} specifies how the spacetime points in the figure are associated with the in/output systems of the agents $A$, $B$, $C$ and $D$. When the control is in the state zero, the target follows the blue trajectory through the spacetime (left side in figure, first line of \cref{eq: QS_Minkowski}), going to Alice first and then to Bob and when the control is in the state $\ket{1}$, the target follows the orange trajectory through spacetime (right side in figure, second line of \cref{eq: QS_Minkowski}) going to Bob first and then to Alice. Irrespective of the order, Alice's in and output events are localised at times $t^{A^I}$ and $t^{A^O}$ in her frame and Bob's in and output events are localised at time $t^{B^I}$ and $t^{B^O}$ in his frame. However, the target degree of freedom is nevertheless not localised in the spacetime as the protocol implements a quantum controlled superposition of the two trajectories, to Alice and Bob this appears as non-localised in space but localised in time.}
    \label{fig: QS_protocol}
\end{figure}

We propose a new realisation of QS in Minkowski spacetime with the property that the in and output events are time localised for both Alice and Bob in their respective frames. The associated protocol requires Alice and Bob to be in relative motion with respect to each other, at a constant velocity. Previous quantum switch protocols in a fixed spacetime typically involve spatial localisation and time non-localisation, in contrast, ours will involve spatial non-localisation and time localisation (with respect to the local reference frames). It also demonstrates that Corollary~\ref{corollary: time} no longer holds when only requiring time localisation with respect to each local frame, as opposed to a single global reference frame.

\bigskip

The network $\mathfrak{N}_{\mathcal{W}_{QS},U,V}$ is realised in Minkowski spacetime according to the same embedding as \cref{eq: embedding_QS} and \cref{eq: QS_Minkowski}, and thus corresponds to the same fine-grained network as that of \cref{fig: PMQS_fg} at a purely information-theoretic level. The only distinction is that we no longer require $A^I_1$ and $B^I_1$ to be localised at time $t_1$ and $A^I_2$, $B^I_2$ at time $t_2>t_1$ as in \cref{fig: PMQS_fg}. Instead, this realisation has the property that $P_1$ and $P_2$ (associated with Alice's inputs) have the same time co-ordinate $t^{A^I}$ in while $Q_1$ and $Q_2$ (associated with Alice's outputs) have the same time co-ordinate $t^{A^O}>t^{A^I}$ in Alice's frame, and similarly $R_1$ and $R_2$ (associated with Bob's inputs) have the time co-ordinate $t^{B^I}$ while $T_1$ and $T_2$ (associated with Bob's outputs) have the time co-ordinate $t^{B^O}>t^{B^I}$ in Bob's frame. Thus the regions associated with Alice and Bob's systems are time localised regions (cf. \cref{definition: coordinates}). Further details of the protocol can be found in in Appendix~\ref{appendix: CBQS} and the intuition behind the proposed realisation and its main features are illustrated in \cref{fig: QS_protocol}.

\section{Towards demystifying indefinite causation}
\label{sec: discussions}

The concept of indefinite causal structures (in the sense of indefinite order of information-theoretic operations) has introduced valuable mathematical tools for studying quantum processes. However, its physical interpretation has remained debated. Consensus is lacking regarding their relationship to spatio-temporal causality notions, the operational meaning of indefinite causality witnesses like causal inequalities, and on their implications for the quantum nature of causal structures. In order to facilitate formal discussions and consensus on these matters, we have developed a general theoretical framework that formalizes different causality notions, connecting them through operational formulations of fundamental principles, under relatively minimal physical and mathematical assumptions. 
More generally, our graph theoretic-formulation of physical considerations makes it possible to analyse how other models of spacetime must structurally deviate from a fixed acyclic spacetime in order to evade the no-go theorems we have derived. We discuss the concrete implications of our work for interpreting indefinite causality within and beyond fixed background spacetimes and outline different directions for future research relating to this discussion. In \cref{sec: conclusions}, we also outline future research directions of more pragmatic relevance.

\subsection{Different notions of events}

Causality is closely related to the concept of events, with each type of causal structure implying a notion of events associated with its the nodes. In spatio-temporal causal structures, events correspond to spacetime points or regions, while in information-theoretic structures, events are linked to the in/output systems of operational procedures, reflecting where agents can intervene. Both notions are pertinent in physical experiments. Instantiating the abstract process matrix description in a specific physical regime, like a background spacetime, can provide agents access to more degrees of freedom not present in original coarse-grained process description. This can potentially expand the set of information-theoretic events that are physically relevant.

\bigskip

This is exemplified in our analysis of the quantum switch where Alice, in the process matrix, can only apply $\mathcal{U}^A$ to her input system $A^I$ independently of whether she acts before or after Bob, whereas in the spacetime realization we have analysed, she can apply different unitaries $\mathcal{U}^A_1$ or $\mathcal{U}^A_2$ based on the time order (associated with the background spacetime structure, cf. \cref{eq: qs_fg}). She must restrict these unitaries to be identical in order to realize the QS transformation of \cref{eq: QS}. Here Alice is associated with a single pair of in/output events in the process matrix, and two pairs in/output events in the spacetime realization. The former corresponds to a \emph{causally non-separable process} with one operational event for Alice and one for Bob, while the latter picture is described by a \emph{fixed-order, causally separable process} involving more operational events. This emphasizes the importance  of distinguishing whether (a) the original indefinite causal structure's behaviour is replicated by restricting the agents' operations in a fine-grained fixed-order processes which physically allows for a larger set of operations, or (b) even when considering all physical interventions that agents can perform in a given realisation of a process in some regime, no deviations from the original causally non-separable process can be observed. Our results indicate that in fixed and acyclic background spacetimes, case (a) always holds. The possibility of realising the transformation of \cref{eq: qs_fg} in quantum gravitational realisations of QS, and the associated notion(s) of events in such scenarios has been discussed in \cite{Moller2023}. More generally, formalizing our framework's requirements in regimes beyond fixed spacetime and exploring whether (a) or (b) is satisfied in different realizations is an intriguing avenue for future research, especially given our demonstration that (b) is impossible in a fixed spacetime.

\subsection{Relating the process matrix assumptions to those of our no-go theorems}

As discussed in Section~\ref{sec: PM}, in \cite{Oreshkov2012}, causal inequalities are shown to be necessary conditions on correlations generated by protocols satisfying the assumptions: free choice (FC), local order (LO), closed labs (CL) and that events are localised in a global causal structure (CS). Thus, a violation of causal inequalities under FC, LO and CL would indicate a violation of (CS), which is often interpreted as certifying the indefiniteness of the causal structure. 

\bigskip

However, a more careful look reveals that (CS) is essentially two assumptions: (CS1) there exists a global partial order in which the process is realised such that signalling is only possible from past to future with respect to this partial order and (CS2) the input/output events of every agent in the process are localised in this partial order. Within our framework, a fixed spacetime inherently defines this partial order, and (CS1) is automatically satisfied for any process realisation which respects our relativistic causality condition (Definition~\ref{def: rel_causality_network}) in a fixed spacetime. (CS2) then becomes an additional constraint on the spacetime embedding, requiring that each system $S$ is embedded at a single spacetime point $\cE(S)\in \cT$ (rather than a non-trivial region). This constraint implies the acyclicity of the region causal structure. Then it immediately follows from \cref{theorem: nogo_main} and \cref{corollary: localisation} that a process realised according to (CS)=(CS1)+(CS2) in any partial order must be a fixed order process and hence can not lead to correlations violating causal inequalities when the agents perform freely chosen operations. Interestingly our definitions of \emph{spacetime realisation} and \emph{relativistic causality} automatically account for the (LO) and (CL) assumptions once the localisation condition of (CS2) is satisfied. Specifically, (LO) follows because the signalling relations of the process network always include $\{A^I\}\longrightarrow \{A^O\}$ for every agent $A$, and relativistic causality together with (CS2) ensure that $\cE(A^I)\prec \cE(A^O)$ has the required local input-output ordering relative to the embedding partial order. (CL) is satisfied since each agent is associated with a single in/output pair of events (both in the informational and spatio-temporal sense) and hence acts only once. This extends the statement of \cite{Oreshkov2012}, showing that formalising (CS) as done in our framework, (CS) alone (/(CS)+(FC)) is sufficient to imply that the process being realised is a fixed order process (/does not generate correlations that violate causal inequalities).

\subsection{Operational meaning of causal inequalities}
\label{ssec: causal_ineq}

The previous sub-section discussed the implications of imposing (CS) = (CS1) + (CS2). By formalising these assumptions independently, our frameworks enables us to explore scenarios where (CS1) holds but not (CS2). This would correspond to fixed spacetime realisations of processes associated with a cyclic region causal structure, violating 3. in \cref{theorem: nogo_main} while satisfying 1. and 2. One can then ask: can a process realisation satisfying (CS1), (FC), (LO) and (CL) violate causal inequalities? Before answering this question, it is essential to note that, even if such a violation of causal inequalities were possible, it would not imply the indefiniteness of the causal structure, as \cref{theorem: PMFinegrain} implies that (CS1) alone guarantees a fine-grained explanation in terms of a fixed causal order process that all agents agree on. A violation of causal inequalities under the assumptions (CS1), (FC), (LO) and (CL) (if it were possible) would however provide a device-independent certification of the non-localization of information (i.e., violation of (CS2)) in a fixed and acyclic background causal structure. 

\bigskip

Having clarified the operational meaning of a potential violation of causal inequalities in a fixed spacetime, we will investigate in a follow-up work whether such violations are indeed possible. This requires formally defining the (LO) and (CL) assumptions for spacetime realizations where (CS1) holds but not (CS2), where an agent may receive a superposition of vacuum and non-vacuum states at a given spacetime location.  To the best of our knowledge, such a formalization is missing in the previous literature, but has been discussed in recent works involving one of the authors (the unpublished abstract \cite{us_QPL}, pre-print \cite{Salzger2024} and the master's thesis \cite{Salzger}) and will be explicitly addressed in forthcoming works based on this. Moreover the results of \cite{Salzger} suggest that under the conditions of (CS1) (or equivalently, satisfying relativistic causality in a fixed spacetime) + (LO) + (CL), the only realizable processes are effectively those corresponding to quantum circuits with quantum-controlled superposition of orders \cite{Wechs2021}, which include QS, and do not violate causal inequalities when combined with local operations satisfying (FC). Interestingly, through a different formulation of (CL) and (LO) and of non-localisation of information in time, \cite{Wechs2022} reaches the conclusion that causal inequality violations are physically realisable. There is no mathematical contradiction between these two results; the (CL) and (LO) assumptions of \cite{Wechs2022} operate at the level of the coarse-grained causal structure of the process matrix (where vacuum states do not feature), and this does not imply our formulation of these assumptions which operate at the fine-grained level of the spacetime realization (which includes vacuum and non-vacuum states). In the latter case, one wishes to ensure that each agent $A$ acts on a single non-vacuum message of dimension $d_{A^I}$ (dimension of $A$'s input system in the original process matrix description) in the spacetime realisation, which can be physically verified using coherent counters for each agent (this is the case for the fine-graining of QS presented in \cref{sec:qs_expts} and \cref{appendix: CBQS}). In our view, such a formulation of these constraints at the level of the fine-grained spacetime realisation is necessary for realising a process in a ``faithful'' or ``loophole free'' manner.  Establishing mathematical connections between the different formulations of (LO) and (CL), at the coarse and fine-grained levels, would inform a rigorous discussion in the community about what would constitute a ``loophole free'' violation of causal inequalities. 

\bigskip

Finally, we note an intriguing recent work \cite{Lugt2023} that derives a different kind of causal inequality than those discussed above, by including an additional ``relativistic causality'' assumption. Such inequalities can in fact be violated by an extended quantum switch protocol. In particular, this involves a quantum switch between Alice$_1$ and Alice$_2$ and a Bob who is space-like separated to the two Alices, who measures a system which is entangled with the control of the quantum switch. The relativistic causality assumption relates to a non-signalling condition between Bob and the two Alices. In this case as well, if a loophole free violation of such an inequality were observed in a fixed spacetime implementation of the quantum switch, then it would not certify any causal indefiniteness (at the fine-grained level) as implied by our results, but can instead be seen as a certification of the violation of (CS2) or witnessing non-classicality in the spacetime localisation of systems. This is further discussed in our associated letter \cite{us_short}.

\subsection{Cyclic causal structures and insights from causal inference}
\label{ssec: disc_cyclic}


The lack of a definite acyclic causal structure need not imply that the causal structure is indefinite (as commonly interpreted in the indefinite causality literature), but can also mean that the causal structure is definite but cyclic. Indeed, connections between indefinite and cyclic causal structures have been previously observed \cite{Chiribella2013, Oreshkov2012,Baumeler_2016, Araujo2017, Barrett2020}. While these results provide significant insights on the information-theoretic causal structure of process matrices, they do not directly address the questions regarding physical realisability of quantum processes in a spacetime or the relation between such cyclic causal structures and relativistic causality in acyclic spacetimes, which we have addressed here.  Building on these previous insights, our work shows a tight correspondence (cf. \cref{theorem: indef_cyclic}) between non-fixed order process matrices and cyclic causal structures as characterised by operationally observable signalling relations. 

\bigskip

Interestingly, cyclic information-theoretic causal structures can capture both physical scenarios with feedback \cite{Portmann2017, Bongers2021, VilasiniColbeckPRA} as well model physics in the presence of exotic closed timelike curves (CTCs) \cite{LloydPRD2011, LloydPRL2011, Araujo2017, Tobar2020, VilasiniColbeckPRL}. The physical distinction between these two situations come from how the cyclic causal structure is linked with spatio-temporal degrees of freedom \cite{VilasiniColbeckPRA, VilasiniColbeckPRL}, e.g., whether the nodes of the information-theoretic structure are embedded into spacetime regions or spacetime events (as discussed in \cref{fig: demand_price}). 
Linking indefinite causal structures of the process framework and cyclic quantum networks has allowed us to provide a clear physical interpretation to their realisations in an acyclic spacetime, which is analogous to the interpretation of cyclic causal models used in classical data sciences to model physical processes with feedback (which are not regarded as exotic CTCs). This interpretation is possible due to our relativistic causality condition which is a necessary constraint on the signalling relations of spacetime realised network needed to ensure that there are \emph{no superluminal (information-theoretic) causal influences}. Interestingly, in recent works \cite{VilasiniColbeckPRA, VilasiniColbeckPRL} involving one of the authors, it is shown that using a weaker relativistic causality principle which is necessary for ruling out \emph{superluminal signalling} (as opposed to causation, which is distinct from signalling), it is not possible to rule out genuine causal loops from being embedded in Minkowski spacetime. Moreover, there can be a certifiable gap between the information processing possibilities in theories constrained by these distinct relativistic causality principles \cite{VilasiniColbeck2024}. 
 In \cref{appendix: relation_to_VC}, we further discuss the relation between the distinct relativistic causality/compatibility conditions of \cite{VilasiniColbeckPRA, VilasiniColbeckPRL} and the present work, as well as their physical implications for causal loops. Characterising the general relation between such classes of causal loops found in the causal modelling framework of \cite{VilasiniColbeckPRA, VilasiniColbeckPRL} and process matrices remains an open problem.

\bigskip

More broadly, on the topic of cyclic causal structures, there are at least two interesting and inter-related research directions for future research. Firstly, the study of fine-graining of cyclic causal networks as initiated here can also provide a pathway for investigating the simulation of CTCs in a fixed acyclic spacetime by unravelling them into acyclic structures. Such simulations have been previously considered using experimental post-selection \cite{LloydPRL2011}. In general, we can simulate cyclic causal structures using acyclic ones even without invoking post-selection (the quantum switch network being an example, see \cref{sec: QS}). Which are the set of cyclic networks that admit acyclic fine-grainings without post-selection? How is this set related to the set of networks of CPTPMs which describe process matrices? 

\bigskip

The second direction relates to causal inference in the presence of cycles. It is known that several causal inference results which hold in classical and quantum acyclic models (such as the \emph{d-separation} theorem \cite{Pearl2009, Henson2014, Barrett2020}) fail in cyclic models, already in the classical case, and there is active research in characterising the domain on applicability of the \emph{d-separation} theorem in classical cyclic models \cite{Bongers2021}. A particular consequence of the \emph{d-separation} theorem is that if two nodes in a causal graph associated with classical variables, say $a$ and $y$ have no directed paths between each other, then we must have $P(ya)=P(y)P(a)$ in the resulting correlations. This does not hold in general cyclic models, but our results of \cref{appendix: equiv_signalling} imply that this would hold for the subset of cyclic networks corresponding to valid process matrices. An upcoming work \cite{Ferradini} further develops the causal modelling aspects of cyclic quantum networks, providing insights on graph separation properties such as d-separation. The precise subset of cyclic quantum networks that respect the \emph{d-separation} theorem, and their relation to the set of valid process networks remain open problems. Addressing them would simultaneously shed light on open questions in classical causal inference and the causal decompositions of process matrices, and would also provide useful graph-theoretic tools for addressing the first question of this paragraph relating to spacetime realisations.

\subsection{Beyond classical spacetimes and classical reference frames}
\label{ssec:QRF} 
We have modelled a fixed spacetime structure as a global partial order, which is the same in the all of reference frames (cf. \cref{definition: coordinates}). This is the case, for instance in special relativity, where reference frames are classical, the light cone structure of the spacetime is invariant under Lorenz transformations which relate such classical frames. Are there any potential features of information-theoretic causal structures which are impossible to achieve in a fixed spacetime but may be possible in scenarios that deviate from this model of fixed spacetime? We discuss this further in light of our no-go results.

\bigskip

Firstly, our model of a fixed spacetime imposes an agent independent notion of spacetime localisation/non-localisation. That is, if a system $S$ is localised at a spacetime point $P\in \cT$, all agents agree that $S$ is localised at $P$, and if $S$ is in a superposition of being at $P_1,P_2\in \cT$, all agents agree that $S$ is not localised and is non-localised between $P_1$ and $P_2$ (even if they may describe the points using different co-ordinates). This property is indeed satisfied in standard special and general relativity where classical reference frames, sharing a common origin, are used. This is also the case in most existing quantum experiments (such as the quantum switch experiments discussed in \cref{sec:qs_expts}). However, this property can fail (even classically) if different agents only have access to reference frames that are mutually unsynchronised, such that they cannot unambiguously order events. This highlights that the underlying assumptions about reference frames and about what information/degrees of freedom are accessible to agents play a crucial role in the identification and ordering of events (already classically).

\bigskip

More fundamentally, we may no longer expect an absolute notion of localisation to hold when spacetime co-ordinates are described using quantum reference frames (QRFs) since superposition and entanglement can become frame dependent in such settings, even using perfect QRFs \cite{Giacomini2019, Hohn_2021}. As we have seen in Figure~\ref{fig: QSexperiments}, realisations of the quantum switch in quantum gravitational spacetimes have been considered, and analysed using quantum clocks (temporal QRFs) \cite{Zych2019, Castro_Ruiz_2020}. Here it is possible that Alice sees her operation (and systems) to be localised in her proper time while she sees Bob's operation in a superposition of occurring earlier and later than her (and similarly for Bob). However, our quantum switch protocol in Minkowski spacetime (\cref{sec: QS_Minkowski}) illustrates that this property of agent-dependent time localisation is not specific to quantum gravitational or QRF based realisations, but can be achieved in a fixed spacetime. What is impossible in our model of fixed spacetime is agent-dependent spacetime localisation. 

\bigskip

To characterise whether agent-dependent spacetime localisation is possible in quantum gravitational realisations of processes, further analysis is needed which accounts for spatial and temporal information, and sheds light on how spacetime distances could be measured in these scenarios. For example, in the quantum gravitational realisation of QS, if the agents measure spatial distances with respect to the gravitating mass, then by construction, their spatial co-ordinate would depend on the branch of the superposition they are in (cf.\ Figure~\ref{fig: QSexperiments}) and they would not be spatially localised. On the other hand, if each agent localises their input event at the origin of their spatial co-ordinates, then each agent can possibly describe their own in/output events as being spatially and temporally localised while describing the in/output events of other agents as not being localised, in the gravitational QS. In a similar spirit, we note that a protocol to distinguish optical realizations of the quantum switch in Minkowski spacetime from theoretical quantum gravitational realizations was previously proposed \cite{Paunkovic2019}. The protocol involves an additional agent, the Friend $F$ to whom Alice and Bob send a photon every time their act on a non-trivial target system, and $F$ performs a non-demolition measurement to test a property relating to the time of arrival of the photon. In \cref{appendix: GQS}, we show that this protocol can be generalised to arbitrary fixed spacetime realisations by formulating the main property in order-theoretic terms, and showing it is impossible to achieve in any fixed spacetime. On the other hand, \cite{Paunkovic2019} suggests there can exist theoretical quantum gravitational models that achieve the property.

\bigskip

Our results for fixed spacetime together with this discussion highlights the following point: whether spatio-temporal co-ordinates are described in terms of classical or quantum RFs, whether the RFs used by different agents are synchronised etc. play an important role in the fine-grained description of these realisations as well as the physical interpretation of the associated causal structure. We envisage that extending our framework to allow for agent-dependent partial orders and fine-grained causal structures can enable a description of both notions of causality and their interplay in scenarios beyond the current model of fixed spacetime, where spacetime localization becomes subjective and where the spacetime may exhibit certain quantum effects. Generalizing our framework and testing the applicability of our no-go results in quantum gravitational realizations of quantum processes and QRF-based descriptions of spacetime coordinates presents another inter-disciplinary avenue for future research, which can have applications for understanding proposed table-top quantum gravity experiments \cite{Bose2017, Marletto2017}.

\bigskip

\section{Conclusions and outlook}
\label{sec: conclusions}

We have developed a general theoretical framework for studying quantum and relativistic causality, which satisfies the 4 desiderata laid out in the introduction. Applying the formalism to indefinite causal order processes, we derived several no-go results regarding their physical realisations in a background acyclic spacetime. In the context of the long-standing debate \cite{Portmann2017,Vilasini_thesis,Paunkovic2019, Oreshkov2019, Ormrod2022, Kabel2024}  regarding the interpretation of experimental realisations \cite{Procopio2015, Rubino19, Goswami2018} of the quantum switch in Minkowski spacetime,  this work furnishes formal theorems in support of the argument that these experiments do not implement a fundamentally indefinite causal structure. Rather, they are useful for simulating the behaviour of such exotic structures (see \cref{sec:qs_expts}) using definite and acyclic causal structures (in the spatio-temporal \emph{and} information theoretic sense).


\bigskip

Our work also highlights that such experiments still remain an interesting subject of further study, as they involve the non-localisation of information in space and in time. The result of \cref{appendix: framework_recovers_CB}, that any such quantum experiment in a spacetime can be described as \emph{causal boxes} \cite{Portmann2017}, enable their analysis in terms of a well-defined causal and compositional structure, and using fine-grained tools which account for information-theoretic and spatio-temporal aspects. Causal boxes also have natural applications for studying composable security in practical relativistic quantum cryptographic protocols \cite{Vilasini_crypto}. Obtaining a fine-grained characterisation of the spatio-temporal quantum resources and correlations involved in these experiments, certifying their non-classicality and developing their applications for relativistic quantum information processing still remain important future directions that can be investigated by building on this work and the related literature (e.g., \cite{Chiribella_2019, Rubino_2021}). 



\bigskip

More generally, apart from process matrices and causal boxes, our formalism also provides a platform for studying and relating other information-theoretic approaches to causality: such as functional causal models on cyclic graphs, classical and quantum Bayesian networks, and quantum split-node causal models (see \cref{appendix: examples_fg}). There is much scope for further developing the information-theoretic aspects of our framework for specific purposes. As discussed in \cref{ssec: disc_cyclic}, there are intriguing open questions regarding graph separation properties in cyclic quantum causal models which are related to ongoing research directions in classical causal inference \cite{Bongers2021}. Moreover, the concept of fine-graining introduced here, and our associated results, also highlight interesting parallels between information-theoretic and spatio-temporal notions of causality.

 \bigskip


Our compatibility condition, utilized in formulating relativistic causality, has broader scope. It allows to investigate the compatibility between detectable information flow in quantum networks with a partial order or acyclic graph, where the former is formulated independently of spacetime and without assuming a unique causal direction (allowing for cycles). This general approach to both information-theoretic and spacetime structures suggests avenues for extending techniques from the quantum causality literature towards the exploration of how spacetime structure may emerge from properties of quantum information, which is an active research area within the quantum gravity community (see, for example, \cite{Raamsdonk2010, Maldacena2013, Jahn2021, Kempf2021}). Finally, there have been recent works on generalising higher-order transformations (such as process matrices) in a quantum-theory independent manner \cite{Wilson2023}, as well as work on linking information-theoretic and spatio-temporal causality in a theory-independent manner \cite{VilasiniColbeckPRA, VilasiniColbeckPRL}. Building on these insights, it would be interesting to consider whether our approach would generalise to information-theoretic structures in post-quantum generalised probabilistic theories (GPTs), and which GPTs would still be constrained by similar no-go theorems.

 \bigskip

In the natural process of science, we may often need to update our preconceived notions in light of new experiments. Bell's theorem \cite{Bell1964} has set an unprecedented example in highlighting the power of no-go theorems in this scientific process--- establishing what is impossible to achieve within certain physical regimes tells us how physics in new regimes challenges our prior intuitions and how we can exploit these new physical phenomena for useful practical tasks. 
A theory of quantum gravity may further challenge our current understanding of causality in stronger ways than Bell's theorem, but we cannot fully anticipate how. The approach of disentangling different causality notions and carefully reconnecting them, as developed in \cite{VilasiniColbeckPRA, VilasiniColbeckPRL} and the present paper, can help better prepare for such challenges. We have shown how this approach provides concrete theorems for reinterpreting existing experiments and also to analyse how causality in more exotic physical regimes (such as quantum gravity) may differ. As we have outlined throughout this text, there are several fascinating questions that still remain open and areas where consensus is lacking regarding the interpretation of quantum and relativistic causal structures. We hope that our work, by formalising connections between different notions and approaches to causality, facilitates the process of building a consensus and also addressing some of these important open questions that span different research directions.

\bigskip
\textit{Note} Shortly after the first arXiv version of this manuscript was posted, an independent work \cite{Ormrod2022} appeared which also proposes a related concept of fine-graining for quantum causal structures, specifically in the context of the quantum switch. While our framework and definition of fine-graining have mathematical and conceptual differences from those of this work, the conclusions drawn about experimental realisations of the quantum switch are similar.

\bigskip
\textit{Acknowledgements} VV thanks Lin-Qing Chen, Augustin Vanrietvelde, \v{C}aslav Brukner, \"{A}min Baumeler, Esteban Castro-Ruiz, Hl\'{e}r Kristj\'{a}nsson, L\'{i}dia del Rio, Nick Ormrod, Lorenzo Maccone and Ognyan Oreshkov for interesting discussions on quantum and relativistic causality. VV's research has been supported by an ETH Postdoctoral Fellowship. VV and
RR acknowledge support from the ETH Zurich Quantum Center, the Swiss National Science Foundation via project No.\ 200021\_188541 and the QuantERA programme via project No.\ 20QT21\_187724.

\newpage
\appendix

\section{Overview of symbols and notation}
\label{appendix: notation}

Here, we provide an overview of the notations
used in the paper, with a particular focus on the graph-theoretic notation as the framework involves various types of order relations.

\bigskip

\noindent
\textbf{General notation}

 \begin{itemize}
     \item  $\mathcal{H}^S$: Hilbert space of a system $S$
     \item  $\mathcal{L}(\cH^S)$: Density operators on $\cH^S$ or state space of $S$
     \item $\mathcal{I}^S$: The identity channel on system $S$
     \item $\mathds{1}^S$: Identity matrix on system $\cH^S$ used for denoting states
      \item   $\oplus$: Addition modulo 2 
      \item $\Sigma(\cS)$: Powerset of a set $\cS$
     \item $P(X)$: Probability distribution of a random variable $X$
     \item $P(x)$: Probability of an event $x$ (such as a measurement outcome)
 \end{itemize}

\noindent
\textbf{Graphs, arrows and order relations}

\begin{itemize}
    \item $\mathcal{G}$:  Directed graph
    \item $\hookrightarrow$: $S\hookrightarrow S'$ denotes that an output system $S$ is loop composed to an input system $S'$ of some CPMs in a network of CPMs
    \item $\longrightarrow$: Order relation on sets of systems which indicates (information-theoretic) \emph{signalling} between them
    \item $\longrsquigarrow$: Order relation on systems which indicates (information-theoretic) \emph{causal influence} between them
    \item $\cT$: Partially ordered set used to denote spacetime structure
    \item $\prec$, $\preceq$, $\succ$, $\succeq$: Partial order relation of a spacetime $\cT$, with $P\prec Q$ (equivalently $Q\succ P$) denoting that $P$ and $Q$ are distinct with $P$ being strictly in the causal past of $Q$ while $P\preceq Q$ (equivalently $Q\succeq P$) implies $P\prec Q$ or $P=Q$
    \item $\not\preceq\not\succeq$: Used to denote that two spacetime locations are unordered relative to the partial order of the spacetime
    \item $\xrightarrow{R}$: Order relation associated with a region causal structure $\cG^{\mathrm{reg}}_{\cT}$ of a spacetime $\cT$
    \item $\xrightarrow{G}$: Edges of a generic directed graph $\cG$
\end{itemize}

 

\section{Further details of loop composition}

\label{appendix: framework_loop}
\begin{restatable}{lemma}{CPMcomp}[Closedness under composition]
\label{lemma: CPMcomp}
    CPMs are closed under arbitrary parallel and loop composition. 
\end{restatable}

\begin{proof}
    Note that parallel composition $\cM\otimes\cN$ of two CPMs can be equivalently written as the following sequential composition 
    \begin{equation}
    \label{eq: ParComp}
    \cM\otimes\cN= (\cI^{\text{Out}(\cM)}\otimes\cN )\circ(\cM\otimes \cI^{\text{In}(\cN)}).
    \end{equation}

Therefore $\cM\otimes\cN\otimes \cI^A$ for any system $A$, is equivalent to $(\cI^{\text{Out}(\cM)}\otimes\cN \otimes \cI^A )\circ(\cM\otimes \cI^{\text{In}(\cN)}\otimes \cI^A)$.
Complete positivity of the parallel composition $\cM\otimes\cN$ requires  that $\cM\otimes\cN\otimes \cI^A$ maps positive states to positive states. This follows from the fact that $(\cM\otimes \cI^{\text{In}(\cN)}\otimes \cI^A)$ maps positive states to positive states (by the CP property of $\cM$) and $(\cI^{\text{Out}(\cM)}\otimes\cN \otimes \cI^A )$ also maps positive states to positive states (by the CP property of $\cN$).

\bigskip

To see that loop composition preserves complete positivity, we show that loop composition connecting an output subsystem $S$ to an input subsystem $S'$ of the CPM $\cM$ can be equivalently represented as follows: a preparation of a maximally entangled state $\ket{\Phi^+}^{S'R}:=\sum_i\ket{ii}^{S'R}$ on the input $S'$ and an isomorphic ancilla $R$ with $\cH^{S'}\cong \cH^R$, followed by action of $\cM$, and finally a post-selection of the same maximally entangled state $\ket{\Phi^+}^{SR}:=\sum_i\ket{ii}^{SR}$ on the output $S$ along with the ancilla $R$.\footnote{The fact that we use non-normalised maximally entangled states does not affect the arguments regarding the CP property, since the overall normalisation constant that would result from using normalised states in the pre and post selection would be $\frac{1}{d_S}$, which is positive.} 
Explicitly, this results in an overall map acting on input systems $\text{In}\backslash \{S'\}$ and having output systems $\text{Out}\backslash \{S\}$, which acts on arbitrary input states $\rho^{\text{In}\backslash \{S'\}}$ as follows.
\begin{align}
\begin{split}  &\tr_{SR}\Big[\ket{\Phi^+}\bra{\Phi^+}^{SR}\Big(\cM\otimes \cI^R(\rho^{\text{In}\backslash \{S'\}}\otimes \ket{\Phi^+}\bra{\Phi^+}^{S'R})\Big)\ket{\Phi^+}\bra{\Phi^+}^{SR}\Big]\\
=&\bra{\Phi^+}^{SR}\Big(\cM\otimes \cI^R(\rho^{\text{In}\backslash \{S'\}}\otimes \ket{\Phi^+}\bra{\Phi^+}^{S'R})\Big)\ket{\Phi^+}^{SR}.
\end{split}    
\end{align}
It is clear that the above map acts as a CPM on the input $\rho$ as it is formed from the sequential composition of CPMs (preparation of $\ket{\Phi^+}$ on a disjoint set of subsystems, application of $\cM$ and $\cI$, projection on $\ket{\Phi^+}$ and tracing out, are all CPMs). In the previous paragraph, we have already argued why the sequential composition of the CPMs in \cref{eq: ParComp} is itself a CPM, it is easy to see that this argument holds for arbitrary CPMs and sequential compositions. Finally, we show that the map defined above is in fact the same as the loop composed map $\cM^{S\hookrightarrow S'}$, which will complete the proof and establish that loop compositions map CPMs to CPMs.

\begin{align}
    \begin{split}
 &\bra{\Phi^+}^{SR}\Big(\cM\otimes \cI^R(\rho^{\text{In}\backslash \{S'\}}\otimes \ket{\Phi^+}\bra{\Phi^+}^{S'R})\Big)\ket{\Phi^+}^{SR}\\
 =& \sum_{i,j,k,l} \bra{ii}^{SR}\Big(\cM\otimes \cI^R(\rho^{\text{In}\backslash \{S'\}}\otimes \ket{jj}\bra{kk}^{S'R})\Big)\ket{ll}^{SR}\\
 =& \sum_{i,j,k,l} \bra{i}^{S}\Big(\cM(\rho^{\text{In}\backslash \{S'\}}\otimes \ket{j}\bra{k}^{S'})\Big)\ket{l}^{S} \bra{i}^R\cI^R\ket{j}\bra{k}^R\ket{l}^R\\
  =& \sum_{i,k} \bra{i}^{S}\Big(\cM(\rho^{\text{In}\backslash \{S'\}}\otimes \ket{i}\bra{k}^{S'})\Big)\ket{k}^{S}\\
  =& \cM^{S\hookrightarrow S'} (\rho^{\text{In}\backslash \{S'\}}).
    \end{split}
\end{align}

\end{proof}

\begin{restatable}{lemma}{OrderIndep}[Composition order independence]
\label{lemma: OrderIndep}
Consider a CPM $\cM$ such that $S',R'\in \text{In}(\cM)$ and $S,R\in \text{Out}(\cM)$ with $\cH^{S}\cong \cH^{S'}$ and $\cH^{R}\cong \cH^{R'}$. Then we have the following
\begin{equation}
\cM^{\{S\hookrightarrow S', R\hookrightarrow R'\}}:=  \Big(\cM^{S\hookrightarrow S'}\Big)^{R\hookrightarrow R'}= \Big(\cM^{R\hookrightarrow R'}\Big)^{S\hookrightarrow S'}.
\end{equation}
  That is, the order of loop compositions do not matter.  
\end{restatable}

\begin{proof}
For simplicity and without loss of generality, we take $R'$ to be the first input subsystem and $S'$ to be the last input subsystem of $\cM$.  Then we have for arbitrary input state $\rho^{\text{In}\backslash \{R',S'\}}$ 
\begin{equation}
       \Big( \cM^{S\hookrightarrow S'}\Big)^{R\hookrightarrow R'}(\rho^{\text{In}\backslash \{R',S'\}})=\sum_{k,\ell}\sum_{i,j}\bra{k}^{R}\bra{i}^{S}\Big( \cM(\ket{k}\bra{\ell}^{R'}\otimes\rho^{\text{In}\backslash \{R',S'\}}\otimes \ket{i}\bra{j}^{S'})\Big)\ket{j}^S\ket{\ell}^R
\end{equation}

\begin{equation}
       \Big( \cM^{R\hookrightarrow R'}\Big)^{S\hookrightarrow S'}(\rho^{\text{In}\backslash \{R',S'\}})=\sum_{i,j}\sum_{k,\ell}\bra{k}^{R}\bra{i}^{S}\Big( \cM(\ket{k}\bra{\ell}^{R'}\otimes\rho^{\text{In}\backslash \{R',S'\}}\otimes \ket{i}\bra{j}^{S'})\Big)\ket{j}^S\ket{\ell}^R
\end{equation}
These two expressions only differ in the order of the summation, but this order does not matter (these are finite sums) and we can equivalently write $\sum_{i,j}\sum_{k,\ell}=\sum_{k,\ell}\sum_{i,j}=\sum_{i,j,k,\ell}$ and we have the following, which completes the proof.
\begin{align}
\begin{split}
   &\Big( \cM^{S\hookrightarrow S'}\Big)^{R\hookrightarrow R'}(\rho^{\text{In}\backslash \{R',S'\}})=  \Big( \cM^{R\hookrightarrow R'}\Big)^{S\hookrightarrow S'}(\rho^{\text{In}\backslash \{R',S'\}})\\
=&\sum_{i,j,k,\ell}\bra{k}^{R}\bra{i}^{S}\Big( \cM(\ket{k}\bra{\ell}^{R'}\otimes\rho^{\text{In}\backslash \{R',S'\}}\otimes \ket{i}\bra{j}^{S'})\Big)\ket{j}^S\ket{\ell}^R\\:=&\cM^{\{S\hookrightarrow S',R\hookrightarrow R'\}}(\rho^{\text{In}\backslash \{R',S'\}}).
\end{split}  
\end{align}
We note that the composition order independence under multiple loop compositions was first shown in \cite{Portmann2017} for so-called causal boxes which are CPMs with spacetime labels, satisfying relativistic notions of causality in the spacetime. The relation between our framework and causal boxes is discussed in \cref{appendix: framework_recovers_CB}. The above proof closely mirrors the original proof, but we have provided it here for completeness. We note that the original proof also applies to the infinite dimensional case (as causal boxes can act on infinite dimensional Fock spaces as input), where the loop composition is defined in a different manner (in terms of the infinite dimensional Choi-Jamiolkowski representation) but the definition  reduces to the finite dimensional definition provided here. For this paper, the finite dimensional case will suffice. 

\end{proof}

Applying this lemma recursively, we obtain the following corollary.
\begin{restatable}{corollary}{CorOrderIndep}
\label{corollary: CorOrderIndep}
Let $\cM$ be a CPM, let $\{S_k\}_{k\in \{0,1,...,K\}}$ be any $K$ distinct elements of $\text{Out}(\cM)$ and let $\{S'_k\}_{k\in \{0,1,...,K\}}$ be any $K$ distinct elements of $\text{In}(\cM)$ such that we have $\cH^{S_k}\cong \cH^{S'_k}$ for all $k\in \{0,1,...,K\}$. Then, 
denoting $\mathfrak{M}=\{S_k\hookrightarrow S_k'\}_{k\in \{0,1,...,K\}}$, we have, for any two permutations $\pi$ and $\sigma$ of the set $\{0,1,...,K\}$, 
\begin{align}
\begin{split}
 \cM^{\{S_k\hookrightarrow S'_k\in \mathfrak{M}\}}:=&\Bigg(\Big(\big(\cM^{S_{\pi(1)}\hookrightarrow S'_{\pi(1)}}\big)^{S_{\pi(2)}\hookrightarrow S'_{\pi(2)}}\Big)^{...}\Bigg)^{S_{\pi(K)}\hookrightarrow S'_{\pi(K)}}\\=&\Bigg(\Big(\big(\cM^{S_{\sigma(1)}\hookrightarrow S'_{\sigma(1)}}\big)^{S_{\sigma(2)}\hookrightarrow S'_{\sigma(2)}}\Big)^{...}\Bigg)^{S_{\sigma(K)}\hookrightarrow S'_{\sigma(K)}}   
\end{split}    
\end{align}
\end{restatable}

\section{Illustrative examples of fine-graining and link to causal models}
\label{appendix: examples_fg}

The causal modelling paradigm \cite{Pearl2009} provides an information-theoretic approach for connecting observable correlations and effects of interventions to underlying causal explanations. It was originally developed for scenarios that can be described by classical variables, and has found useful applications across (classical) data sciences. In recent years, there has been great progress in generalising this approach to define quantum and non-classical causal models \cite{Leifer2013,Henson2014,Wood2015,Pienaar2015,Ried_2015,Costa2016,Fritz_2015,Allen2017,Barrett2020A,Pienaar_2020}. We briefly outline some prominent causal modelling frameworks and how their main objects of study can be seen as special cases of the quantum information-theoretic networks that can be described in our formalism. 

\bigskip

In a classical deterministic (or functional) causal models, the nodes of the causal structure (a directed graph, as before) are random variables and each variable $X$ with a non-empty set $\text{par}(X)$ of parents is obtained by applying a deterministic function $f_X: \text{par}(X) \mapsto X$ to the parental variables. For parentless variables $X$, a probability distribution $P(X)$ is specified in the model. More generally, the formalism of classical Bayesian networks allows for an arbitrary stochastic map (or equivalently, a classical channel) from the parental set of each variable to the variable. For the case of acyclic causal structures, this has also been generalised to quantum and post-quantum probabilistic theories by considering the information-theoretic channels of the theory \cite{Henson2014}. Every quantum Bayesian network (which includes functional causal models and classical Bayesian networks as special cases) uniquely identifies an acyclic quantum circuit constructed by the parallel and sequential composition of quantum channels (CPTPMs) which correspond to networks of CPTPMs (since CPTPMs are closed under parallel and sequential composition). Therefore our framework can describe these as special case. Moreover, \cite{Barrett2020, Barrett2020A} introduces frameworks for split-node quantum causal models for acyclic and cyclic graphs. Here each node $N$ is not a system, but rather a ``slot'' associated with an input system $N_I$ and output system $N_O$ between which a quantum operation $\cM_N: N_I\mapsto N_O$ can be plugged in. Indefinite causal order process matrices can be described as cyclic causal models in this formalism. Objects arising in such split-node causal models can also be studied in our framework analogously to process matrices (cf. \cref{sec: PM_network}), by defining an extended local map in each slot. 

\bigskip

Further details on the general mapping between these approaches will appear in future work \cite{Ferradini}. Here, we focus on intuitive examples that illustrate how our framework can capture cyclic functional models, and how our concept of fine-graining apply to them. We will discuss two types of fine-grainings that can be considered. The procedure would be analogous for the quantum cases, as illustrated for the cyclic causal structure of the quantum switch in \cref{sec: QS} and \cref{appendix: CBQS}. 

\begin{example}[Fine-graining by splitting into smaller subsystems]
\label{example: fine-graining1}
Consider the directed cyclic graph of \cref{fig: eg1} and the following functional causal model on this graph where the nodes $A$, $B$, and $D$ correspond to random variables with cardinality 4, 2 and 2 respectively. Let $P(D)$ be some probability distribution on $D$ and the functions of the model be $f_A: (D,B)\mapsto A$ which maps $(D,B)=(0,0),(0,1),(1,0),(1,1)$ to $A=0,1,2,3$ respectively, and $f_B: A\mapsto B$ sets $B=0$ whenever $A\in\{0,1\}$ and $B=1$ whenever $A\in\{2,3\}$. This causal model can equivalently be described as a cyclic network (\cref{fig: eg1_map}), where we have two CPTPMs $\cM_1: \cL(D)\otimes \cL(B')\mapsto \cL(A)$ and $\cM_2: \cL(A')\mapsto \cL(B)$ which are connected by loop compositions $A\hookrightarrow A'$ and $B\hookrightarrow B'$, along with a preparation on $D$, which can also be seen as a CPTPM. The in/output systems of $\cM_1$ and $\cM_2$ encode the RVs in the computational basis and these CPTPMs implement the functions $f_A$ and $f_B$ respectively (in the computational basis).

\bigskip

Now, notice that we can split $A$ into two bits $A_1$, $A_2$ by identifying $A=0,1,2,3$ with $(A_1,A_2)=(0,0),(0,1),(1,0),(1,1)$ through an isomorphism. Then $D$ specifies the first bit $A_1$ while $B$ specifies the second bit $A_2$, and in addition $B$ is itself the first bit $A_1$. Thus we can equivalently
describe the situation through the functional relations $A_1=D$, $B=A_1$ and $A_2=B$, with the same distribution $P(D)$ over $D$. This specifies a causal model on the acyclic causal structure \cref{fig: eg1_fg}. Similarly, this defines an acyclic network as shown in \cref{fig: eg1_map_fg} which is a fine-graining of the cyclic network of \cref{fig: eg1_map} associated with the systems fine-graining $\cF(A)=\{A_1,A_2\}$, $\cF(B)=\{B\}$ and $\cF(D)=\{D\}$ and with a fine-grained map $\cM^f_i$ associated with each $\cM_i$ of the original network. The relevant encoders and decoders simply act as identities on the $B$ and $D$ systems and map between the 4-dimensional system $A$ and the two 2-dimensional systems $A_1$ and $A_2$ through the isomorphism mentioned above.\footnote{By construction of this example, the only difference between the coarse- and fine-grained networks is in the isomorphism relating the states of the 4-dimensional to those of the two 2-dimensional systems. Therefore it is immediate that the encoders and decoders mentioned in the text would satisfy both properties of \cref{definition: fg_map} for relating the induced map of every sub-network of the original network to the corresponding induced map of the fine-grained network. That is, the acyclic network defined here is indeed a fine-graining of the original cyclic network, as per \cref{definition: fg_network}. }
\end{example}

\begin{figure}[t!]
    \centering
 \subfloat[\label{fig: eg2_map}]{\includegraphics[scale=1.0]{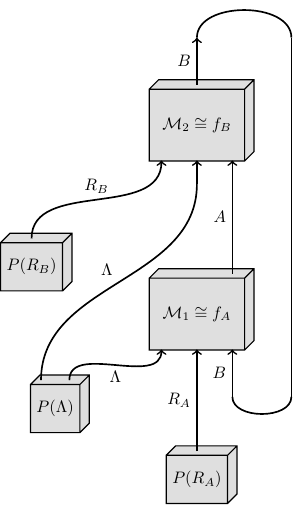}}\qquad\qquad\qquad\qquad \subfloat[\label{fig: eg2_map_fg}]{\includegraphics[scale=1.0]{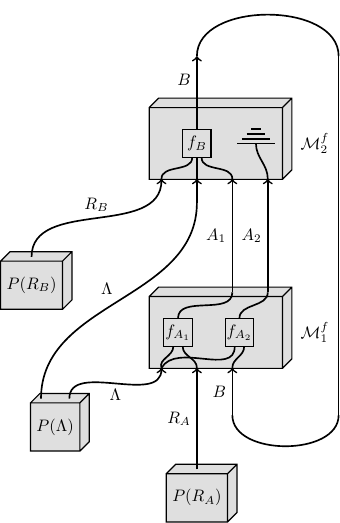}}

\vspace{7mm}
    
\subfloat[\label{fig: eg2}]{\includegraphics[scale=1.0]{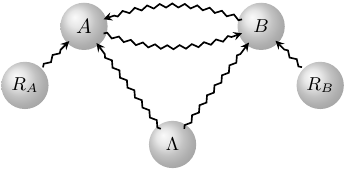}}\qquad\qquad\qquad\subfloat[\label{fig: eg2_fg}]{\includegraphics[scale=1.0]{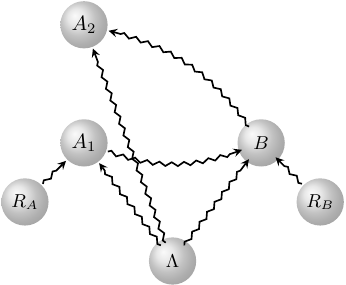}}

    \caption{(a) and (b) illustrate the original and fine-grained networks of  Example~\ref{example: fine-graining2}, while (c) and (d) illustrate the corresponding causal structures respectively. The maps $\cM_1$ and $\cM_2$ implement the deterministic functions $f_A$ and $f_B$ described in the text of \cref{example: fine-graining2}. }
    \label{fig:fg_example2}
\end{figure}

Our definitions of fine-graining do not place any contraints on how the dimensions of a system $S$ correspond to the dimensions of the set of fine-grained systems $\cF(S)$. In the above example, we fine-grained the network by splitting the 4-dimensional system $A$ into two smaller subsystems $A_1$ and $A_2$ of 2-dimensions each. We now present a simple example of another type of fine-graining which can be associated with the uncertainty in location of a message in a background causal structure, where the dimensions of the coarse-grained system does not correspond to a product of the dimensions of the corresponding fine-grained systems.

\bigskip

Suppose that Alice and Bob share a classical channel and a common source of randomness $\Lambda$ and they execute the following protocol.
Whenever $\Lambda=0$, Alice independently generates a bit $A$ and sends it to Bob who outputs a bit $B=A$ and whenever $\Lambda=1$, Bob independently generates a bit $B$ and sends it to Alice who outputs $A=B$. 
Whenever $\Lambda=0$, Alice acts before Bob and whenever $\Lambda=1$, Alice acts after Bob and whenever $\Lambda$ is unknown, there appears to be no definite acyclic order between Alice and Bob's operations. This is however a physically plausible protocol, which we can interpret as follows in terms of a definite acyclic causal structure, by breaking down Alice's operations (and associated system $A$) in two steps (associated with $A_1$ and $A_2$) and distinguishing when she exchanges a bit valued message with Bob and when she exchanges no messages with Bob (equivalently modelled as a trivial message $\Omega\neq 0,1$ exchanged with Bob). Then, when $\Lambda=0$,  $A_1$ contains a bit valued message which is sent from Alice to Bob while $A_2=\Omega$ and when $\Lambda=1$, $A_1=\Omega$ while $A_2$ equals the bit valued message that Alice receives from Bob. Thus the bit valued message which was earlier denoted by $A$ and related to $B$ through a non-definite causal order is now a bit that is ``non-localised'' in a well defined acyclic causal structure over a greater number of nodes $A_1$, $A_2$ and $B$ with $A_1$ preceding $B$ preceding $A_2$. The following example formalises this protocol as a cyclic causal model (associated with a cyclic network in our framework) which can be fine-grained to an acyclic causal model (associated with an acyclic network). For this, consider the following two acyclic causal models associated with two different causal orders between the variables $A$ and $B$. 
\begin{itemize}
    \item  CM$_0$: Graph $R_A\longrsquigarrow A \longrsquigarrow B$ with a model where $P(R_A)$ is arbitrary, $A=R_A$, $B=A$.
    \item CM$_1$: Graph $R_B\longrsquigarrow B \longrsquigarrow A$ with a model where $P(R_B)$ is arbitrary, $B=R_B$, $A=B$.
\end{itemize}

The cyclic causal model of the following example can be seen as realising a probabilistic mixture of these two acyclic models.
\begin{example}[Fine-graining through uncertainty in location]
\label{example: fine-graining2}
Consider the cyclic causal structure of Figure~\ref{fig: eg2} and the following causal model over this causal structure. The parentless nodes $R_A$, $R_B$ and $\Lambda$ are distributed according arbitrary distributions $P(R_A)$, $P(R_B)$ and $P(\Lambda)$ where $P(\Lambda)$ is not a deterministic distribution. The non-parentless nodes are related by the functional dependences: $A=(\Lambda\oplus 1)\cdot R_A\oplus \Lambda\cdot B:=f_A(\Lambda, R_A,B)$ and $B=(\Lambda\oplus 1) \cdot A\oplus \Lambda \cdot R_B:=f_B(R_B,\Lambda, A)$, where $\oplus$ denotes modulo-2 addition and $\cdot$ denotes multiplication (or logical AND). 
This causal model can equivalently be described as a cyclic network in our framework, as shown in \cref{fig: eg2_map} where the CPTP maps $\cM_1$ and $\cM_2$ implement the functions $f_A$ and $f_B$ respectively on the corresponding systems (which encode the associated variables). Notice that depending on whether $\Lambda=0$ or $\Lambda=1$, we effectively have the acyclic causal model CM$_0$ or CM$_1$, and in a general (since $P(\Lambda)$ is non-deterministic), a probabilistic mixture of the two.

\bigskip

We now construct a fine-graining of this cyclic causal model, which will be associated with the acyclic causal structure of Figure~\ref{fig: eg2_fg}. Consider a causal model on this acyclic graph, where the parentless nodes  $R_A$, $R_B$ and $\Lambda$ have the same distributions as before. The non-parentless nodes are $B$, $A_1$ and $A_2$. $B$ is binary as before, while $A_1$ and $A_2$ can now take an additional value associated with a symbol $\Omega$, thus making them trinary. The functional dependences are: $A_1=(\Lambda\oplus 1)\cdot R_A\oplus \Lambda\cdot \Omega:=f_{A_1}(\Lambda, R_A)$, $B=(\Lambda\oplus 1)\cdot A_1\oplus \Lambda\cdot R_B:=f_{B}(R_B,\Lambda, A_1)$, $A_2=(\Lambda\oplus 1)\cdot \Omega\oplus \Lambda\cdot B:=f_{A_2}(\Lambda\cdot B)$, where $\Omega\oplus 0=\Omega$, $0\cdot \Omega=0$ and $1\cdot \Omega=\Omega$. As illustrated in \cref{fig: eg2_map_fg}, this defines a fine-graining of the original network, where the fine-graining $\cM_1^f$ of the CPM $\cM_1$ encodes the functions $f_{A_1}$ and $f_{A_2}$ and the fine-graining $\cM_2^f$ of the CPM $\cM_2$ encodes the function $f_B$ (along with a trace on $A_2$). $A_1$ and $A_2$ correspond to qutrits encoding the corresponding variables in the basis $\{\ket{0},\ket{1},\ket{\Omega}\}$. Here we have, whenever $\Lambda=0$, the acyclic model CM$_0$ but with $A_1$ playing the role of $A$ and $A_2=\Omega$ and whenever $\Lambda=1$, the model CM$_1$ with $A_2$ taking the place of $A$ but $A_1=\Omega$. 
\end{example}

Let us describe explicitly, the encoders and decoders involved in the above example. Consider the encoding-decoding scheme that relates $\cM_1: \Lambda\otimes R_A\otimes B\mapsto A$ to $\cM_1^f: \Lambda\otimes R_A\otimes B\mapsto A_1\otimes A_2$. The encoder $\mathrm{Enc}_1$ in this case is a CPTPM from the inputs of $\cM_1$ to the inputs of $\cM_1^f$, since these are the same systems, the encoder is trivial (identity on all systems). The decoder $\mathrm{Dec}_1$ is a map from $A_1$ and $A_2$ to $A$ which must be CPTP on the image of $\cM_1^f\circ \mathrm{Enc}_1$. Notice that this image only has support on the following values of $A_1$ and $A_2$: $\{(0,\Omega),(1,\Omega),(\Omega,0),(\Omega,1)\}$. The decoder performs the mapping $\{(i,\Omega),(\Omega,i)\}\mapsto i$ for every $i\in \{0,1\}$. This can be physically realised through a measurement of $A_1$ and $A_2$ in the basis $\{\ket{0},\ket{1},\ket{\Omega}\}$ followed by reporting only the value which lies in $\{0,1\}$. It is then easy to check that $\cM_1=\mathrm{Dec}_1\circ \cM^f_1\circ \mathrm{Enc}_1$ holds, and also that all signalling relations of $\cM_1$ (each of $\{R_A\}$, $\{\Lambda\}$ and $\{B\}$ signal to $\{A\}$) are preserved in $\cM_1^f$ (each of $\{R_A\}$, $\{\Lambda\}$ and $\{B\}$ signal to $\{A_1,A_2\}$). Now consider the encoding-decoding scheme that relates $\cM_2: \Lambda\otimes R_B\otimes A\mapsto B$ to $\cM_2^f: \Lambda\otimes R_B\otimes A_1\otimes A_2\mapsto B$. In this case, the decoder $\mathrm{Dec}_2$ is trivial. The encoder $\mathrm{Enc}_2: \Lambda\otimes R_B\otimes A\mapsto \Lambda\otimes R_B\otimes A_1\otimes A_2$ acts as identity on $R_B$ and whenever $\Lambda=0$, maps $A$ to $A_1=A$ and $A_2=\Omega$ and whenever $\Lambda=1$, maps $A$ to $A_1=\Omega$ and $A_2=A$. It is then easy to verify that $\cM_2=\mathrm{Dec}_2\circ \cM^f_2\circ \mathrm{Enc}_2$ holds and all signalling relations of $\cM_2$ are preserved in $\cM_2^f$. 

\bigskip

To show that the network of \cref{fig: eg2_map_fg} is indeed a fine-graining of the network of \cref{fig: eg2_map} according to our \cref{definition: fg_network}, we still need to establish that the above encoding and decoding for the individual maps implies the existence of encoding and decoding schemes between every sub-network in the coarse and fine-grained networks.
The remaining sub-networks of our cyclic network are: the network itself (whose induced map is simply a constant number, since it has no in/outputs), the set of sub-networks formed by the sequential composition $\cM_2\circ \cM_1$ (through the $A$ system) composed with any subset of the preparations on $\Lambda$, $R_A$ and $R_B$ and another set of sub-networks formed by the sequential composition $\cM_1\circ \cM_2$ (through the system $B$) composed with any subset of the preparations. The encoders and decoders for the latter two sets of sub-networks are obtained from those constructed above, by simply applying the appropriate sequential composition to both sides of $\cM_1=\mathrm{Dec}_1\circ \cM^f_1\circ \mathrm{Enc}_1$ and $\cM_2=\mathrm{Dec}_2\circ \cM^f_2\circ \mathrm{Enc}_2$ along with composing the same set of preparations (noting that $\Lambda$, $R_A$ and $R_B$ remain unchanged in the two networks). In particular, the sub-network obtained by applying the sequential composition $\cM_2\circ \cM_1$ on all preparations has an induced map with a single in and output system, both isomorphic to $B$. The original cyclic network can be recovered by performing a loop composition on $B$ in this sub-network. The same is true in the fine-grained scenario, the sequential composition $\cM_2^f\circ \cM_1^f$  acting on the initial preparations yields an induced map having a single in and output isomorphic to $B$ and recovers the fine-grained network through loop composition on this in-output pair. Since the coarse and fine-grained sub-networks act identically in this case (as we have argued above), the loop compositions on $B$ in the two cases will yield identical results, and the same induced map for the original network (\cref{fig: eg2_map}) and fine-grained network (\cref{fig: eg2_map_fg}). This shows that the latter network is indeed a fine-graining of the former relative to this encoding-decoding scheme.

\bigskip

These examples highlight that, even in purely classical scenarios, the causal structure that we associate with a protocol, as well its cyclic/acyclic nature depends on the information captured by the nodes of the causal structure and the level of detail to which we specify this information. \cref{example: fine-graining2} illustrates a cyclic functional causal model that is inspired by the classical switch process \cite{Chiribella2013} which implements a classical mixture of the orders $A\prec B$ and $B\prec A$ in the process framework (cf. \cref{sec: PM}). The quantum generalisation of the process, the quantum switch can also be formalised as a cyclic network in our framework, which admits an acyclic fine-graining that has a similar physical interpretation (see \cref{sec: QS} and \cref{appendix: CBQS} for details).

\section{Physical spacetime realisations are causal boxes}
\label{appendix: framework_recovers_CB}

In this appendix, we connect spacetime realisations of quantum networks as defined in our top-down formalism to the previously known framework of \emph{causal boxes} \cite{Portmann2017}. The main result of this section indicates that the most general physical protocols that can be physically realised in a spacetime are those described by causal boxes. Our present formalism and proofs apply to any protocol involving a finite number of finite-dimensional quantum systems, but there is scope for generalising them to the infinite dimensional case in the future (noting that the causal box framework can indeed model infinite dimensional systems).
This result provides a top-down justification for the causal box framework, at least for the case of such finite protocols. For such protocols, we can effectively treat the spacetime poset $\cT$ as being finite, without loss of generality. Only a finite set of spacetime locations (or sufficiently localised regions) will be relevant for describing such protocols and we can take $\cT$ to only include those locations as all our statements only concern spacetime locations that can embed a system of the protocol. We will not review the causal box framework here but simply outline why it is a powerful approach with several useful features. 

\bigskip

Causal boxes describe quantum information processing protocols in an acyclic background spacetime where agents may send/receive messages at a superposition of different spacetime locations, in a manner consistent with relativistic causality. They generalise the standard quantum information formalism to explicitly include spacetime information and also to treat space and time on a more equal footing, in the spirit of special relativity. The standard formalism defines quantum states on fixed time slices (possibly non-localised in space) while causal box formalism allows us to consider states that are not localised in both space and time. Moreover, they preserve the desired mathematical properties of the usual formalism: they are closed under composition, admit a circuit decomposition into a well-defined sequence of isometries that are compatible with the partial order of the spacetime (the sequence representation), as well as admitting Stinespring and Choi matrices and thus being purifiable to unitaries. More generally, the framework is defined on a full infinite-dimensional Fock space and can capture superpositions of different number of messages exchanged between physical devices/agents. The formalism has useful applications for relativistic quantum information processing, in particular for studying composable cryptographic security also for practical relativistic quantum schemes \cite{Vilasini_crypto}. 

\bigskip

Causal boxes are defined relative to a fixed partially order set (which models the spacetime), and they satisfy a certain causality condition relative to this poset. The following theorem connects our physical spacetime realisations of networks in our framework to causal boxes (more precisely these realisations map to the subset of \emph{finite} causal boxes), showing that the causality condition of the causal box formalism can be derived from our relativistic causality condition (or more abstractly, compatibility between signalling structure of CPTPMs and the partial order).

\begin{restatable}{theorem}{CausalBoxGen}
Let $\mathfrak{N}$ be a network of CPTPMs whose signalling structure is compatible with an acyclic causal structure $\mathcal{G}$. Then the induced map of every sub-network of $\mathfrak{N}$ is a causal box relative the partial order induced by $\mathcal{G}$, therefore such a network $\mathfrak{N}$ is itself completely described by the causal box formalism.    
\end{restatable}
\begin{proof}
Consider any acyclic causal structure $\mathcal{G}$. Being a directed acyclic graph, it induces a partial order, where we say that $N_1\prec N_2$ relative to this partial order if and only if there exists a directed path from $N_1$ to $N_2$ in $\cG$. Denote the induced poset as $\cT^{\cG}$. Let $\cE(S)\in \mathrm{Nodes}(\cG)$ define an embedding of the systems $S\in \mathfrak{N}^{sys}$ in  $\cG$ and equivalently, in the partial order $\cT^{\cG}$ i.e., $\cE(S)\in \cT^{\cG}$. We can then associate a ``region'' $\cR^{\cS}$ or subset of locations in $\cT^{\cG}$ with every subset of $\cS\subseteq  \mathfrak{N}^{sys}$, with $\cR^{\cS}:=\cup_{S\in \cS} \cE(S)$. Then define
\begin{equation}
    \mathrm{Past}(\cR^{\cS}):=\{P\in \cT^{\cG}: \exists Q\in \cR^{\cS}, P\prec Q\},
\end{equation}
where $\prec$ is the partial order relation associated with $\cT^{\cG}$. $\mathrm{Past}(\cR^{\cS})$ corresponds to all points that are in the past of the region $\cR^{\cS}$ relative to this partial order. 

\bigskip

Let us now focus on the induced map $\cN$ of $\mathfrak{N}$ with In and Out denoting the set of in and output systems of $\cN$. Consider $\cS_O\subseteq \mathrm{Out}$, and a corresponding $\cS_I\subseteq \mathrm{In}$ defined as follows
\begin{equation}
    \cS_I:=\{S\in \mathrm{In}:\cE(S)\in \mathrm{Past}(\cR^{\cS_O})\}.
\end{equation}

That is, $\cS_I$ consists of all inputs of $\cN$ which are embedded at locations in the past of the region $\cR^{\cS_O}$ associated to $\cS_O$. 
Then, recalling that any poset $\cT$ imposes an order relation $\xrightarrow[]{R}$ on its region $\Sigma(\cT)$ (cf. \cref{def: region_order}),
$\cR^{\text{In}\backslash \cS_I}\cancel{\xrightarrow{R}} \mathcal{R}^{\cS_O}$ follows from the above definition of $\cS_I$. Imposing compatibility between the signalling structure of $\mathfrak{N}$ and the causal structure $\cG$ implies that we must have $\text{In}\backslash \cS_I$ does not signal to $\cS_O$ in $\mathfrak{N}$. Applying \cref{definition: signalling} to the CPTPM $\cN$, this non-signalling relation implies that for all states $\rho\in \cL(\text{In}\backslash \cS_I)$, 
    \begin{equation}
    \label{eq: causalbox_causality1}
     \tr_{\text{Out}\backslash \cS_O}\circ \cN=\tr_{\text{Out}\backslash \cS_O}\circ \cN\circ \tr_{\text{In}\backslash \cS_I}^{\rho},
    \end{equation}
  where $ \tr_{\text{In}\backslash \cS_I}^{\rho}$ is a particular choice of local operation on $\text{In}\backslash \cS_I$ which discards the incoming state on the subsystems $\text{In}\backslash \cS_I$ and replaces it with the fixed state $\rho$. As the embedding $\cE$ associates each system to a node in $\cG$, we can equivalently represent $\tr_{\cS}$ as $\tr_{\cR^{\cS}}$ (where $\cR^{\cS}$ is the region or subset of nodes in $\cG$, associated to the set of systems $\cS$), to capture the fact that it traces out systems embedded in the region $\cR^{\cS}$. 
Then defining a causality function $\chi: \cR \mapsto \mathrm{Past}(\cR)$ for every $\cR\subseteq \cT^{\cG}$, and denoting $  \tr_{\text{Out}\backslash \cS_O}\circ \cN= \cN^{\cR^{\cS_O}}$ to capture that this map only produces outputs in the region $\cR^{\cS_O}$ (when embedded in $\cG$ according to $\cE$) \cref{eq: causalbox_causality1} is equivalent to

 \begin{equation}
    \label{eq: causalbox_causality2}
    \cN^{\cR^{\cS_O}}=\cN^{\cR^{\cS_O}}\circ \tr_{\cT^{\cG}\backslash \chi(\cR^{\cS_O})}^{\rho}.
    \end{equation}
Fixing $\rho$ to be the vacuum state $\ket{\Omega}\bra{\Omega}$ (0-message space of the causal box Fock space), this is precisely the causality condition for regarding $\cN$ as a finite causal box, relative to the causality function $\chi$ \cite{Portmann2017}. It tells us that the outputs of a causal box in a region $\cR$, are independent of the input systems of the causal box that are not in the past of $\cR$, relative to the partial order. We can see that $\chi$ is indeed a valid finite causality function. This requires us to show that for every $P\in \cT^{\cG}$, there exists a finite $n\in \mathbb{N}$ such that $P\not\in \chi^n(\cT^{\cG})$, which ensures that every point in the partial order can be reached in a finite number of steps, through the causality function \cite{Portmann2017}. To show this, notice that the function $\chi$ defined above has the property that $\chi(\cR)\subset \cR$ for all $\cR\subseteq \cT^{\cG}$. This is because $\cT^{\cG}$ being a partial order ensures that every region $\cR$ contains at least one ``latest'' point $L\in \cR$ such that $L\not\prec P$ for all $P\in \cR$. Since $\cT^{\cG}$ is finite, every point $P$ will eventually be excluded from $\chi^n(\cT^{\cG})$ for some finite number $n$ applications of $\chi$. This establishes that the induced map $\cN$ of $\mathfrak{N}$ is a causal box with respect to the partial order induced by $\cG$. The same argument as above can be extended to the induced map of every sub-network of $\mathfrak{N}$ to show that they are all causal boxes relative to $\cG$ (and using the same embedding $\cE(S)\in \mathrm{Nodes}(\cG)$ for $S\in \mathfrak{N}^{sys}$).

\end{proof}

\begin{restatable}{corollary}{CausalBox}[Physical spacetime realisations are causal boxes]
\label{theorem:causal_box}
Consider a spacetime realisation of a network $\mathfrak{N}$ of CPTPMs that satisfies relativistic causality in a spacetime $\cT$ with respect to an embedding $\cE$. Let $\mathfrak{N}^f_{\cT,\cE}$ be the fine-graining in the specification of the  realisation, which is associated with an acyclic region causal structure $\cG^{\text{reg},f}_{\mathfrak{N}}$.
Then the induced maps of all sub-networks of $\mathfrak{N}^f_{\cT,\cE}$ are causal boxes relative to the order relation defined by $\cG^{\text{reg},f}_{\mathfrak{N}}$. Therefore all physical spacetime realisations of networks in our framework can be described within the causal box formalism.
\end{restatable}
\begin{proof}
This follows from the proof of the above theorem, by taking $\mathfrak{N}^f_{\cT,\cE}$ to play the role of the network $\mathfrak{N}$ used in that proof, and the acyclic causal structure $\cG^{\text{reg},f}_{\mathfrak{N}}$ to play the role of $\cG$, while noting that relativistic causality (cf. \cref{def: rel_causality_network}) in particular requires the signalling structure of $\mathfrak{N}^f_{\cT,\cE}$ to be compatible with the causal structure $\cG^{\text{reg},f}_{\mathfrak{N}}$. 
\end{proof}

\section{Further details of the process matrix framework}
\label{appendix: PM}

In \cref{sec: PM}, we described different classes of processes for the bipartite case. In particular, the set of fixed order processes are relevant for our main results and we review the full multi-partite definition for this class of processes below. This is adapted from definitions originally provided in \cite{Araujo2015, Oreshkov2016}, where the only adaptation is to distinguish the in and output systems $A^I$ and $A^O$ of each agent $A$ rather than grouping them together as a single element.

 \begin{definition}[Fixed order processes]
 \label{definition: causallyordered}
 An $N$-partite process matrix $W$ is said to be a fixed order process, if there exists a partial order $\mathcal{K}(\mathcal{S}_{I/O})$ on the set $\mathcal{S}_{I/O}=\{A^I_1,A^O_1,...,A^I_N,A^O_N\}$ of the input and output systems of the $N$ agents and associated with the binary relations $\prec_{\mathcal{K}}$ (first element precedes the second), $\succ_{\mathcal{K}}$ (first element succeeds the second) and $\nprec\nsucc_{\mathcal{K}}$ (the elements are unordered) such that the following conditions are satisfied
 \begin{enumerate}
     \item For any $i\in\{1,...,N\}$, $A^I_i\prec_\mathcal{K}A^O_i$.

     \item For any agent $A_i$ and a subset $A^{\mathcal{S}}$ of the remaining agents, such that $A_i^O\nprec_{\mathcal{K}} A_S^I$, $\forall A_S\in A_{\mathcal{S}}$ (which is denoted in short as $A_i^O\nprec_{\mathcal{K}} A_{\mathcal{S}}^I$) with respect to the partial order $\mathcal{K}(\mathcal{S}_{I/O})$, the joint probability distribution $P(x_1,...,x_N|a_1,...,a_N)$ (cf. Equation~\eqref{eq: prob}) obtained from $W$ for any choice of local measurements of the $N$ agents does not allow the outcome of any of the agents in $A_{\mathcal{S}}$ to depend on the setting of agent $A_i$. That is, taking $x_{\mathcal{S}}$ to denote the set of outcomes $\{x_S\}_{S\in \mathcal{S}}$ of the agents in $A_{\mathcal{S}}$, we have the following whenever $A_i^O\nprec_{\mathcal{K}} A_{\mathcal{S}}^I$ with respect to $\mathcal{K}(\mathcal{S}_{I/O})$
    \begin{equation}
     \label{eq: indep}
         \begin{split}
             P(x_{\mathcal{S}}|a_1,...a_{i-1},a_i,a_{i+1},...,a_N)=&\sum_{x_j\not\in x_{\cS}}P(x_1,...,x_N|a_1,...,a_N)\\=&P(x_{\mathcal{S}}|a_1,...,a_{i-1},a_{i+1},...,a_N).
        \end{split}
     \end{equation}

 \end{enumerate}
 \end{definition}

\bigskip

We now review the process matrix description of the quantum switch discussed in \cref{sec: QS}. The quantum switch can be modelled as a 4-partite process involving the agents $A$, $B$, $C$ and $D$. Where the agent $C$ has a trivial (i.e., 1-dimensional) input system and $D$ has a trivial output system, which we can ignore in the description. $C$'s output decomposes as $C^O=C^O_C\otimes C^O_T$ and $D$'s input as $D^I=D^I_C\otimes D^I_T$ corresponding to the control and target systems respectively. Thus $C$ and $D$ act on the control and target, with $C$ acting in the global past of all other agents (i.e., cannot be signalling to by others) and $D$ in the global future of all other agents (i.e., cannot signal to the others). $A$ and $B$ only act on the target qudit. Therefore the dimensions of the non-trivial systems are $d_{C^O_C}=d_{D^I_C}=2$ and $d_{C^O_T}=d_{D^I_T}=d_{A^I}=d_{A^O}=d_{B^I}=d_{B^O}=d$ where $d\geq 2$. The corresponding process matrix is pure (or equivalently, the process map $\mathcal{W}_{QS}$ is unitary) i.e., $W^{QS}=\ket{W^{QS}}\bra{W^{QS}}$ where
\par
\begin{equation}
\label{eq: w'qs}
\ket{W^{QS}}=|\mathds{1}\rrangle^{C^O_TA^I}|\mathds{1}\rrangle^{A^OB^I}|\mathds{1}\rrangle^{B^OD^I_T}\ket{00}^{C^O_CD^I_C}+|\mathds{1}\rrangle^{C^O_TB^I}|\mathds{1}\rrangle^{B^OA^I}|\mathds{1}\rrangle^{A^OD^I_T}\ket{11}^{C^O_CD^I_C}
\end{equation}

\noindent
The situation is illustrated in Figure~\ref{fig: extendedpm}. If the lab $C$ prepares an initial state $(\alpha\ket{0}+\beta\ket{1})^{C^O_C}\otimes \ket{\psi}^{C^O_T}$ (for arbitrary normalised amplitudes $\alpha$, $\beta$ and qudit state $\ket{\psi}$) and labs $A$ and $B$ perform the respective operations $\mathcal{U}^A: A^I\mapsto A^O$ and $\mathcal{V}^B: B^I\mapsto B^O$ (taken to be unitaries), the final state arriving at lab $D$ is given as follows and recovers the final state of \cref{eq: QS} on the control and target.
\begin{equation}
\label{eq: w'qsuv}
\begin{split}
&\Big((\alpha\bra{0}+\beta\bra{1})^{C^O_C}\otimes \bra{\psi^{*}}^{C^O_T}\otimes \llangle \mathcal{U}^{A^{*}}|^{A^IA^O}\otimes\llangle  \mathcal{V}^{B^{*}}|^{B^IB^O}\Big)\cdot \ket{W^{QS}}\\
=&\alpha\ket{0}^{D^I_C}\otimes (\mathcal{V}^B\mathcal{U}^A\ket{\psi})^{D^I_T}+\beta\ket{1}^{D^I_C}\otimes (\mathcal{U}^A\mathcal{V}^B\ket{\psi})^{D^I_T},
\end{split}
\end{equation}
where $\bra{\psi^{*}}$ denotes the complex conjugate of $\bra{\psi}=\ket{\psi}^{\dagger}$ in the computational basis $\{\ket{0},\ket{1}\}$, such that $\inprod{\psi^{*}}{i}=\inprod{i}{\psi}, i\in \{0,1\}$. $|\mathcal{U}^{A^*}\rrangle^{A^IA^O} = (\mathcal{I}\otimes \mathcal{U}^{A^*})|\mathds{1}\rrangle^{A^IA^I}$ and similarly for $\mathcal{V}^B$, where $*$ denotes the complex conjugate in the chosen orthonormal basis. 

\bigskip

The process matrix $W^{QS}$ is known to be causally non-separable (i.e., cannot be decomposed as in Equation~\eqref{eq: causalsep}) but nevertheless causal (i.e., always produces probabilities that decompose as per Equation~\eqref{eq: PMcausal}) \cite{Araujo2015}. The process map $\mathcal{W}_{QS}: C^O_C\otimes C^O_T\otimes A^O\otimes B^O \mapsto D^I_C\otimes D^I_T\otimes A^I\otimes B^I$ is related to $W_{QS}$ through the Choi isomorphism $W_{QS}=\mathcal{I}^{\mathrm{In}}\otimes \mathcal{W}_{QS}(|\mathds{1}\rrangle\llangle \mathds{1}|^{\mathrm{In}})$, and is obtained from $W_{QS}$ through the inverse Choi isomorphism. Here the superscript In corresponds to the total input system $C^O_CC^O_TA^OB^O$ of the process map. We can also obtain the action of the process map more directly, by considering how it acts on the computational basis $\{\ket{ijkl}^{\text{In}}\}_{ijkl}$ with $i\in\{0,1\}$ and $j,k,l\in\{0,...,d-1\}$, which provides a complete orthonormal basis for its inputs. The process map is unitary, and when the control is $\ket{0}^{C^O_C}$, $\mathcal{W}_{QS}$ behaves as a fixed order process with $A$ before $B$. In this case it forwards the target from, $C^O_T$ to $A^I$, then from $A^O$ to $B^I$ and from $B^O$ to $D^I_T$ (as can be read off from the first term of \cref{eq: w'qs}). When the control is $\ket{1}^{C^O_C}$, it behaves as a fixed order process with $B$ before $A$, forwarding $C^O_T$ to $B^I$, $B^O$ to $A^I$ and $A^O$ to $D^I_T$ and in both cases (as can be read off from the first term of \cref{eq: w'qs}), and the control being forwarded to $D^I_C$. Therefore we have, for all $j,k,l\in \{0,...,d-1\}$,

\begin{align}
\label{eq: QS_processmap}
    \begin{split}
  \mathcal{W}_{QS}:& \quad    \ket{0}^{C^O_C}\ket{j}^{C^O_T}\ket{k}^{A^O} \ket{l}^{B^O} \mapsto    \ket{0}^{D^I_C}\ket{l}^{D^I_T}\ket{j}^{A^I} \ket{k}^{B^I}\\
   \mathcal{W}_{QS}:& \quad    \ket{1}^{C^O_C}\ket{j}^{C^O_T}\ket{k}^{A^O} \ket{l}^{B^O} \mapsto    \ket{1}^{D^I_C}\ket{k}^{D^I_T}\ket{l}^{A^I} \ket{j}^{B^I}
    \end{split}
\end{align}

\section{Equivalence of notions of signalling in process networks}
\label{appendix: equiv_signalling}

In this section we define two notions of signalling which are relevant in process networks, one is formulated at the level of the probabilities of the classical settings and outcomes and the other is formulated at the level of the quantum in and output systems of the process map and local operations. We then show an equivalence between these two notions which holds whenever the process map represents a valid process matrix. 

\begin{definition}[Probabilistic signalling]
\label{def: signalling_prob}
      Let $W$ be an $N$-partite process matrix associated with the probability distribution $P(x_1,...,x_N|a_1,...,a_N)$. We say that an agent $A_i$ probabilistically signals to a set $A_{\cS}$ of the remaining agents if and only if the setting $a_i$ of $A_i$ is correlated with outcomes $x_{\cS}:=\{x_j\}_{j\in \cS}$ of the subset of the remaining $N-1$ agents, associated with the indices $\cS\subseteq \{1,...,i-1,i+1,...,N\}$ i.e., 
      \begin{equation}
          P(x_{\cS}|a_1,...,a_{i-1},a_i,a_{i+1},...,a_N)\neq P(x_{\cS}|a_1,...,a_{i-1},a_{i+1},...,a_N),
      \end{equation}
      where $P(x_{\cS}|a_1,...,a_{i-1},a_i,a_{i+1},...,a_N)=\sum\limits_{x_j: j\not\in \cS}P(x_1,...,x_N|a_1,...,a_N)$.
      
\end{definition}

\begin{definition}[Quantum signalling]
\label{def: signalling_network}
 Let $W$ be an $N$-partite process matrix and $\mathfrak{N}_{\mathcal{W},N}$ be the corresponding network in our framework. Let $A_i$ be an agent and $A_{\cS}$ be any subset of the remaining agents, and $\mathfrak{N}_{\mathcal{W},\overline{i\cup \cS}}$ be the sub-network of $\mathfrak{N}_{\mathcal{W},N}$ obtained from composing $\mathcal{W}$ with the extended local operations of all agents except those in $A_i\cup A_{\cS}$. We say that an agent $A_i$ quantumly signals to a set $A_{\cS}$ of the remaining agents if and only if $\{A^O_i\}$ signals to $\{A^I_j\}_{j\in \cS}$ in the induced map $\cN_{\mathcal{W},\overline{i\cup \cS}}$. 
\end{definition}

\begin{restatable}{theorem}{EquivSignalling}[Equivalence of two notions of signalling]
\label{theorem: equivsignalling}
Whenever $W$ is a valid $N$-partite process matrix, $A_i$ probabilistically signals to $A_{\cS}$ if and only if $A_i$ quantumly signals to $A_{\cS}$.
\end{restatable}
\begin{proof}
We first prove the equivalence for bipartite processes $W$ over agents $A$ (Alice) and $B$ (Bob), and later show how this implies the general $N$-partite result.

\paragraph{ Probabilistic signalling implies quantum signalling} Here, we wish to prove that $P(y|ab)\neq P(y|a)$ implies that $\{A^O\}$ signals to $\{B^I\}$ in the process map $\mathcal{W}$. We will prove the equivalent contrapositive statement which is that $\{A^O\}$ does not signal to $\{B^I\}$ in $\mathcal{W}$ implies  $P(y|ab)= P(y|a)$.

\bigskip

First we show that $\{A^O\}$ does not signal to $\{B^I\}$ in $\mathcal{W}$ implies $\{A^O\}$ does not signal to $\{A^I,B^I\}$ in $\mathcal{W}$. This follows from a result about the decomposition of process matrices proven in \cite{Oreshkov2012} which shows that any bipartite process matrix $W\in A^I\otimes A^O\otimes B^I\otimes B^O$ can be expressed in the Hilber-Schmidt basis in the following general form, where a Hilbert-Schmidt basis for linear operators on a Hilbert space $\mathcal{H}^X$ is given by a set of matrices $\{\sigma^X_{\mu}\}_{\mu=0}^{d^2_X-1}$ with $\sigma^X_0=\mathds{1}^X$, $\tr\sigma_{\mu}^X\sigma_{\nu}^X=d_X\delta_{\mu\nu}$ and $\tr\sigma_{j}^X=0$ for all $j\in \{1,...,d^2_X-1\}$.

\begin{align}
    \begin{split}
   W&= \frac{1}{d_{A^I}d_{B^I}}\Big(\mathds{1}^{A^IA^OB^IB^O}+\sigma^{B\preceq A}+\sigma^{A\preceq B}+\sigma^{A\not\preceq\not\succeq B}\Big),\\
   \sigma^{B\preceq A}&:=\sum_{i,j>0}c_{ij}\sigma_i^{A^I}\otimes \mathds{1}^{A^O}\otimes \mathds{1}^{B^I}\otimes \sigma_j^{B^O}+ \sum_{i,j,k>0}d_{ijk}\sigma_i^{A^I}\otimes \mathds{1}^{A^O}\otimes \sigma_j^{B^I}\otimes \sigma_k^{B^O}\\
   \sigma^{A\preceq B}&:=\sum_{i,j>0}e_{ij}\mathds{1}^{A^I}\otimes \sigma_i^{A^O}\otimes \sigma_j^{B^I}\otimes \mathds{1}^{B^O} + \sum_{i,j,k>0}f_{ijk}\sigma_i^{A^I}\otimes \sigma^{A^O}_j\otimes \sigma_k^{B^I}\otimes \mathds{1}^{B^O}\\
   \sigma^{A\not\preceq\not\succeq B}&:= \sum_{i>0}\mathcal{V}_i\sigma_i^{A^I}\otimes \mathds{1}^{A^O}\otimes \mathds{1}^{B^I}\otimes \mathds{1}^{B^O}+\sum_{i>0}x_i\mathds{1}^{A^I}\otimes \mathds{1}^{A^O}\otimes \sigma_i^{B^I}\otimes \mathds{1}^{B^O}\\& \quad+\sum_{i,j>0}g_{ij}\sigma_i^{A^I}\otimes \mathds{1}^{A^O}\otimes\sigma_j^{B^I}\otimes \mathds{1}^{B^O},
    \end{split}
\end{align}

where $c_{ij},d_{ijk},e_{ij},f_{ijk},\mathcal{V}_i,x_i,g_{ij}\in \mathbb{R}$. We then use the fact that the non-signalling relations $\{A^O\}\not\longrightarrow \{B^I\}$ and $\{A^O\}\not\longrightarrow \{A^I,B^I\}$ in the process map $\mathcal{W}$ can be equivalently expressed as the following conditions on its Choi matrix (which is by construction, the process matrix $W$), 
\begin{align}
\label{eq: Choi_NS}
    \begin{split}
    \tr_{A^I}W&=\tr_{A^IA^O}W\otimes \frac{\mathds{1}^{A^O} }{d_{A^O}} \\
W&=\tr_{A^O}W\otimes \frac{\mathds{1}^{A^O} }{d_{A^O}}.     
    \end{split}
\end{align}
Then we can immediately see for each term of $W$, except the term $\sigma^{A\preceq B}$, both conditions of the above equation are satisfied when replacing $W$ with the term under consideration. Thus a violation of the second condition by a given $W$ could only come from the presence of terms of the form of $\sigma^{A\preceq B}$. We see that the term $\sigma^{A\preceq B}$ always violates both conditions, whenever there are any non-zero $e_{ij}$ or $f_{ijk}$ coefficients. This is because $\sigma_j^{A^O}\neq \mathds{1}^{A^O}$ by construction (since the $j$ values start from 1, and for all such $\sigma_j^{A^O}$, the trace is 0 which is distinct from the trace of the identity). Thus, tracing and replacing the $A^O$ system will change the term, and violate the second condition, and by the same argument we can see that it will also change the term even if we have already traced out $A^I$ , hence the first condition will also be violated.

\bigskip

Thus we have shown that in each term of a valid $W$, if there is a violation of the second condition of \cref{eq: Choi_NS}, then there will also be a violation of the first condition. This implies (by linearity) that for all valid $W$'s this implication holds, and consequently that $\{A^O\}$ signals to $\{A^I,B^I\}$ implies $\{A^O\}$ signals to $\{B^I\}$.  Moreover, $\{A^O\}$ signals to $\{A^I,B^I\}$ which is equivalent to the second condition of \cref{eq: Choi_NS} is also shown to be equivalent to $P(y|ab)=P(y)$ in \cite{Oreshkov2016}. Therefore, $P(y|ab)\neq P(y|b)$ implies that $\{A^O\}$ signals to $\{A^I,B^I\}$ in $\mathcal{W}$, which in turn implies that $\{A^O\}$ signals to $\{B^I\}$ in $\mathcal{W}$ and altogether, this establishes that $A$ probabilistically signals to $B$ implies $A$ quantumly signals to $B$.

 \paragraph{ Quantum signalling implies probabilistic signalling} Suppose that we have the quantum signalling relation $\{A^O\}$ signals to $\{B^I\}$ in $\mathcal{W}$. For any CPTP map $\mathcal{W}$, this is equivalent \cite{Schumacher2005} to the condition that there exists a state $\sigma^{B^O}$ on $B^O$ and distinct states $\rho^{A^O}$ and $\tilde{\rho}^{A^O}$ on $A^O$ such that
$$\tr_{A^I}\circ\mathcal{W}(\rho^{A^O}\otimes \sigma^{B^O})\neq \tr_{A^I}\circ\mathcal{W}(\tilde{\rho}^{A^O}\otimes \sigma^{B^O}).$$ 

In other words, the state on $B^I$ when $\mathcal{W}$ acts on $\rho^{A^O}\otimes \sigma^{B^O}$ is distinct from the state on $B^I$ when it acts on $\tilde{\rho}^{A^O}\otimes \sigma^{B^O}$ and hence there exists a measurement $\{\mathcal{M}^{B}_{y|b}\}_{y}$ that $B$ can perform to distinguish these states with a non-zero probability. Then, we can define $A$'s local operation $\mathcal{M}^{A}$ to be such that whenever the setting $a=0$ is input on $A^s$, it discards the input on $A^I$ and prepares the state $\rho^{A^O}$ to send to $\mathcal{W}$ and whenever the setting $a=1$ is input on $A^s$, it discards the input on $A^I$ and prepares the state $\tilde{\rho}^{A^O}$ to send to $\mathcal{W}$. This in turn implies that $P(y|a=0,b)\neq P(y|a=1,b)$ or $P(y|ab)\neq P(y|b)$, and establishes the claim.

\paragraph{ Generalisation to the multi-partite case} For the multi-partite case, we consider the signalling relation $A_i$ signals to the set $A_{\cS}$ in terms of the two notions of signalling. We can map this to the proof for the bipartite case which we have shown above, by treating $A_i$ as $A$ in the above proof, and the set $A_{\cS}$ of agents together as $B$. In this case, instead of signalling in $\mathcal{W}$, we would consider signalling in the induced map $\mathcal{N}_{\mathcal{W},\overline{i\cup\cS}}$ of the sub-network $\mathfrak{N}_{\mathcal{W},\overline{i\cup\cS}}$ (note that in the bipartite case, this sub-network reduces to the network $(\mathcal{W},\emptyset)$ whose induced map is $\mathcal{W}$). Then the proof of the statement that probabilistic implies quantum signalling carries forth unchanged, we can again prove the equivalent contrapositive statement using the same arguments, showing that $\{A^O_i\}$ does not signal to $\{A^I_j\}_{j\in \cS}$ in $\mathcal{N}_{\mathcal{W},\overline{i\cup\cS}}$  (i.e., no quantum signalling) implies that $\{A^O_i\}$ does not signal to $\{A^I_j\}_{j\in i\cup \cS}$ in $\mathcal{N}_{\mathcal{W},\overline{i\cup\cS}}$  which implies that $A_i$ does not probabilistically signal to $A_{\cS}$.

\bigskip

For the statement that quantum signalling implies probabilistic signalling, the only aspect that needs additional consideration is that, when we treat Bob as a collection of more than one agent $A_{\cS}$  in the multi-partite case, then he can only apply extended local maps that act independently on each input $A^I_j$ for $j\in \cS$ and not arbitrary joint operations. Bob's role, as the receiver of the signal is to be able to distinguish any two distinct states that arise on the subsystems $\{A^I_j\}_{j\in \cS}$ which are the outputs of the CPTP map $\cN_{\mathcal{W},\overline{i\cup\cS}}$, which arise as a result of different local operations that Alice may perform on the input $A^O_i$ of the same map. Quantum theory satisfies the property that any two states of a multi-partite system can be distinguished through local measurements performed on the subsystems i.e., there will exist a local measurement on each subsystem such that the outcome probabilities for the two states differ when performing those local measurements simultaneously on all the systems. This is known as the \emph{local distinguishability} or \emph{local discriminability} property of quantum theory \cite{Chiribella2010, Chiribella2011}. This property ensures that all the arguments used above for proving the equivalence in the bipartite case also carry over to the general multi-partite case. 
\end{proof}

\section{Fine-grained acyclic network describing a spacetime realisation of QS}
\label{appendix: CBQS}

Here we describe a spacetime realisation of the QS network $\mathfrak{N}_{\mathcal{W}_{QS},U,V}$ that satisfies relativistic causality, and the corresponding fine-grained network which will be associated with a definite and acyclic (information-theoretic) causal structure (as shown in \cref{theorem: PMFinegrain} and \cref{corollary: nogoB}). The realisation is associated with the spacetime embedding given in \cref{eq: embedding_QS} and \cref{eq: QS_Minkowski} of the main text and the fine-grained description will correspond to a \emph{causal box} \cite{Portmann2017}, as expected from our results of \cref{appendix: framework_recovers_CB}. We first detail the fine-grained causal box description, and then show that this is indeed a fine-graining of the original network $\mathfrak{N}_{\mathcal{W}_{QS},U,V}$ as per \cref{definition: fg_network}. 

\subsection{Causal box description of the fine-grained network}
\begin{figure}[t]
\centering
\includegraphics[width=0.8\textwidth]{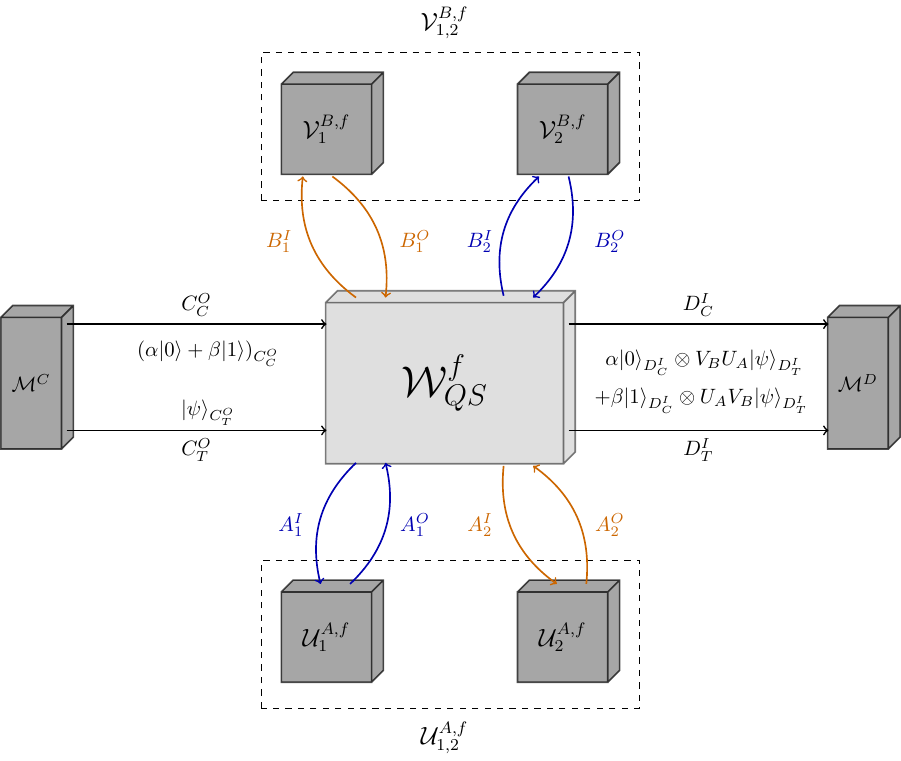}
	\caption{An equivalent representation of the fine-grained network of \cref{fig: PMQS_fg}: This representation makes explicit that $\mathcal{W}_{QS}^f$ itself is a CPTPM which is composed with the fin-grained local operations through feedback loops. The blue path and orange wires represent the paths of the non-vacuum target when the control is $\ket{0}$ and $\ket{1}$ respectively and correspond to the blue and orange paths in \cref{fig: PMQS_fg}. Although there are feedback loops involved, $\mathcal{W}_{QS}^f$ is a causal box and decomposes into a sequence of CPTPMs as shown in \cref{fig: qsseq}. This means that every causal influence between the systems involved in this network, is that such whenever $S$ is a cause of $S'$, then $S$ is embedded in the past of $S'$ according to the spacetime embedding associated with this realisation (\cref{eq: embedding_QS} and \cref{eq: QS_Minkowski}). As the spacetime forms a partial order, this implies that the causal structure of this network is acyclic as shown in \cref{fig: PMQS_fg}. 
	}
	\label{fig: cbqs}
\end{figure}

Consider the quantum switch process map $\mathcal{W}_{QS}$ described in \cref{appendix: PM}, which is known to be a unitary, along with local unitaries $\mathcal{U}^A$ and $\mathcal{V}^B$ of Alice and Bob. Recall from the main text that the corresponding fine-grained process map $\mathcal{W}^f_{QS}$ associated with the spacetime embedding of \cref{eq: embedding_QS} and \cref{eq: QS_Minkowski} has the in and output systems, In $=\{C^O_C,C^O_T,A^O_1,A^O_2,B^O_1,B^O_2\}$ and Out $=\{D^I_C,D^I_T,A^I_1,A^I_2,B^I_1,B^I_2\}$, where $d_{C^O_C}=d_{D^I_C}=2$, $d_{C^O_T}=d_{D^I_T}=d$ ($d\geq 2$) and $d_{A^I_1}=d_{A^I_2}=d_{A^O_1}=d_{A^O_2}=d_{B^I_1}=d_{B^I_2}=d_{B^O_1}=d_{B^O_2}=d+1$. The $d+1$ dimensional systems are associated with the Hilbert space $\mathbb{C}^d\oplus \ket{\Omega}$, where $\ket{\Omega}$ is the vacuum state. The fine-grained local operations $\mathcal{U}^{A,f}_{1,2}: A^I_1\otimes A^I_2\mapsto A^O_1\otimes A^O_2$ and $\mathcal{V}^{B,f}_{1,2}: B^I_1\otimes B^I_2\mapsto B^O_1\mapsto B^O_2$ are given as 
\begin{align}
\label{eq: qs_fg_localmaps}
    \begin{split}
    \mathcal{U}^{A,f}_{1,2}=\mathcal{U}^{A,f}_1\otimes \mathcal{U}^{A,f}_2,\quad     \mathcal{V}^{B,f}_{1,2}=\mathcal{V}^{B,f}_1\otimes \mathcal{V}^{B,f}_2,
    \end{split}
\end{align}
where $\mathcal{U}_i^{A,f}$ and $\mathcal{V}_i^{B,f}$ act on vacuum and non-vacuum states as in \cref{eq: vacuum} (leaving the vacuum $\ket{\Omega}$ invariant and applying a qudit unitary $\mathcal{U}_i^A$ or $\mathcal{V}_i^B$ on a non-vacuum input). We repeat this action here for convenience.

\begin{align}
\label{eq: vacuum_appendix}
    \begin{split}
        \mathcal{U}^{A,f}_i\ket{\Omega}^{A^I_i}&=\ket{\Omega}^{A^O_i}\\
        \mathcal{U}^{A,f}_i\ket{\psi}^{A^I_i}&=\mathcal{U}_i^A\ket{\psi}^{A^O_i}
    \end{split}
\end{align}
The resultant fine-grained network is identical to that of \cref{fig: PMQS_fg} at the information-theoretic level (maps and compositions) and can be equivalently represented as shown in \cref{fig: cbqs}.

\bigskip

\begin{figure}[t]
	\centering
\includegraphics[scale=0.8]{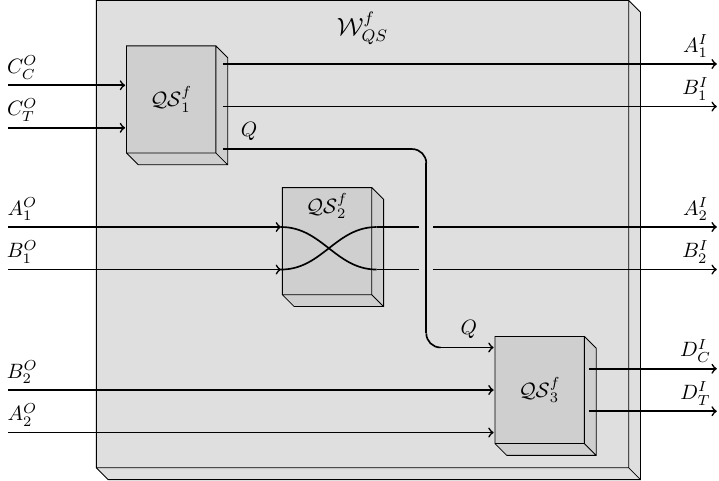}
	\caption{Sequence representation of the causal box $\mathcal{W}^f_{QS}$ of Figure~\ref{fig: cbqs} }
	\label{fig: qsseq}
\end{figure}

We now proceed to describe the internal decomposition of $\mathcal{W}^f_{QS}$. This is identical to the description of the quantum switch originally given in the causal box framework \cite{Portmann2017} and we repeat it here in our notation. $\mathcal{W}^f_{QS}$ can be decomposed into three internal CPTP maps, $\mathcal{QS}_1^f$, $\mathcal{QS}_2^f$ and $\mathcal{QS}_3^f$, which are illustrated in \cref{fig: qsseq}. When $\mathcal{W}^f_{QS}$ is composed with the fine-grained local operations $\mathcal{U}^{A,f}_{1,2}$ and $\mathcal{V}^{B,f}_{1,2}$ described above, we have the following well-defined sequence of operations in the network.

\begin{enumerate}
    \item $\mathcal{QS}_1^f$ Receives the initial state of control and target $(\alpha\ket{0}+\beta\ket{1})^{C^O_C}\otimes \ket{\psi}^{C^O_T}$ from $C$, stores the control in an internal memory $Q$, and forwards $\ket{\psi}$ to $A^I_1$ and $\ket{\Omega}$ to $B^I_1$ when the control is in state $\ket{0}$, and $\ket{\Omega}$ to $A^I_1$ and $\ket{\psi}$ to $B^I_1$ when the control is in the state $\ket{1}$.
    \item $\mathcal{U}^{A,f}_1$ and $\mathcal{V}^{B,f}_1$ are applied on the inputs $A^I_1$ and $B^I_1$ and the corresponding outputs $A^O_1$ and $B^O_1$ are fed back to $\mathcal{W}^f_{QS}$.
     \item $\mathcal{QS}_2^f$ Forwards the state in $A^O_1$ to $B^I_2$ and the state in $B^O_1$ to $A^I_2$.
     \item $\mathcal{U}^{A,f}_2$ and $\mathcal{V}^{B,f}_2$ are applied on the inputs $A^I_2$ and $B^I_2$ and the corresponding outputs $A^O_2$ and $B^O_2$ are fed back to $\mathcal{W}^f_{QS}$.
     \item $\mathcal{QS}_3^f$ consults the control qubit stored in $Q$, if it is in the state $\ket{0}$, forwards the state in the output $B^O_2$ to $D^I_T$ and it the control is in the state $\ket{1}$, forwards the output $A^O_2$ to $D^I_T$, and in both cases forwards the control in $Q$ to $D^I_C$.
\end{enumerate}

One can easily verify (see \cite{Portmann2017} for details) that this implements the transformation of \cref{eq: qs_fg} from $C^O_C\otimes C^O_T$ to $D^I_C\otimes D_O^T$ whenever the local maps act on the vacuum as given in \cref{eq: vacuum_appendix} and in the case where $\mathcal{U}_1^{A,f}=\mathcal{U}_2^{A,f}$ and $\mathcal{V}_1^{B,f}=\mathcal{V}_2^{B,f}$ this implements the quantum switch transformation of \cref{eq: QS}. Moreover, as shown in \cite{Portmann2017}, one can additionally include counters with the local operations, initialised to $\ket{0}$ and which increment by 1 whenever the local operation is applied on a non-vacuum state. At the end of the protocol, both Alice and Bob's counters would then record 1, which verifies that in this network, each of Alice and Bob's operations act exactly once on a non-vacuum $d$-dimensional state.

\bigskip

Moreover, this sequence decomposition ensures that the fine-grained causal structure of the network is acyclic as shown in \cref{fig: PMQS_fg}, with a causal influence $S\longrsquigarrow S'$ only if the fine-grained systems $S$ and $S'$ are associated with spacetime locations $P^S$ and $P^{S'}$ (through the embedding $\cE$ of \cref{eq: embedding_QS}) where $P^S\prec P^{S'}$. Consequently, the signalling relations of the fine-grained network are compatible with the partial order of the spacetime. We can view the fine-grained map $\mathcal{W}^f_{QS}$ as a process map associated with the agents $A_1$, $A_2$, $B_1$, $B_2$, $C$ and $D$. Notice that the in and output systems of these 6 agents are localised in the spacetime according to the embedding \cref{eq: embedding_QS}, and satisfy relativistic causality with respect to this embedding. This implies, by \cref{corollary: localisation} that $\mathcal{W}^f_{QS}$ must correspond to a fixed order process. The fine-grained causal structure of \cref{fig: PMQS_fg} along with the embedding \cref{eq: embedding_QS} make it clear that the order involves $C$ and $D$ being in the global past and future respectively of other agents, $A_1$ before $B_2$ and $B_1$ before $A_2$ (with $A_1$ and $A_2$, $B_1$ and $B_2$ not necessarily ordered among themselves).

\bigskip

Here, we have only considered the 0-message (vacuum) and 1-message spaces of $d$ dimensions, for Alice and Bob. More generally, all fine-grained systems can be modelled as Fock spaces which allow for multiple messages of a given dimension and the above fine-grained operations can be consistently extended to the full Fock space as shown in \cite{Portmann2017}. In follow up work based on the master's thesis \cite{Salzger}, we show that such causal box descriptions associated with an acyclic causal structure, and their extensions to the full Fock space can also be obtained for a much general class of processes called QCQCs \cite{Wechs2021}, which realise arbitrary quantum controlled superpositions of orders.

\subsection{Encoders and decoders relating the fine and coarse grained networks}
We now describe the encoder and decoder that allow us to recover the coarse-grained network $\mathfrak{N}_{\mathcal{W}_{QS},U,V}$ from the fine-grained causal box description of the previous sub-section, and show that these respect the conditions of \cref{definition: fg_network}. Having described the coarse and fine-grained systems of the network, we know that the encoder and decoder relating $\mathcal{W}_{QS}$ and $\mathcal{W}_{QS}^f$ must be maps of the following form.
\begin{align}
    \begin{split}
        &\mathrm{Enc}_{QS}: \quad C^O_C\otimes C^O_T\otimes A^O\otimes B^O \mapsto C^O_C\otimes C^O_T\otimes A^O_1\otimes A^O_2 \otimes B^O_1\otimes B^O_2,\\
        &\mathrm{Dec}_{QS}:\quad  D^I_C\otimes D^I_T\otimes A^I_1\otimes A^I_2\otimes B^I_2\otimes B^I_2 \mapsto D^I_C\otimes D^I_T\otimes A^I\otimes B^I.
    \end{split}
\end{align}

A first requirement of \cref{definition: fg_network} is that $\mathrm{Enc}_{QS}$ must be CPTP on the whole input space while $\mathrm{Dec}_{QS}$ must be CPTP on the image of $\mathcal{W}^f_{QS}\circ \mathrm{Enc}_{QS}$. We completely specify the action of $\mathrm{Enc}_{QS}$ by its action on a complete orthonormal basis of $C^O_C\otimes C^O_T\otimes A^O\otimes B^O$, which is given by the computational basis $\{\ket{ijkl}\}_{i\in \{0,1\},j,k,l\in \{0,...,d-1\}}$. This is given as follows, for any $j,k,l\in \{0,...,d-1\}$
\begin{align}
\label{eq: QS_encoder}
    \begin{split}
   &\mathrm{Enc}_{QS}:\quad \ket{0}^{C^O_C}\ket{j}^{C^O_T}\ket{k}^{A^O} \ket{l}^{B^O} \mapsto     \ket{0}^{C^O_C}\ket{j}^{C^O_T}\ket{k}^{A^O_1}\ket{\Omega}^{A^O_2} \ket{\Omega}^{B^O_1}\ket{l}^{B^O_2},\\
   &\mathrm{Enc}_{QS}:\quad \ket{1}^{C^O_C}\ket{j}^{C^O_T}\ket{k}^{A^O} \ket{l}^{B^O} \mapsto     \ket{0}^{C^O_C}\ket{j}^{C^O_T}\ket{\Omega}^{A^O_1}\ket{k}^{A^O_2} \ket{l}^{B^O_1}\ket{\Omega}^{B^O_2}.
    \end{split}
\end{align}

We can see that such an encoder is in fact unitarity (and therefore a CPTPM). We could physically realise it as follows, including an ancilla $E_A$ for Alice and ancilla $E_B$ for Bob, both initialised in the state $\ket{\Omega}$, then performing a controlled swap which sends $A^O$ to $A^O_1$ and $E_A$ to $A^O_2$ when $\ket{0}^{C^O_C}$ and $A^O$ to $A^O_2$ and $E_A$ to $A^O_1$ when  $\ket{1}^{C^O_C}$, and another analogous controlled swap for Bob. Then we can see that the image of $\mathcal{W}^f_{QS}\circ \mathrm{Enc}_{QS}$ (using that $\mathcal{W}^f_{QS}$ can be decomposed as in \cref{fig: qsseq}) is always such that it lies within the subspace defined by the set of projectors $\{\Pi^f_{0jkl},\Pi^f_{1,jkl}\}_{jkl\in \{0,...,d-1\}}$ associated with the following states
\begin{align}
\label{eq: qs_decoder_subspace}
    \begin{split}
        \ket{\Psi^f_{0jkl}}^{\mathrm{Out}}:=& \ket{0}^{D^I_C}\ket{j}^{D^I_T}\ket{k}^{A^I_1}\ket{\Omega}^{A^I_2} \ket{\Omega}^{B^I_1}\ket{l}^{B^I_2},\\
        \ket{\Psi^f_{1jkl}}^{\mathrm{Out}}:=&\ket{1}^{D^I_C}\ket{j}^{D^I_T}\ket{\Omega}^{A^I_1}\ket{k}^{A^I_2} \ket{l}^{B^I_1}\ket{\Omega}^{B^I_2}.
    \end{split}
\end{align}

Then consider a decoder that acts as follows for all $i,j,k$
\begin{align}
\label{eq: QS_decoder}
    \begin{split}
        \mathrm{Dec}_{QS}:&\quad \ket{\Psi^f_{0jkl}}^{\mathrm{Out}}\mapsto \ket{0}^{D^I_C}\ket{j}^{D^I_T}\ket{k}^{A^I}\ket{l}^{B^I},\\
        \mathrm{Dec}_{QS}:&\quad \ket{\Psi^f_{1jkl}}^{\mathrm{Out}} \mapsto \ket{1}^{D^I_C}\ket{j}^{D^I_T}\ket{k}^{A^I}\ket{l}^{B^I}.
    \end{split}
\end{align}
Such a decoder will be unitary (and therefore CPTP) on the image of $\mathcal{W}^f_{QS}\circ \mathrm{Enc}_{QS}$. We could physically realise this by sending the state in $A^I_1$ or $A^I_2$ to $A^I$ depending on $D^I_C$ and tracing out the other state, and similarly for Bob. Note that when applied on the image of $\mathcal{W}^f_{QS}\circ \mathrm{Enc}_{QS}$, the state being traced out will always be the vacuum $\ket{\Omega}$ and will decouple from the rest such that the decoder is unitary on the subspace and maps pure states to pure states as described. 

\bigskip

The next step is to show that $\mathcal{W}_{QS}=\mathrm{Dec}_{QS}\circ \mathcal{W}_{QS}^f\circ \mathrm{Enc}_{QS}$ holds for the encoder and decoder defined above. $\mathcal{W}_{QS}$ is unitary and its action is as given in \cref{eq: QS_processmap}. From the causal box description of $\mathcal{W}_{QS}^f$ given in the previous sub-section, we can see that it is unitary on the image of $\mathrm{Enc}_{QS}$ and acts as follows (noting from \cref{eq: QS_encoder} that the image of the encoder is a subspace of $C^O_C\otimes C^O_T\otimes A^O_1\otimes A^O_2\otimes B^O_1\otimes B^O_2$ that is spanned by states of the form given below)
\begin{align}
\label{eq: CBQS_processmap}
    \begin{split}
  \mathcal{W}_{QS}^f:& \quad    \ket{0}^{C^O_C}\ket{j}^{C^O_T}\ket{k}^{A^O_1}\ket{\Omega}^{A^O_2} \ket{\Omega}^{B^O_1}\ket{l}^{B^O_2} \mapsto    \ket{0}^{D^I_C}\ket{l}^{D^I_T}\ket{j}^{A^I_1}\ket{\Omega}^{A^I_2} \ket{\Omega}^{B^I_1}\ket{k}^{B^I_2},\\
   \mathcal{W}_{QS}^f:& \quad    \ket{1}^{C^O_C}\ket{j}^{C^O_T}\ket{\Omega}^{A^O_1}\ket{k}^{A^O_2} \ket{l}^{B^O_1}\ket{\Omega}^{B^O_2} \mapsto    \ket{1}^{D^I_C}\ket{k}^{D^I_T}\ket{\Omega}^{A^I_1}\ket{l}^{A^I_2} \ket{j}^{B^I_1}\ket{\Omega}^{B^I_2}.
    \end{split}
\end{align}
Notice that the final states above correspond to the states $\ket{\Psi^f_{0ljk}}$ and $\ket{\Psi^f_{1klj}}$ of \cref{eq: qs_decoder_subspace}.
Then it is immediate from combining the description of the encoder (\cref{eq: QS_encoder}), the description of the fine-grained process map (\cref{eq: CBQS_processmap}) and that of the decoder (\cref{eq: QS_decoder}), that we recover the action of the coarse-grained process (\cref{eq: QS_processmap})  i.e., $\mathcal{W}_{QS}=\mathrm{Dec}_{QS}\circ \mathcal{W}_{QS}^f\circ \mathrm{Enc}_{QS}$ holds.  Moreover, it is easy to see that all the signalling relations are preserved, in particular the signalling relations $\{A^O\}\longrightarrow \{B^I\}$ and $\{B^O\}\longrightarrow \{A^I\}$ between the main agents in $\mathcal{W}_{QS}$ is preserved in the corresponding signalling relations $\{A^O_1,A^O_2\}\longrightarrow \{B^I_1,B^I_2\}$ and $\{B^O_1,B^O_2\}\longrightarrow \{A^I_1,A^I_2\}$ in $\mathcal{W}^f_{QS}$.

\bigskip

Recall that in order to regard the causal box realisation of the process network $\mathfrak{N}_{\mathcal{W}_{QS},U,V}$ as its fine-graining, we must compare all corresponding sub-networks of the two descriptions and find corresponding encoders and decoders to translate between their induced maps. $\mathcal{W}_{QS}$ that we have analysed above
is the induced map of one such sub-network of $\mathfrak{N}_{\mathcal{W}_{QS},U,V}$. Other sub-networks of $\mathfrak{N}_{\mathcal{W}_{QS},U,V}$ are associated with the individual local maps $\mathcal{U}^A$, $\mathcal{V}^B$ (without any composition) as well as the 
networks formed by composing $\mathcal{W}_{QS}$ with one or both of $\mathcal{U}^A$ and $\mathcal{V}^B$. The latter type of networks either correspond to fully connecting a local operation to $\mathcal{W}_{QS}$ by composing the inputs and outputs of the operation with $\mathcal{W}_{QS}$, or sequentially composing it to $\mathcal{W}_{QS}$ (in any order) which would correspond to connecting only its input system or only its output system to the corresponding system of $\mathcal{W}_{QS}$ with the same label. In all of these cases, we can construct the corresponding encoders and decoders in an entirely analogous manner as we did for $\mathcal{W}_{QS}$ in the above paragraphs, and we therefore do not repeat the construction for each case. For instance, consider the sub-network of $\mathfrak{N}_{\mathcal{W}_{QS},U,V}$ obtained by composing $\mathcal{V}^B$ but not $\mathcal{U}^A$ with $\mathcal{W}_{QS}$. The induced map then maps $C^O_C\otimes C^O_T\otimes A^O$ to $D^I_C\otimes D^I_T\otimes A^I$ and the corresponding induced map in the fine-grained network maps $C^O_C\otimes C^O_T\otimes A^O_1\otimes A^O_2$ to $D^I_C\otimes D^I_T\otimes A^O_1\otimes A^O_2$. The encoder and decoder in the case would be constructed in the same manner as \cref{eq: QS_encoder} and \cref{eq: QS_decoder} to ensure that the non-vacuum and vacuum states end up in the right fine-grained system for Alice depending on the control. They would also satisfy all the required properties by the same arguments as before.

\bigskip

The only remaining sub-networks are those corresponding to the individual local operations $\mathcal{U}^A$ and $\mathcal{V}^B$. Consider the fine-grained local maps of \cref{eq: qs_fg_localmaps}, which act on the vacuum state according to \cref{eq: vacuum}. Further suppose that one the fine-grained operations $\mathcal{U}^A_i$ acting on the non-vacuum subspace, say for $i=1$, is identical to the coarse-grained operations $\mathcal{U}^A$, and similarly for Bob i.e., $\mathcal{U}^A_1=\mathcal{U}^A$ and $\mathcal{V}^B_1=\mathcal{V}^B$. Here, the encoders and decoders are maps of the following form. 
\begin{align}
    \begin{split}
 &\mathrm{Enc}_A: A^I\mapsto A^I_1\otimes A^I_2,\quad        \mathrm{Enc}_B: B^I\mapsto B^I_1\otimes B^I_2,\\
 &\mathrm{Dec}_A:  A^O_1\otimes A^O_2\mapsto A^O,\quad  \mathrm{Dec}_B: B^O_1\otimes B^O_2\mapsto B^O.
    \end{split}
\end{align}

Then it is sufficient to take encoders are simply identity channels from the coarse-grained system (e.g., $A^I$) to the first fine-grained system (e.g., $A^I_1$) and prepare an arbitrary, independent state on the second fine-grained system. Similarly, it is sufficient to consider decoders that act as identity channels from the first fine-grained system to the coarse-grained system and trace out the second fine-grained system. Then it is clear that the necessary properties to regard the map $\mathcal{U}^{A,f}_{1,2}$ as a fine-graining of $\mathcal{U}^A$ and the map 
 $\mathcal{V}^{B,f}_{1,2}$ as a fine-graining of $\mathcal{V}^B$ are trivially satisfied. 

\bigskip

From the above, it follows that the causal box description of the previous sub-section is indeed a fine-graining of the process network (\cref{definition: fg_network}) and preserves all of it signalling relations, even though it admits an acyclic causal structure (unlike the original coarse-grained process network which has a cyclic causal structure).

\begin{remark}[Local distinguishability of the order]
\label{remark: order_distinguis_qs}    The fixed spacetime realisation of QS described in Figure~\ref{fig: PMQS_fg} and in this section is such that the spacetime location at which Alice or Bob's operation is applied (on a non-vacuum state) is perfectly correlated with the control. Therefore, Alice and Bob can perfectly and locally distinguish the two orders by measuring the time of arrival (or more generally, spacetime location of arrival) of a non-vacuum state to their lab. This would of course collapse the superposition of orders. Alternatively, they can use the location of arrival as a control parameter based on which they apply different unitary operations depending on the order (cf. \cref{eq: qs_fg}). In \cite{Goswami2018, Goswami_2020}, a different type of experimental realisation of QS in Minkowski spacetime has been proposed, which has the interesting property that local measurements by Alice and Bob of the (space)time location at which they receive a non-vacuum state would not reveal significant information about the control since there is a large uncertainty in this spacetime location even when the operations are applied in a fixed order (i.e., where the control is in one of the computational basis states). Our results are general and apply to both these type of realisations, and independently of the agents' ability to perfectly distinguish the orders locally. For the realisations of \cite{Goswami2018, Goswami_2020}, the fine-grained description would not correspond to that of Figure~\ref{fig: PMQS_fg} but one where Alice and Bob act on a larger number of fine-grained systems (associated with more than two spacetime locations) such that the location at which they receive a non-vacuum state would not tell them much about the order. This means that we would need to consider a larger number of agents than 6 in order to describe the experiment using a fixed order processes (cf.\ Theorem~\ref{theorem: PMFinegrain}), and this also means that we may not be able to perfectly implement the transformation of \cref{eq: qs_fg} where the unitary applied by each agent is perfectly correlated with the order in which they act. However, this does not affect the conclusions of Theorem~\ref{theorem: PMFinegrain} that the fine-grained description of such an experiment would correspond to a fixed order process that is compatible with a definite and acyclic causal structure.
\end{remark}

\section{Comparing fixed spacetime and quantum gravitational realisations}
\label{appendix: GQS}

In \cite{Paunkovic2019}, a protocol has been proposed for distinguishing between the physically realised optical realisations and theoretically proposed gravitational realisations of QS in a manner that does not disturb the coherence between the different branches of the superposition. For the definite spacetime case, \cite{Paunkovic2019} focuses on a QS implementation in Minkowski spacetime described with respect to a single global frame where the in/output events of Alice and Bob are spatially localised but not temporally localised. Here we show that their core argument can be generalised to arbitrary fixed spacetime realisations. 

\bigskip

The main property that the protocol of \cite{Paunkovic2019} seeks to distinguish is whether the realisation involves a superposition of orders between two spacetime events (one for Alice and one for Bob) or whether at least one agent is associated with more than one spacetime event. The protocol introduces an additional agent $F$, hereby known as ``Friend'' to whom Alice and Bob send out photons in addition to performing their usual operations in the QS scenario. $F$ is assumed to be spatially localised\footnote{In a classical background spacetime, which is what we consider here, the notion of spatial localisation is well defined. To model a quantum spacetime, \cite{Paunkovic2019} considers a region of the manifold within which there can be a superposition of metrics while there is a well defined classical geometry outside this region where localisation is well-defined as in classical spacetimes. The friend $F$ is assumed to be spatially localised in such a classical region of the spacetime.} and can measure information regarding the times of arrival of the photons arriving from Alice and Bob to decide whether or not the local operation in each lab was a single spacetime event. In the optical implementation of QS, at least one of Alice and Bob must act (on a non-vacuum state) at an earlier time or a distinct later time depending coherently on the control qubit (while the other agent may act at a fixed spacetime location). In this case, at least one agent is associated with two spacetime events which translates into $F$ receiving at least one of the agents' photons arriving to them at a coherent superposition of different times, in the optical realisation in Minkowski spacetime. On the other hand, \cite{Paunkovic2019} proposes (theoretical) quantum gravitational realisations where the superposition of spacetime metrics can used to ensure that any photon from Alice always arrives to $F$ at the same time $t^A$ and any photon from Bob always arrives to $F$ at the same time $t^B$. A non-demolition measurement is then performed by $F$ to distinguish these two scenarios without collapsing the superposition of orders (see \cite{Paunkovic2019} for details).

\bigskip

We can model this protocol by considering a new 5-partite process map $\mathcal{W}_{QS}^{+F}$ obtained from the 4-partite quantum switch map $\mathcal{W}_{QS}$ by including the agent $F$ with input systems $F^I_A$ and $F^I_B$ and a trivial output space. We can give additional outputs $A^O_F$ and $B^O_F$ to Alice and Bob for the photons being sent to $F$ and the process vector $\ket{W_{QS}^{+F}}$ is then identical to $\ket{W_{QS}}$ of Equation~\eqref{eq: w'qs}, but with the additional factor $|\mathds{1}\rrangle^{A^O_F F^I_A} |\mathds{1}\rrangle^{B^O_F F^I_B}$ in both terms of the superposition, representing the identity channels from $A^O_F$ to $F^I_A$ and $B^O_F$ to $F^I_B$. \cite{Paunkovic2019} argue that in the Minkowski spacetime realisation, assuming that Alice and Bob communicate to $F$ using photons (i.e., light speed communication), then it cannot be the case that $F$ (assumed to be spatially localised) receives Alices's photon at time $t^A$ and Bob's at time $t^B$ independently of the order in which they act. This is because such a realisation of QS involves at least one agent (say Alice) acting at multiple spacetime events, and the light-like surfaces emanating from the two events would intersect $F$'s spatially localised worldline at two distinct times. That is, it is impossible for the systems $F^I_A$ and $F^I_B$ to both be localised in the spacetime if we assume light speed communication from $A$ and $B$ to $F$, in a realisation satisfying relativistic causality. Notice that without this assumption, it is in-principle possible to adjust the communication speed according to the order in which Alice acts such that her communication reaches $F$ at the same time (e.g., if Alice acts earlier, she uses free space communication and if she act later, she uses a fibre optic cable with a suitable refractive index). Here, we suggest a way to generalise the arguments of \cite{Paunkovic2019} to arbitrary spacetime realisations, without assuming light speed communication.


\bigskip

For this, consider an arbitrary but definite and acyclic spacetime $\mathcal{T}$, and that $F^I_A$ is embedded at a single spacetime location $P^{F^I_A}$ and $F^I_B$ at the location $P^{F^I_B}$, taking $P^{F^I_B}\prec P^{F^I_A}$ (the argument for $P^{F^I_A}\prec P^{F^I_B}$ is analogous). Denoting $\mathrm{Past}(P):=\{Q\in \mathcal{T}| Q\prec P\}$ to be the past of a spacetime point $P$, we have $\mathrm{Past}(P^{F^I_B})\subset \mathrm{Past}(P^{F^I_A})$. Then, relativistic causality, along with the fact that Alice and Bob must in general, signal to $F$ in the protocol implies that $\mathcal{R}^{A^O}\xrightarrow[]{R} P^{F^I_A}$ and 
$\mathcal{R}^{B^O}\xrightarrow[]{R} P^{F^I_B}$. Now if we can ensure that $P^{F^I_B}$ is outside the lightcone of all points in $\mathcal{R}^{A^O}$ i.e., that it is impossible for Alice to have communicated to $F$ such that Alice's message could arrive to $F$ at the location $P^{F^I_B}$ in any relativistically causal protocol, then we would have $\mathcal{R}^{A^O}\subseteq \mathrm{Past}(P^{F^I_A})\backslash \mathrm{Past}(P^{F^I_B})$. Together with $\mathcal{R}^{B^O}\subseteq \mathrm{Past}(P^{F^I_B})$, this would imply that $\mathcal{R}^{A^O}\cancel{\xrightarrow{R}} \mathcal{R}^{B^O}$. However, since we have $\{A^O\}\longrightarrow \{B^O\}$ in the QS protocol, this would be a violation of relativistic causality.

\bigskip

This shows that the following properties cannot be simultaneously achieved in any fixed spacetime realisation of \cite{Paunkovic2019}'s modified QS protocol that satisfies relativistic causality: (1) the spatially localised agent $F$ receives the communication from Alice at a single spacetime location $P^{F^I_A}=(\vec{r}^F,t^{F^I_A})$ and from Bob at a distinct, single location $P^{F^I_B}=(\vec{r}^F,t^{F^I_B})$ and (2) for at least one agent, say $A$, the spacetime location associated with the other agent, $P^{F^I_B}$ is such that $P^{F^I_B}$ is outside the future light-cone of $A$'s output locations i.e, $\mathcal{R}^{A^O}\cancel{\xrightarrow{R}} P^{F^I_B}$. Notice that (2) is a weaker condition that requiring $A$ and $B$ to communicate to $F$ only at light speed.
If consider Minkowski spacetime and we restrict communication from $A$ and $B$ to $F$ to be light speed communication, then we can recover the argument of \cite{Paunkovic2019} as a special case. In this case, assuming (1) and relativistic causality, would imply that every point $Q\in \cR^{A^O}$ must lie on the past light-like surface of $P^{F^I_A}$. Taking  $P^{F^I_B}\prec P^{F^I_A}$ w.l.o.g, and recalling that $P^{F^I_B}$ and $ P^{F^I_A}$ have the same spatial co-ordinate $\vec{r}^F$, it is clear that the past light-like surface of $ P^{F^I_A}$ cannot have any overlap with the past light-cone of $ P^{F^I_B}$. Therefore $\mathcal{R}^{A^O}\subseteq \mathrm{Past}(P^{F^I_A})\backslash \mathrm{Past}(P^{F^I_B})$ and consequently $\mathcal{R}^{A^O}\cancel{\xrightarrow{R}} P^{F^I_B}$ holds. However, in a gravitational implementation where the spacetime locations are described with respect to local quantum reference frames \cite{Castro_Ruiz_2020}, it might nevertheless be possible to satisfy all these conditions in a physical implementation of QS, as suggested by the gravitational realisations proposed in \cite{Paunkovic2019}. It would be interesting to study, in the future, how (2) can be imposed and justified in quantum gravitational realisations.

\section{Relation to causal loops without superluminal signalling}
\label{appendix: relation_to_VC}

In recent works \cite{VilasiniColbeckPRA, VilasiniColbeckPRL} involving one of the authors, a formalism for describing cyclic and non-classical information-theoretic causal models as well as their compatibility with spacetime has been developed. While the present paper is inspired by the spirit of this work, namely to disentangle the two causality notions and relate them through compatibility conditions (which capture relativistic causality constraints), there are some important differences between the two formalisms as they were developed for describing different types of scenarios. For instance, \cite{VilasiniColbeckPRA, VilasiniColbeckPRL} can model certain beyond-quantum theories and correlations known as \emph{jamming non-local theories/correlations} \cite{Grunhaus1996}, however it focused on interventions on observed classical nodes in information-theoretic causal models rather than interventions directly on the underlying non-classical systems. On the other hand, the present formalism focuses on quantum information-theoretic networks where agents can intervene on quantum systems, which enables us to explicitly model indefinite causal structures and process matrices. In the following, we discuss the relation between the two formalisms, and in particular, clarify the link between our results about the acyclic fine-grained causal structure of spacetime realised networks and the surprising result of \cite{VilasiniColbeckPRL} regarding the possibility of causal loops without superluminal signalling in Minkowski spacetime. 

\bigskip

As highlighted in \cref{sec:signalling} and in previous works \cite{Wood2015, Barrett2020, VilasiniColbeckPRA, VilasiniColbeckPRL}, information-theoretic causation and signalling are distinct notions. It is possible to have causal influence ($\longrsquigarrow$) without signalling ($\longrightarrow$), which implies that \emph{no superluminal causation} and \emph{no superluminal signalling} correspond to two distinct relativistic causality principles. Let us illustrate this more concretely with the CPTPM $\cM:I\mapsto O_1\otimes O_2$ from \cref{fig:jamming_example}, where $\{I\}\longrightarrow \{O_1,O_2\}$ but $\{I\}\not\longrightarrow \{O_1\}$ and $\{I\}\not\longrightarrow \{O_2\}$. Suppose that the in/output systems are embedded at distinct spacetime locations $P^S$, $S\in \{I,O_1,O_2\}$. Then in any decomposition of $\cM$, $I$ must be a cause of at least one of $O_1$ or $O_2$, the decompositions illustrated in \cref{fig:jamming_example} have this property.\footnote{For if it were not (i.e., we had a decomposition where there is no internal wire connection $I$ to either of $O_1$ or $O_2$, then $\{I\}$ signals to $\{O_1,O_2\}$ would be impossible.} Therefore, a necessary condition for ensuring that information-theoretic causal influences flow from past to future in the spacetime (i.e, there is no superluminal causation) is that at least one of $P^I\prec P^{O_1}$ or $P^I\prec P^{O_2}$ must hold. This is what our relativistic causality condition of \cref{def: rel_causality_sig} imposes. On the other hand, noting that the signalling $\{I\}\longrightarrow \{O_1,O_2\}$ can only be verified when $O_1$ and $O_2$ are jointly accessed, imposing that the joint future (intersection of the future light-cones) of $P^{O_1}$ and $P^{O_2}$ is contained within the future of $I$ would be sufficient for ensuring no superluminal signalling in this scenario \cite{VilasiniColbeckPRA}.\footnote{Provided that the systems $I$, $O_1$ and $O_2$ are the only ones that are accessible to agents (systems involved in the internal decomposition of $\cM$ being outside their access).} This condition is strictly weaker than requiring $P^I\prec P^{O_1}$ or $P^I\prec P^{O_2}$ to hold, as either of these would imply that the joint future of the outputs would be contained in the inputs but the converse is not true: the joint future condition can be satisfied even when $P^I$, $P^{O_1}$ and $P^{O_2}$ are space-like separated \cite{Grunhaus1996}. The compatibility condition of \cite{VilasiniColbeckPRA, VilasiniColbeckPRL} which is necessary and sufficient for ensuring no superluminal signalling in that framework, imposes these weaker conditions. 

\bigskip

\begin{figure}[t!]
            \centering
\subfloat[\label{figure: VC_causal_structure}]{\includegraphics[scale=1.0]{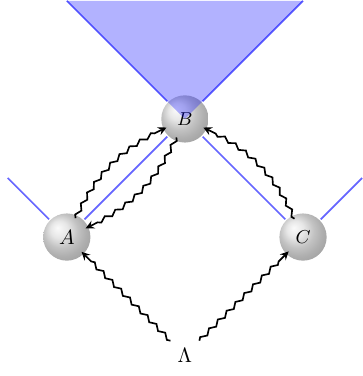}}\qquad\qquad\subfloat[\label{figure: VC_network}]{ \includegraphics[scale=1.0]{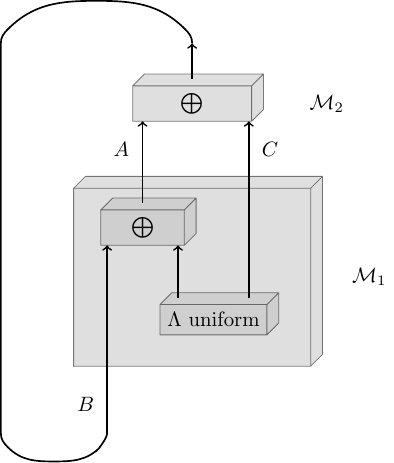}}
            \caption{(a) The directed cyclic graph over the nodes $A$, $B$, $C$, $\Lambda$ represents the information-theoretic causal structure of the causal loop of \cite{VilasiniColbeckPRL}. The loop corresponds to a classical functional causal model (reviewed in \cref{appendix: examples_fg}) on this causal structure where all nodes are binary variables, the parentless node $\Lambda$ is associated with the uniform distribution $P(\Lambda)$, and the remaining nodes depend on their parent nodes through the functional relations: $A=\Lambda\oplus B$, $C=\Lambda$ and $B=A\oplus C$. Here only $A$, $B$ and $C$ are accessible to agents, $\Lambda$ is a latent node and cannot be intervened upon. The signalling relations of the causal model are such that $B$ signals jointly to $A$ and $C$ but not individually to $A$ or $C$. The causal model is embedded in (1+1)-Minkowski spacetime by assigning a single spacetime event to each of its nodes as shown, the 45 degree straight lines (blue) denote light-like surfaces. In the given embedding, it is shown in \cite{VilasiniColbeckPRL} that this causal loop does not lead to superluminal signalling, since the signalling from $B$ to $\{A,C\}$ can only be verified in the blue shaded region which is contained in the future of $B$. Notice however that the causal influence from $B$ to $A$ is superluminal in this embedding. (b) Equivalent representation of the causal loop of \cite{VilasiniColbeckPRL} as a cyclic network in our framework. Notice that the map $\cM_1$ acts identically to the map $\cM_1$ of \cref{fig:jamming_example1}, with $B$, $A$ and $C$ playing the role of $I$, $O_1$ and $O_2$ respectively. If this network is embedded in spacetime, our relativistic causality condition (\cref{def: rel_causality_sig}) would require at least one of $A$ or $C$ to be embedded in the future light-cone of $B$ (since $\{B\}$ signals to $\{A,C\}$), the spacetime embedding of (a) for this network therefore violates the relativistic causality condition of the present paper, although it respects the strictly weaker relativistic causality condition of \cite{VilasiniColbeckPRA, VilasiniColbeckPRL} and does not lead to superluminal signalling as discussed in the text (and shown in \cite{VilasiniColbeckPRL}. }
            \label{fig: VCloop}
        \end{figure}

The difference between the two compatibility conditions, which correspond to two distinct (but related) relativistic causality principles as outlined above, leads to different characterisations of the set of information processing protocols which can be realised in a fixed spacetime. Interestingly, only demanding no superluminal signalling (the weaker condition) does not rule out the possibility of genuine causal loops being embedded in Minkowski spacetime as shown in \cite{VilasiniColbeckPRL}. In the language of our formalism, such genuine loops would correspond to those where the maximally fine-grained description is also cyclic i.e., the causal loop of \cite{VilasiniColbeckPRL} is a counter-example which shows that \cref{theorem: PMFinegrain} would not hold under the weaker compatibility condition. \cite{VilasiniColbeckPRL} does not suggest that such loops are physical in any way, rather, the main message of \cite{VilasiniColbeckPRL} is that the principle of no superluminal signalling is not sufficient for ruling out causal loops in Minkowski spacetime, as commonly believed. In light of this, it remains important to understand the physical principles that can rule them out. Our work suggests that the stronger relativistic causality constraint (which is necessary for ruling out superluminal causation) is sufficient for ruling out such genuine causal loops in a fixed spacetime, at least in quantum theory. Whether this is also a necessary condition for ruling out such loops (even in quantum and classical theories) in a fixed spacetime, and whether the sufficiency result generalises to post-quantum theories remain open questions. In \cref{fig: VCloop} we illustrate the information-theoretic causal model of the causal loop proposed in \cite{VilasiniColbeckPRL}, the cyclic network by which we can model it within the present framework and the spacetime embedding proposed in \cite{VilasiniColbeckPRL} whereby the causal loop does not lead to superluminal signalling. While we have discussed some of the relationships and distinctions between the two formalisms, characterising the precise relationship(s) between the classes of causal loops proposed in \cite{VilasiniColbeckPRA, VilasiniColbeckPRL}  and the cyclic quantum networks proposed here, specifically the subset that describes indefinite causal structure processes, remains an interesting open question.

\bigskip

We stress that although our compatibility condition is stronger than the condition of \cite{VilasiniColbeckPRA, VilasiniColbeckPRL} and is necessary for ensuring the absence of superluminal causation, it is formulated only at the level of the operational signalling structure of the network and without referring to the underlying causal decompositions of the maps involved. This is because the signalling, although distinct from causation, does allow us to operationally infer the presence of certain causal influences.
For the questions of interest in the present work, namely for understanding the physical and experimental realisations of indefinite causal structures in spacetime, the stronger compatibility condition that we have used is the more relevant one, as we are not yet aware of the existence of any physical mechanisms for superluminal causation in Minkowski spacetime. However, one way to evade the consequences of our main no-go theorems would be through a theoretical model which incorporates such superluminal causal mechanisms (which, could in-principle be possible without violating the principle of superluminal signalling). However, such causal explanations are typically disfavoured as they would need to be fine-tuned \cite{Wood2015, VilasiniColbeckJamming}.

\section{Proofs of all results from the main text}
\label{appendix: proofs}

\EmbeddingSig*
\begin{proof}

In our framework, a signalling structure $\cG^{\text{sig}}_{\mathfrak{N}}$ of some network $\mathfrak{N}$ (which will simply be denoted as $\cG^{\text{sig}}$ in the rest of the proof) is in general a directed graph where the nodes are subsets of in/output systems in the network, i.e., $\mathrm{Nodes}(\cG^{\text{sig}})\subseteq \Sigma(\mathfrak{N}^{sys})$. An embedding $\cE$ of $\cG^{\text{sig}}$ in a spacetime $\cT$ corresponds to an assignment of spacetime regions to each system in $\mathfrak{N}^{sys}$. This immediately implies an embedding for all systems in $\Sigma(\mathfrak{N}^{sys})$: for any subset $\cS$ of $\mathfrak{N}^{sys}$ the spacetime region assigned to $\cS$ through the embedding $\cE$ is simply the union $\cR^{\cS}=\bigcup_{S\in \cS}\cR^S$ of all the spacetime regions assigned to the individual elements $S\in \cS$ under the embedding $\cE$. Thus, in order to establish the theorem statement, we first find an embedding for systems in $\mathfrak{N}^{sys}$, such that the signalling relations in $\cG^{\text{sig}}$ over the nodes corresponding to systems in $\mathfrak{N}^{sys}$ respects relativistic causality. We will later see that this immediately implies an embedding of all the nodes of $\cG^{\text{sig}}$ such that relativistic causality is still preserved. Therefore, for the purpose of the next few paragraphs, we will treat $\cG^{\text{sig}}$ as a directed graph $\cG$ over the nodes $\mathfrak{N}^{sys}$ with directed edges $\longrightarrow$, and generalise the result to the powerset $\Sigma(\mathfrak{N}^{sys})$ at the end.

\bigskip

We first need to set out some nomenclature. For every node $N$ of a directed graph $\cG$, let the set $\mathrm{Par}(N):=\{N'\in \mathrm{Nodes}(\cG): N'\longrightarrow N \in \cG\}$ denote the set of all parents of the node $N$, the set $\mathrm{Ch}(N):=\{N'\in \mathrm{Nodes}(\cG): N\longrightarrow N' \in \cG\}$ denote the set of all children of the node $N$, the set $\mathrm{Anc}(N):=\{N'\in \mathrm{Nodes}(\cG): \exists \text{ directed path } N'\longrightarrow...\longrightarrow N \in \cG\}$ denote the set of all ancestors of $N$ in $\cG$ and the set $\mathrm{Desc}(N):=\{N'\in \mathrm{Nodes}(\cG): \exists \text{ directed path } N\longrightarrow...\longrightarrow N' \in \cG\}$ denote the set of all descendants of $N$ in $\cG$. Then the set of all nodes in $\cG$ that are involved in at least one cycle is defined as
$$\mathrm{Cycle}(\cG)=\{N\in \mathrm{Nodes}(\cG): N\in \mathrm{Anc}(N)\}.$$

That is, every node $N$ that is its own ancestor belongs to the set $\mathrm{Cycle}(\cG)$. Note that since $N_1\in \mathrm{Anc}(N_2)$ is equivalent to $N_2\in \mathrm{Desc}(N_1)$, we could have equivalently defined this set as $\mathrm{Cycle}(\cG)=\{N\in \mathrm{Nodes}(\cG): N\in \mathrm{Desc}(N)\}$. In particular this implies that $N\in \mathrm{Anc}(N)\cap \mathrm{Desc}(N)$ (or $N$ is contained in its own strongly connected component \cite{Bongers2021}), for every $N\in \mathrm{Cycle}(\cG)$. Moreover, $\cG$ is a directed acyclic graph if and only if $\mathrm{Cycle}(\cG)=\emptyset$.

\bigskip

The rest of the proof proceeds as follow. We will show that we can fine-grain (cf. \cref{definition: fg_graph}) any signalling structure $\cG$ into a directed acyclic graph $\cG'$ by ``splitting'' nodes in $\mathrm{Cycle}(\cG)$, such that when the split nodes in $\cG'$ are recombined, we get back the original graph $\cG$. 
Once we have a directed acyclic graph $\cG'$, we can always embed it in a partially ordered spacetime $\cT$ though an embedding $\mathcal{E'}$ that respects relativistic causality. We can then coarse-grain the embedding $\mathcal{E'}$ of $\cG'$ to an embedding $\cE$ of the original structure $\cG$ that also respects relativistic causality, which will establish the theorem statement. 

\bigskip

Construct a new directed graph $\cG'$ from the original directed graph $\cG$ as follows. If $\mathrm{Cycle}(\cG)=\emptyset$, set $\cG'=\cG$. If $\mathrm{Cycle}(\cG)\neq \emptyset$, then split every node $N\in \mathrm{Cycle}(\cG)$ into two nodes $N_1$ and $N_2$ such that $\mathrm{Par}(N_1)=\mathrm{Par}(N)\backslash \mathrm{Cycle}(\cG)$ and $\mathrm{Ch}(N_1)=\mathrm{Ch}(N)$ and $\mathrm{Par}(N_2)=\mathrm{Par}(N)$ and $\mathrm{Ch}(N_2)=\mathrm{Ch}(N)\backslash \mathrm{Cycle}(\cG)$ i.e., $N_1$ contains no incoming arrows from nodes in $\mathrm{Cycle}(\cG)$ but all the same outgoing arrows as $N$ does in $\cG$ while $N_2$ contains no outgoing arrows to nodes in $\mathrm{Cycle}(\cG)$ but all the same incoming arrows as $N$ does in $\cG$. Nodes that do not belong to $\mathrm{Cycle}(\cG)$, and other edges not featuring in the above construction remain unaffected. This fully defines $\cG'$. $\cG'$ constructed in this way is a fine-graining of $\cG$ (cf. \cref{definition: fg_graph}). More precisely, when we recombine $N_1$ and $N_2$ back into a single node $N$, for each pair of split nodes and without altering the edge structure, we recover the original graph $\cG$ since $\mathrm{Par}(N_1\cup N_2)=\mathrm{Par}(N)$ and $\mathrm{Ch}(N_1\cup N_2)=\mathrm{Ch}(N)$ for every $N\in \mathrm{Cycle}(\cG)$, and the nodes $N\not\in \mathrm{Cycle}(\cG)$ were not split or altered in going from $\cG$ to $\cG'$. This implies that we can define a graph fine-graining $\cF^{\text{graph}}: \mathrm{Nodes}(\cG)\mapsto \Sigma(\mathrm{Nodes}(\cG'))$ such that $\cF^{\text{graph}}(N)=\{N_1,N_2\}$ for all $N\in \mathrm{Cycle}(\cG)$ and $\cF^{\text{graph}}(N)=N$ otherwise, with respect to which $\cG'$ is a fine-graining of $\cG$.

\bigskip

Now, we can show that $\cG'$ is a directed acyclic graph. For this, observe that a node of a graph can belong to the cycle of the graph only if it has a parent and a child that belong to the cycle of the graph.
This is because $N\in \mathrm{Cycle}(\cG)$ implies that there is a directed path from $N$ to itself in $\cG$ which implies a directed path from $N$ to at least one $N'\in \mathrm{Par}(N)$ and this directed path together with $N'\longrightarrow N$ tells us that $N'\in \mathrm{Cycle}(\cG)$. Similarly, $N\in \mathrm{Cycle}(\cG)$ implies a directed path from at least one $N''\in \mathrm{Ch}(N)$ to $N$ and this directed path together with $N\longrightarrow N''$ implies that $N''\in \mathrm{Cycle}(\cG)$. Next, notice that if $N\not\in \mathrm{Cycle}(\cG)$, then by construction, we have $N\not\in \mathrm{Cycle}(\cG')$. Moreover, for each $N\in \mathrm{Cycle}(\cG)$, we have a pair of nodes $N_1$ and $N_2$ in $\cG'$ where all parents of $N_1$ and all children of $N_2$ are nodes of the form $N\not\in \mathrm{Cycle}(\cG)$ (and hence $N\not\in \mathrm{Cycle}(\cG')$). Then, it follows that no node $N$ in $\cG'$ can have both a parent and a child in $\mathrm{Cycle}(\cG')$, and therefore that $N$ itself cannot belong to $\mathrm{Cycle}(\cG')$. In other words, this establishes that $\mathrm{Cycle}(\cG)=\emptyset$ or that $\cG'$ is acyclic.



\bigskip

Since $\cG'$ is a directed acyclic graph, there exists an embedding $\cE': \mathrm{Nodes}(\cG')\mapsto \cT$ of $\cG'$ in a partially ordered set $\cT$ (associated with the order relation $\prec$) such that $N_i\longrightarrow N_j$ in $\cG$ $\Rightarrow$ $\cE'(N_i)\prec \cE'(N_j)$. By virtue of being a partial order, $\cT$ satisfies our minimal definition of spacetime structure, according to Definition~\ref{def: spacetime}.
Then the required embedding $\cE$ of $\cG$ in the spacetime $\cT$ simply associates two spacetime locations with each node $N\in \mathrm{Cycle}(\cG)$, the two locations being precisely those assigned by $\cE'$ to each of the split nodes i.e., $\cE(N):=\{\cE'(N_1),\cE'(N_2)\}$. For all nodes $N\not\in \mathrm{Cycle}(\cG)$, $\cE(N)=\cE'(N)$ noting that these nodes never got split. Then it is clear that the embedding $\cE$ of $\cG$ respects relativistic causality whenever the embedding $\cE'$ of $\cG'$ respects relativistic causality (which it does by construction).

\bigskip

We now describe how the proof generalises to case where $\mathrm{Nodes}(\cG)=\Sigma(\mathfrak{N}^{sys})$. For this, note that our above proof covers all cases where $\cG$ has the property that $\cS_1\longrightarrow \cS_2$ for two subsets $\cS_1, \cS_2$ of $\mathfrak{N}^{sys}$, then there exists $S_1\in \cS_1$ and $S_2\in \cS_2$ such that $S_1\longrightarrow S_2$. However, suppose that we have a signalling relation $\cS_1\longrightarrow \cS_2$ in $\cG$ such that there is no signalling relation between individual elements of these two sets (this is indeed possible, see for instance \cref{fig:jamming_example}). The relativistic causality condition implied by this signalling relation on the corresponding spacetime embedding is that $\cR^{\cS_1}\xrightarrow[]{R} \cR^{\cS_2}$ (cf.\ Definition~\ref{def: rel_causality_sig}). Since $\cR^{\cS_1}=\bigcup_{S_1\in\cS_1}\cR^{S_1}$ (and similarly for $\cS_2$), this is equivalent to saying that there exists $S_1\in \cS_1$ and $S_2\in \cS_2$ such that the corresponding spacetime regions satisfy $\cR^{S_1}\xrightarrow[]{R} \cR^{S_2}$. In other words, the relativistic causality constraints on the spacetime embedding of $\cG$ are the same irrespective of whether or not $\cG$ satisfies the aforementioned property. Thus the above proof also applies to establish the theorem statement for signalling structures $\cG$ not satisfying this property and hence applies to all signalling structures.

\bigskip

Finally, we note that according to Definition~\ref{def: spacetime} any partially ordered set corresponds to a spacetime. This rather minimal definition allows us to derive general results that only depend on the order relation between spacetime points and does not require the spacetime to have any further symmetries, or a smooth differentiable structure. However, under this minimal definition, one might regard two different partially ordered sets $\cT$ and $\cT'$ as two different ``spacetimes''. On the other hand, if we consider the more standard method of modelling spacetime as a differentiable manifold $\cM$, as done in relativistic physics, we could sample different sets of points on the same manifold to generate different partially ordered sets\footnote{That is, if the manifold is globally hyperbolic, in more exotic spacetimes with closed timelike curves, we can also obtain pre-ordered sets from sampling suitable points.} $\cT$ and $\cT'$ from the same spacetime. If we model spacetime as a globally hyperbolic manifold that ensures the absence of closed timelike curves, then the statement of the present theorem would instead become ``For every signalling structure $\cG^{\text{sig}}$ and every globally hyperbolic manifold $\cM$, there exists an embedding $\cE$ of $\cG^{\text{sig}}$ in a region causal structure
$\cG^R_{\cM}$ of $\cM$ that respects relativistic causality, where each node of $\cG^R_{\cM}$ is a finite set of points in $\cM$.'' This can be shown as follows. If $\cG$ is a directed acyclic graph, then it can be embedded in any globally hyperbolic manifold $\cM$ through an embedding $\cE:\mathrm{Nodes}(\cG)\mapsto \cM$ that assigns a point in $\cM$ to each node of $\cG$ (see \cite{Paunkovic2019} for an explicit construction of such an embedding for the acyclic case). This is because the graph has a finite number of nodes and we can always sample a suitable set of points in the manifold having the required order relations. One can apply this embedding to the acyclic graph $\cG'$ constructed in the proof above, this would define the embedding $\mathcal{E'}:\mathrm{Nodes}(\cG)\mapsto \cM$. The rest of the proof will be the same as the above case for partially ordered sets $\cT$.
\end{proof}

\ProbRule*
\begin{proof}

Recall from \cref{eq: process_network} that the induced map of the network $\mathfrak{N}_{\mathcal{W},N}$ which we denote by $\mathcal{P}_{\mathcal{W}}$ is given as follows
\begin{equation}
    \mathcal{P}_{\mathcal{W}}:=\Big(\mathcal{W}\bigotimes_{k=1}^N \cM^{A_k}\Big)^{\{A^I_k\hookrightarrow A^{'I}_{k},A^O_k\hookrightarrow A^{'O}_{k}\}_{k=1}^N},
\end{equation}
where $\mathcal{W}$ is an $N$-partite process map with inputs $\{A^{'O}_{k}\}_k$ and outputs $\{A^I_k\}_k$, while $\cM^{A_k}$ is the extended local map of the agent $A_k$, associated with a quantum input $A^{'I}_{k}$ and classical input $A^s_k$ (for the setting $a_k$) and quantum output $A^O_k$ and a classical output $A^o_k$ (for the outcome $x_k$). Here the primed and corresponding unprimed systems are isomorphic to each other, which allows them to be loop composed together. Then the map $\mathcal{P}_{\mathcal{W}}$ obtained this way has $N$ classical inputs $\{A^s_k\}_k$ and $N$ classical outputs $\{A^o_k\}_k$. The above equation can be read as a parallel composition of the process map with the $N$ local maps followed by $2N$ loop compositions.

In the following, for brevity, we detail the proof for the bipartite case. However, the proof readily generalises to the $N$ agent case.
In the bipartite case, taking the agents to be $A$ and $B$ with local settings associated with input systems $A^s$, $B^s$ and outcomes associated with output systems $A^o$, $B^o$, the parallel composition yields the map $\mathcal{M}^A\otimes\mathcal{M}^B\otimes\mathcal{W}$ with input systems $\{A^{'I},A^s,B^{'I},B^s,A^{'O},B^{'O}\}$ and output systems $\{A^O,A^o,B^O,B^o,A^I,B^I\}$. 
Applying the loop formula of \cref{eq: loops} to compose the corresponding prime and unprimed systems, we obtain the following expression for the induced map $\mathcal{P}_{\mathcal{W}}$ acting on an input state $\ket{a}\bra{a}^{A^s}\otimes\ket{b}\bra{b}^{B^s}$.

\begin{align*}
    \begin{split}
&\mathcal{P}_{\mathcal{W}}(\ket{a}\bra{a}^{A^s}\otimes\ket{b}\bra{b}^{B^s})\\    &=\sum_{k...r}\bra{kmoq}^{A^IA^OB^IB^O}\Bigg(\mathcal{M}^A_a\big(\ket{k}\bra{l}^{A^{'I}}\big)\otimes\mathcal{M}^B_b\big(\ket{o}\bra{p}^{B^{'I}}\big)\otimes\mathcal{W}\big(\ket{mq}\bra{nr}^{A^{'O}B^{'O}}\big)\Bigg)\ket{lnpr}^{A^IA^OB^IB^O}\\
          &=\sum_{k...r}\bra{knor}^{A^IA^OB^IB^O}\Bigg(\big[\mathcal{M}^A_{a}\big(\ket{k}\bra{l}\big)\big]^T\otimes\big[\mathcal{M}^B_{b}\big(\ket{o}\bra{p}\big)\big]^T\otimes\mathcal{W}\big(\ket{mq}\bra{nr}\big)\Bigg)\ket{lmpq}^{A^IA^OB^IB^O},
    \end{split}
\end{align*}

where we have used the notation $\mathcal{M}^A_a\big(\ket{k}\bra{l}^{A^{'I}}\big):=\mathcal{M}^A\big(\ket{a}\bra{a}^{A^s}\otimes\ket{k}\bra{l}^{A^{'I}}\big)$ (and similarly for $B$'s operation) and suppressed the system labels where they are evident from context. Denoting the factor $\Bigg(\big[\mathcal{M}^A_{a}\big(\ket{k}\bra{l}^{A^{'I}}\big)\big]^T\otimes\big[\mathcal{M}^B_{b}\big(\ket{o}\bra{p}^{B^{'I}}\big)\big]^T\otimes\mathcal{W}\big(\ket{m}\bra{n}^{A^{'O}}\otimes\ket{q}\bra{r}^{B^{'O}}\big)\Bigg)$ by $(...)$, introducing factors of the identity $\mathcal{I}=\sum_j\ket{j}\bra{j}$, and then rearranging the resulting inner products we have
\begin{align*}
\begin{split}
       &\mathcal{P}_{\mathcal{W}}(\ket{a}\bra{a}^{A^s}\otimes\ket{b}\bra{b}^{B^s})\\&=\sum_{ijst}\sum_{k...r}\bra{k}^{A^I}\inprod{n}{j}\bra{j}^{A^O}\bra{o}^{B^I}\inprod{r}{s}\bra{s}^{B^O}\Bigg(...\Bigg)\ket{i}^{A^I}\inprod{i}{l}^{A^I}\ket{m}^{A^O}\ket{t}^{B^I}\inprod{t}{p}^{B^I}\ket{q}^{B^O}\\
        &=\sum_{ijst}\sum_{k...r}\bra{i}^{A^I} \ket{l}\bra{k}^{A^I} \bra{j}^{A^O}\bra{t}^{B^I}\ket{p}\bra{o}^{B^I} \bra{s}^{B^O}\Bigg(...\Bigg) \ket{i}^{A^I}\ket{m}\bra{n}^{A^O}\ket{j}^{A^O}\ket{t}^{B^I}\ket{q}\bra{r}^{B^O}\ket{s}^{B^O}\\
        &=\tr_{A^IA^OB^IB^O}\Bigg[\sum_{k...r}\ket{l}\bra{k}^{A^I}\otimes \ket{p}\bra{o}^{B^I}\otimes\Bigg(...\Bigg)\otimes \ket{m}\bra{n}^{A^O}\otimes \ket{q}\bra{r}^{B^O}\Bigg]\\
        &=\tr_{A^IA^OB^IB^O}\Bigg[\Bigg(\sum_{kl}\ket{l}\bra{k}^{A^I}\otimes\big[\mathcal{M}^A_{a}\big(\ket{k}\bra{l}^{A^I}\big)\big]^T\Bigg)\otimes \Bigg(\sum_{op}\ket{p}\bra{o}^{B^I}\otimes\big[\mathcal{M}^B_{b}\big(\ket{o}\bra{p}^{B^I}\big)\big]^T\Bigg)\\
        &\qquad\qquad\otimes\Bigg(\sum_{mnqr}\ket{mq}\bra{nr}^{A^OB^O}\otimes\mathcal{W}\big(\ket{mq}\bra{nr}^{A^OB^O}\big)\Bigg)\Bigg]
        \end{split}
\end{align*}
Now, we wish to calculate the probability that the output of $\mathcal{P}_{\mathcal{W}}(\ket{a}\bra{a}^{A^s}\otimes\ket{b}\bra{b}^{B^s})$ is $\ket{x}^{A^o}\otimes\ket{y}^{B^o}$ i.e., the outcomes $x$ and $y$ are obtained by Alice and Bob upon measuring the settings $a$ and $b$. This is given by \cref{eq: prob_composition1}. We now show that the numerator of this expression i.e, $\tr\Bigg[\Big(\ket{x}\bra{x}^{A^o}\otimes\ket{y}\bra{y}^{B^o}\Big)\Big(\mathcal{P}_{\mathcal{W}}(\ket{a}\bra{a}^{A^s}\otimes\ket{b}\bra{b}^{B^s})\Big)\Bigg]$ equals the process matrix probability rule of \cref{eq: prob}. 
Using the expression for $\mathcal{P}_{\mathcal{W}}(\ket{a}\bra{a}^{A^s}\otimes\ket{b}\bra{b}^{B^s})$ derived above, along with \cref{eq: projection} to absorb the outcome projectors into the definition of the maps $\mathcal{M}_{x|a}^A$ and $\mathcal{M}_{y|b}^B$, we immediately obtain the following which coincides with \cref{eq: prob}.
\begin{equation}
P(xy|ab)
=\tr\left[\left(M_{x|a}^{A^IA^O}\otimes M_{y|b}^{B^IB^O}\right)W\right],
\end{equation}
 where $W=\mathcal{I}\otimes\mathcal{W}|\mathds{1}\rrangle\llangle\mathds{1}|$ is the process matrix and $M_{x|a}^{A^IA^O}=\Big[\mathcal{I}\otimes\mathcal{M}^A_{x|a}\Big(|\mathds{1}\rrangle\llangle\mathds{1}|\Big)\Big]^T$ is the Choi matrix of the local map $\mathcal{M}^{A}_{x|a}$ (and similarly for $B$'s operation) as defined in Section~\ref{sec: PM}. Whenever $W$ is  valid process matrix, this distribution is normalised, which implies that $\sum_x\tr\left[\left(M_{x|a}^{A^IA^O}\otimes M_{y|b}^{B^IB^O}\right)W\right]=1$, which in turn implies that the denominator of \cref{eq: prob_composition1} equals unity.
 
\end{proof}

\ReducedPM*
\begin{proof}

The proof method is very similar to that of Lemma~\ref{lemma: probabilities} but we provide it here for completeness and follow the same notation as the previous proof. Again, we restrict to the bipartite case for simplicity but the proof easily generalises to the multi-partite case. Consider a bipartite process map $\mathcal{W}$ associated with the agents $A_1\equiv A$ and $A_2\equiv B$, with input systems $A^{'O}$, $B^O$ and output systems $A^I$, $B^I$. Let $\mathcal{M}^A_a: A^{'I}\mapsto A^O$ be a local operation of the agent $A$ associated with a particular setting $a$, and where all the primed systems are isomorphic to the corresponding unprimed ones (we only add the primes on $A$'s systems as we will consider only their composition in the next steps).
The sub-network $\mathfrak{N}_{\mathcal{W},1<2}$ (which we denote here as $\mathfrak{N}_{\mathcal{W},A}$) formed by composing $\mathcal{W}$ only with the operation of $A$, through $A^I\hookrightarrow A^{'I}, A^O\hookrightarrow A^{'O}$ will have an induced map $\cN_{\mathcal{W},A}$ which is associated with the input $B^O$ and output $B^I$. We now show that for every choice of setting $a$, the Choi matrix of the induced sub-network map $\cN_{\mathcal{W},A}$ is the reduced process matrix $\overline{W}(M_a^{A^IA^O})$ over $B$ (cf.\ Equation~\eqref{eq: ReducedPM}).
This will complete the proof as the same arguments, under exchange of $A$ and $B$ apply for showing that the Choi matrix of the submetwork $\cN_{\mathcal{W},B}$ is the reduced process matrix $\overline{W}(M_b^{B^IB^O})$ on $A$, and these two are the only non-trivial reduced processes here. The Choi matrix of $\cN_{\mathcal{W},A}$ is given as
$$\sum_{m,n}\ket{m}\bra{n}^{B^O}\otimes \cN_{\mathcal{W},A}(\ket{m}\bra{n}^{B^O}).$$ Using the composition operation, we can write $\cN_{\mathcal{W},A}(\ket{m}\bra{n}^{B^O})$ as

\begin{align*}
    \begin{split}
     \cN_{\mathcal{W},A}(\ket{m}\bra{n}^{B^O})&=  \sum_{i,j,k,l}\bra{i}^{A^I}\bra{k}^{A^O} \Big(\mathcal{M}^A_a\otimes\mathcal{W}\Big)\Big(\ket{i}\bra{j}^{A^{'I}}\otimes\ket{km}\bra{ln}^{A^{'O}B^O}\Big)\ket{j}^{A^I}\ket{l}^{A^O}\\
        &= \sum_{i,j,k,l}\bra{k}^{A^O}\mathcal{M}^A_a\Big(\ket{i}\bra{j}^{A^{'I}}\Big)\ket{l}^{A^O}\otimes\bra{i}^{A^I}\mathcal{W}\Big(\ket{km}\bra{ln}^{A^{'O}B^O}\Big)\ket{j}^{A^I}
    \end{split}
\end{align*}

Plugging this back into the Choi matrix, inserting factors of the identity and rearranging, we can write the Choi matrix of  $\cN_{\mathcal{W},A}$ as

\begin{align*}
    \begin{split}
&\sum_{i,j,k,l,m,n}\ket{m}\bra{n}^{B^O}\otimes\bra{k}^{A^O}\mathcal{M}^A_a\Big(\ket{i}\bra{j}^{A^{'I}}\Big)\ket{l}^{A^O}\otimes\bra{i}^{A^I}\mathcal{W}\Big(\ket{km}\bra{ln}^{A^{'O}B^O}\Big)\ket{j}^{A^I}\\
=&\sum_{i,j,k,l,m,n}\ket{m}\bra{n}^{B^O}\otimes\bra{l}^{A^O}\Big[\mathcal{M}^A_a\Big(\ket{i}\bra{j}^{A^{'I}}\Big)\Big]^T\ket{k}^{A^O}\otimes\bra{i}^{A^I}\mathcal{W}\Big(\ket{km}\bra{ln}^{A^{'O}B^O}\Big)\ket{j}^{A^I}\\
=&\sum_{i,j,k,l,m,n,p,q}\ket{m}\bra{n}^{B^O}\otimes\bra{l}^{A^O}\ket{p}\bra{p}^{A^O}\Big[\mathcal{M}^A_a\Big(\ket{i}\bra{j}\Big)\Big]^T\ket{k}^{A^O}\otimes\bra{i}^{A^I}\ket{q}\bra{q}^{A^I}\mathcal{W}\Big(\ket{km}\bra{ln}\Big)\ket{j}^{A^I}\\
=&\sum_{i,j,k,l,m,n,p,q}\ket{m}\bra{n}^{B^O}\otimes\bra{p}^{A^O}\Big[\mathcal{M}^A_a\Big(\ket{i}\bra{j}\Big)\Big]^T\ket{k}^{A^O}\inprod{l}{p}^{A^O}\otimes\bra{q}^{A^I}\mathcal{W}\Big(\ket{km}\bra{ln}\Big)\ket{j}^{A^I}\inprod{i}{q}^{A^I}\\
=&\sum_{p,q}\bra{pq}^{A^OA^I}\Bigg(\mathds{1}^{B^IB^O}\otimes\sum_{i,j}\ket{j}\bra{i}\otimes\Big[\mathcal{M}^A_a\Big(\ket{i}\bra{j}\Big)\Big]^T\Bigg)\Bigg(\sum_{k,l,m,n}\ket{km}\bra{ln}\otimes\mathcal{W}\Big(\ket{km}\bra{ln}\Big)\Bigg)\ket{pq}^{A^OA^I}\\
    =&\Tr_{A^IA^O}\Bigg(\Big(\mathds{1}^{B^IB^O}\otimes M^{A^IA^O}_a\Big).W\Bigg)=\overline{W}(M^{A^IA^O}_a),
    \end{split}
\end{align*}
where we have used the definition of the Choi matrices of the local operations and the process map (see Section~\ref{sec: PM}) in the last line, along with that of the reduced process matrix (Equation~\eqref{eq: ReducedPM}). Moreover, we have identified the primed and unprimed versions of the systems $A^I$ and $A^O$ as they get composed to form effectively a single system. Moreover the reduced process matrix $\overline{W}(M^{A^IA^O}_a)$ being a valid process matrix implies that the map $ \cN_{\mathcal{W},A}$ of which it is the Choi matrix is a CPTP map \cite{Araujo2015}. This completes the proof.    
\end{proof}

\PMNetworkCPTP*
\begin{proof}
The sub-networks of $\mathcal{N}_{\mathcal{W},N}$ are formed by taking subsets of its maps and compositions (cf. \cref{eq: process_network}). Subsets of $\mathcal{N}_{\mathcal{W},N}^{maps}$ are of two types: either the subset includes $\mathcal{W}$ or it does not. In the latter case, none of the compositions of $\mathcal{N}_{\mathcal{W},N}^{comp}$ can be applied, as all compositions involve $\mathcal{W}$. Thus such sub-networks are always guaranteed to be CPTP since they correspond to the parallel composition (i.e., tensor product) of CPTP maps, and we only need to consider sub-networks formed by composing $\mathcal{W}$ with some subset of the extended local operations through some subset of the compositions in $\mathcal{N}_{\mathcal{W},N}^{comp}$. W.l.o.g, consider the subset of the first $l<N$ extended local operations. The compositions associated with these operations and $\mathcal{W}$ are $\{A^I_k\hookrightarrow A^{'I}_{k},A^O_k\hookrightarrow A^{'O}_{k}\}_{k=1}^l$. If we were to perform all these compositions, we obtain the sub-network $\mathfrak{N}_{\mathcal{W}, l<N}$, whose induced map is CPTP as it corresponds to a reduced process matrix associated with the remaining $N-l$ agents, as we have shown in \cref{lemma: reducedPM}.

\bigskip

Suppose we perform only a subset of the compositions in $\{A^I_k\hookrightarrow A^{'I}_{k},A^O_k\hookrightarrow A^{'O}_{k}\}_{k=1}^l$. Consider all the agents $A_k$ for which both compositions $\{A^I_k\hookrightarrow A^{'I}_{k},A^O_k\hookrightarrow A^{'O}_{k}\}$ are included in this subset. Again, w.l.o.g, we can take these to be the first $p<l$ agents. As the order in which we perform the compositions does not matter (cf. \cref{lemma: OrderIndep}), we can perform all the operations in the subset under consideration by first performing the compositions $\{A^I_k\hookrightarrow A^{'I}_{k},A^O_k\hookrightarrow A^{'O}_{k}\}_{k=1}^p$ and then the rest. By \cref{lemma: reducedPM} if again follows that performing the compositions $\{A^I_k\hookrightarrow A^{'I}_{k},A^O_k\hookrightarrow A^{'O}_{k}\}_{k=1}^p$ results in a CPTP map $\cN_{\mathcal{W},p<N}$. Now by construction, the remaining compositions, include at most one of $A^I_k\hookrightarrow A^{'I}_{k}$ or $A^O_k\hookrightarrow A^{'O}_{k}$ for each agent in $\{A_{p+1},...,A_l\}$. It is easy to see that each of these compositions would correspond to a regular sequential composition of the CPTP map $\cN_{\mathcal{W},p<N}$ with a CPTP map $\cM^{A_k}$ for an agent $A_k$ in $\{A_{p+1},...,A_l\}$. The sequential composition of two CPTP maps (in any given order) is always a CPTP map. This shows that the remaining compositions also yield a CPTP map. We have covered all possible sub-networks in the above and shown their induced maps to be CPTP. This establishes that our original network is a network of CPTP maps, as required. 
\end{proof}

\IndefCyclic*
\begin{proof}
Observe that in the network $\mathfrak{N}_{W,N}$, we have $\{A_k^I\}$ signals to $\{A_k^O\}$ for every agent $A_k$ since the extended local map for each agent encodes all possible operations between their these in and output systems and in particular the identity channel which clearly signals from the input to output. This implies that when we compatibly embed (\cref{definition: graph_compatibility}) the systems of $\mathfrak{N}_{W,N}$ in a causal structure $\mathcal{G}$ with directed edges $\xrightarrow{\cG}$ (thereby allowing us to treat each system as a node of a directed graph with these edges), then we must have $A^I_k\xrightarrow{\cG} A^O_k$ for each $k\in \{1,...,N\}$. By \cref{definition: causallyordered} of fixed ordered processes and \cref{def: signalling_prob}, we can see that a process $W$ is a fixed order process if any only if there exists a partial order on the in/output systems of all agents where $A^O_i$ precedes $A^I_j$ in the partial order for some $j\in \cS$ whenever an agent $A_i$ probabilistically signals to a set of agents $A_{\cS}$, and $A^I_k$ precedes $A^O_k$ in this partial order for every agent $A_k$. By \cref{theorem: equivsignalling}, $A_i$ probabilistically signals to $A_{\cS}$ is equivalent to $\{A^O_i\}$ signals to $\{A^I_j\}_{j\in \cS}$ in the network. Recall that compatibility of 
a signalling relation $\{A^O_i\}$ signals to $\{A^I_j\}_{j\in \cS}$ with an embedding in a causal structure $\mathcal{G}$ implies that there exists $j\in \cS$ such that there is a directed path $A^O_i \xrightarrow{\cG}... \xrightarrow{\cG}  A^I_j$ with respect to the order induced by $\cG$ on the embedded systems (cf. \cref{definition: graph_compatibility}). Then the above arguments, taken together imply that $W$ is a fixed order process if and only if there exists a directed acyclic graph $\cG$ with which the signalling relations of the process network $\mathfrak{N}_{\mathcal{W},N}$ are compatible. Equivalently $W$ is not a fixed order process if and only if any directed graph $\mathcal{G}$ with which the signalling relations of $\mathfrak{N}_{\mathcal{W},N}$ are compatible is a directed cyclic graph i.e., the signalling relations of $\mathfrak{N}_{\mathcal{W},N}$ being compatible with an embedding in some $\mathcal{G}$ certify the cyclicity of $\mathcal{G}$ whenever $W$ is not a fixed order process.

\end{proof}

\NogoA*
\begin{proof}
Recall that a spacetime realisation of a network $\mathfrak{N}_{\mathcal{W},N}$ is specified by a spacetime embedding $\mathfrak{N}_{\mathcal{W},N, \cT, \cE}$ of the network (obtained by associating a spacetime region $\cE(S)\subseteq \cT$ for each system $S$ in the network) along with a fine-graining $\mathfrak{N}_{\mathcal{W},N, \cT, \cE}^f$. For this theorem, the fine-graining and its properties will be irrelevant, we only need to focus on the spacetime embedded network $\mathfrak{N}_{\mathcal{W},N, \cT, \cE}$ to prove the result. Observe that the relativistic causality condition of \cref{def: rel_causality_network} implies, in particular, that the signalling structure of $\mathfrak{N}_{\mathcal{W},N, \cT, \cE}$ must be compatible with the region causal structure $\cG^{\text{reg}}_{\cT}$ induced by the embedding $\cE$. Moreover, the signalling structure of $\mathfrak{N}_{\mathcal{W},N, \cT, \cE}$ is, by construction, 
the same as the signalling structure of $\mathfrak{N}_{\mathcal{W},N}$, since the only difference in the two networks is in the relabelling of each system $S$ to the corresponding spacetime embedded system $S^{\cE(S)}$. This implies that the signalling structure of $\mathfrak{N}_{\mathcal{W},N}$ must be compatible with the causal structure $\cG^{\text{reg}}_{\cT}$ if condition 2 of the theorem must hold. Now if we impose condition 1, that the associated process $W$ is not a fixed order process, it immediately follows from \cref{theorem: indef_cyclic} that $\cG^{\text{reg}}_{\cT}$ must contain a directed cycle and therefore that it is not an acyclic graph, which would violate condition 3. This proves the theorem. 
\end{proof}

\Time*
\begin{proof}
    This corollary follows from noting that the first two assumptions are identical to the first two assumptions of Theorem~\ref{theorem: nogo_main}, while the third assumption here implies the third assumption of the theorem. More explicitly, assumption 3 here requires that for each system $S$, the corresponding spacetime region $\cR^S\subseteq \cT$ assigned to $S$ by the embedding $\mathcal{E}$ is such that all spacetime points in $\cR^S$ have the same time coordinate, say $t^S$ in a global reference frame. Then, Definitions~\ref{def: spacetime} and~\ref{def: region_order} tell us that $\cR^{S_1}\xrightarrow[]{R} \cR^{S_2}$
implies $t^{S_1} < t^{S_2}$. Since we can never have a sequence of times in some global reference frame such that $t^{S_1} < t^{S_2}<...<t^{S_n}<t^{S_1}$, it follows that the set of spacetime regions satisfying assumption 3 of this corollary can never contain a sequence of regions such that $\cR^{S_1}\xrightarrow[]{R}\cR^{S_2}\xrightarrow[]{R}...\xrightarrow[]{R}\cR^{S_n}\xrightarrow[]{R}\cR^{S_1}$, i.e., the regions satisfy assumption 3 of Theorem~\ref{theorem: nogo_main}.
\end{proof}

\PMFinegrain*

\begin{proof}
    We prove the bipartite case here for simplicity, but the proof method readily generalises to the $N$-partite case. Let $W$ be a bipartite process matrix associated with the agents $A$ and $B$ and $\mathfrak{N}_{\mathcal{W},N}$ be the associated network with $N=2$. Consider a spacetime embedding  $\mathfrak{N}_{\mathcal{W},N, \cT, \cE}$ of $\mathfrak{N}_{\mathcal{W},N}$ in a fixed acyclic spacetime $\cT$. If the region causal structure $\cG^{\text{reg}}_{\cT}$ induced by the embedding $\cE$ is acyclic, then this spacetime embedding fully specifies a spacetime realisation of the network (\cref{definition: spacetime_realisation}). In this case, if we impose the first condition of the theorem, namely that relativistic causality condition of \cref{def: rel_causality_network} is satisfied, then it follows that the signalling structure of $\mathfrak{N}_{\mathcal{W},N, \cT, \cE}$ is compatible with the causal structure $\cG^{\text{reg}}_{\cT}$ (which is acyclic in the case under consideration). This signalling structure is the same as the signalling structure of $\mathfrak{N}_{\mathcal{W},N}$ (since the spacetime embedded network only differs from the original network in the fact that each system $S$ is additionally labelled by the corresponding spacetime region $\cE(S)$), which implies that the signalling structure of $\mathfrak{N}_{\mathcal{W},N}$ is compatible with an acyclic causal structure, which is only possible if $W$ is a fixed order process (cf. \cref{theorem: indef_cyclic}).

\bigskip

    Therefore we consider the case where $\cG^{\text{reg}}_{\cT}$ is cyclic. In this case, we require a fine-graining $\mathfrak{N}_{\mathcal{W},N, \cT, \cE}^f$ of $\mathfrak{N}_{\mathcal{W},N}$, associated with an acyclic region causal structure $\cG^{\text{reg},f}_{\cT}$ in order to fully specify the spacetime realisation. Moreover, $\cG^{\text{reg},f}_{\cT}$ would be a fine-graining of the region causal structure $\cG^{\text{reg}}_{\cT}$ induced by the original embedding $\cE$. Invoking the input-output correspondence between the regions in the embedding, we can denote by $\cR^{A^I}\in \mathrm{Nodes}(\cG^{\text{reg}}_{\cT})$ the region assigned to $A$'s input systems $A^s$ and $A^I$ by $\cE$ and by $\cR^{A^O}\in \mathrm{Nodes}(\cG^{\text{reg}}_{\cT})$ the region assigned to $A$'s output systems $A^o$ and $A^O$, and similarly for $B$. Further, since  $\cG^{\text{reg},f}_{\cT}$ is a fine-graining of  $\cG^{\text{reg}}_{\cT}$, there is a fine-graining map that maps $\cR^{A^I}\in \mathrm{Nodes}(\cG^{\text{reg}}_{\cT})$ to a set of nodes $\{\cP^I_1,...,\cP^I_n\}\subseteq \mathrm{Nodes}(\cG^{\text{reg},f}_{\cT})$ which partition it, $\cR^{A^I}=\bigcup_{j=1}^n \cP^I_j$. The input-output correspondence ensures that $\cR^{A^O}\in \mathrm{Nodes}(\cG^{\text{reg}}_{\cT})$ is mapped to a corresponding set of nodes $\{\cP^O_1,...,\cP^O_n\}\subseteq \mathrm{Nodes}(\cG^{\text{reg},f}_{\cT})$ under the fine-graining, which partition it as $\cR^{A^O}=\bigcup_{j=1}^n \cP^O_j$ and where $\cP^I_i\xrightarrow[]{R} \cP^O_i$ for each $i\in \{1,...,n\}$. Similarly, we have the set of regions $\{\cQ^I_1,..,\cQ^I_m\}\subseteq \mathrm{Nodes}(\cG^{\text{reg},f}_{\cT})$ associated with Bob's input $B^I$ under the fine-graining, and a corresponding set of regions $\{\cQ^O_1,..,\cQ^O_m\}\subseteq \mathrm{Nodes}(\cG^{\text{reg},f}_{\cT})$ associated with $B^O$, where $\cQ^I_j\xrightarrow[]{R} \cQ^O_j$ for each $j\in \{1,...,m\}$, with $\cR^{B^I}=\bigcup_{j=1}^m \cQ^I_j$ and $\cR^{B^O}=\bigcup_{j=1}^m \cQ^O_j$.

    \bigskip

    Then in the fine-grained network $\mathfrak{N}_{\mathcal{W},N, \cT, \cE}^f$, we have $n$ input and $n$ output systems associated with Alice (an input system for each $\cP^I_i$ and an output system for each $\cP^O_i$) and $m$ input and $m$ output systems associated with Bob (an input system for each $\cQ^I_j$ and an output system for each $\cQ^O_j$). Moreover, since $\mathfrak{N}_{\mathcal{W},N, \cT, \cE}^f$ is a fine-graining of $\mathfrak{N}_{\mathcal{W},N}$, each sub-network of the former must be a fine-graining of a corresponding sub-network of the latter (\cref{definition: fg_network}). As the process map $\mathcal{W}$ itself corresponds to a sub-network of $\mathfrak{N}_{\mathcal{W},N}$, the fine-grained network $\mathfrak{N}_{\mathcal{W},N, \cT, \cE}^f$ must be associated with a fine-grained process map $\mathcal{W}^f_{\cT,\cE}$ whose inputs are the $n$ output systems of Alice, along with the $m$ output systems of Bob, and whose outputs are the $n$ input systems of Alice and the $m$ input systems of Bob. We then split Alice into $n$ and Bob into $m$ agents, associated with the corresponding in/output systems in order to model $\mathcal{W}^f_{\cT,\cE}$ as an $n+m$-partite process map. In the following we show that whenever the spacetime realisation satisfies the relativistic causality condition, $\mathcal{W}^f_{\cT,\cE}$ is indeed a valid $n+m$-partite process map which corresponds to a fixed order process.

    \bigskip

   In \cref{theorem:causal_box} we have shown that spacetime realisations of any network that respect the relativistic causality condition correspond to so-called \emph{causal boxes} which are CPTPMs that are closed under arbitrary compositions.
Imposing relativistic causality for the spacetime realisation of the original network $\mathfrak{N}_{\mathcal{W},N}$, this result implies that the induced map of each sub-network of $\mathfrak{N}_{\mathcal{W},N, \cT, \cE}^f$ must be a causal box and in particular that $\mathcal{W}^f_{\cT,\cE}$ is a causal box. This implies that composing $\mathcal{W}^f_{\cT,\cE}$ together with any choice of extended local operations (which are themselves CPTP maps) between each input-output pair from the $n+m$ inputs associated with $\{\cP^I_1,...,\cP_n^I,\cQ^I_1,...,\cQ^I_m\}$ and $n+m$ outputs associated with $\{\cP^O_1,...,\cP_n^O,\cQ^O_1,...,\cQ^O_m\}$ can only lead to valid normalised probabilities. This composition is also a valid composition of causal boxes as the input regions are ordered before the corresponding output regions relative to the partial order induced by the acyclic region causal structure $\cG^{\text{reg},f}_{\cT}$. From this argument, we conclude that $\mathcal{W}^f_{\cT,\cE}$ is a valid $n+m$-partite process map whenever the given spacetime realisation of $\mathfrak{N}_{\mathcal{W},N}$ satisfies the relativistic causality condition. Then it immediately follows from \cref{theorem: indef_cyclic}, just as in the first paragraph of this proof, that $\mathcal{W}^f_{\cT,\cE}$ must be a fixed-order process as its signalling structure is compatible with an acyclic causal structure $\cG^{\text{reg},f}_{\cT}$ (which is implied by the relativistic causality condition).

\bigskip

Finally, we note that $\mathfrak{N}_{\mathcal{W},N, \cT, \cE}^f$ being a fine-graining of $\mathfrak{N}_{\mathcal{W},N}$ also implies that the extended local maps $\cM^A$ and $\cM^B$ of Alice and Bob in the original network get fine-grained to extended local maps $\cM^{A,f}$ and $\cM^{B,f}$ acting on $n$ and $m$ in/output systems respectively.
As we have not placed any constraints on the fine-grained local maps, these need not factorise as $\cM^{A,f}=\bigotimes_{i=1}^n \cM^{A,f}_i$ with each $\cM^{A,f}_i$ acting between the in/output systems associated with $\cP^I_i$ and $\cP_i^O\xleftarrow[]{R} \cP^I_i$. Thus the $n+m$ agents in the fine-grained process can perform non-product local operations by communicating to each other. The fact that the spacetime realisation satisfying relativistic causality can be described as a causal box guarantees that the process $\mathcal{W}^f_{\cT,\cE}$ will lead to valid normalised probabilities even when composed with such non-product local operations. This completes the proof.

\bigskip

As an additional remark, we note that above argument using causal boxes also guarantees that 
each fine-grained local operation will admit a decomposition into a sequential composition of CPTP maps (the sequence representation), whenever relativistic causality is imposed.  This means that the local operations correspond to quantum combs (these are defined, for instance in \cite{Chiribella2009}) and allows us to recover the product structure of the local operations through a simple trick originally proposed in \cite{Hoffreumon2021}. One can always map a process and set of non-product local operations modelled as quantum combs (on which the action of the process leads to normalised probabilities) to a new process and a set of product local operations such that the composition is identical in the two cases, yielding the same correlations. This can be done by absorbing all communication channels in the non-product local operations into the definition of the new process. 
\end{proof}

\QS*
\begin{proof}
We first show that in $\mathcal{W}_{QS}$, when the initial state of the control and target systems is $\ket{\psi_C}:=\alpha\ket{0}+\beta\ket{1}$ and $\ket{\psi_T}$ respectively where $\alpha$ and $\beta$ are both non-zero amplitudes and $\ket{\psi_T}$ is an arbitrary qudit state, then we have $\{A^O\}\longrightarrow \{B^I\}$ irrespective of the state input on $B^O$ and $\{B^O\}\longrightarrow \{A^I\}$ irrespective of the state input on $A^O$, in $\mathcal{W}$. This implies that both signalling relations can be realised in the network $\mathfrak{N}_{\mathcal{W}_{QS},U,V}$ for any fixed choice of non-trivial operations $\mathcal{U}^A$ and $\mathcal{V}^B$ without needing to consider the extended local operations. This is in contrast to the general case of Theorem~\ref{theorem: nogo_main} where the signalling relation being realised may depend on the choice of local operations, such that not all signalling relations allowed by $\mathcal{W}$ are realised when it is composed with a given fixed set of local operations.

\bigskip

Consider the action of the process map $\mathcal{W}_{QS}:C^O_C\otimes C^O_T\otimes A^O\otimes B^O\mapsto D^I_C\otimes D^I_T\otimes A^I\otimes B^I$ on the input state $\ket{\Psi}^{C^O_CC^O_TA^OB^O}:=(\alpha\ket{0}+\beta\ket{1})^{C^O_C}\otimes \ket{\psi_T}^{C^O_T}\otimes \ket{\psi_A}^{A^O}\otimes \ket{\psi_B}^{B^O}$, where $\ket{\psi_A}$ and $\ket{\psi_B}$ are arbitrary qudit states. We see that $\mathcal{W}_{QS}$ maps the initial state $\ket{\Psi}^{C^O_CC^O_TA^OB^O}$, to the final state $$\alpha\ket{0}^{D^I_C}\ket{\psi_B}^{D^I_T}\ket{\psi_T}^{A^I}\ket{\psi_A}^{B^I}+\beta\ket{1}^{D^I_C}\ket{\psi_A}^{D^I_T}\ket{\psi_B}^{A^I}\ket{\psi_T}^{B^I}:=\ket{\Phi}^{D^I_CD^I_TA^IB^I}.$$

This gives, 
\begin{align}
    \begin{split}
        \tr_{D^I_CD^I_TA^I}\big[\mathcal{W}_{QS}(\ket{\Psi}\bra{\Psi}^{C^O_CC^O_TA^OB^O})\big]&=|\alpha|^2\ket{\psi_A}\bra{\psi_A}^{B^I}+|\beta|^2\ket{\psi_T}\bra{\psi_T}^{B^I}\\
            \tr_{D^I_CD^I_TB^I}\big[\mathcal{W}_{QS}(\ket{\Psi}\bra{\Psi}^{C^O_CC^O_TA^OB^O})\big]&=|\alpha|^2\ket{\psi_T}\bra{\psi_T}^{A^I}+|\beta|^2\ket{\psi_B}\bra{\psi_B}^{A^I}
    \end{split}
\end{align}
Notice that $\tr_{D^I_CD^I_TA^I}\big[\mathcal{W}_{QS}(\ket{\Psi}\bra{\Psi})\big]$ which is the output on $B^I$ does depend on $\psi_A$, the input on $A^O$ but not depend on $\psi_B$, the input on $B^O$. Similarly $\tr_{D^I_CD^I_TB^I}\big[\mathcal{W}_{QS}(\ket{\Psi}\bra{\Psi})\big]$ depends on $\psi_B$ but does not depend on $\psi_A$. This implies that given the knowledge of the initial control and target states, $\alpha\ket{0}+\beta\ket{1}$ and $\ket{\psi_T}$, Alice and Bob can signal to each other by suitable choices of $\ket{\psi_A}$ and $\ket{\psi_B}$ on their respective output systems $A^O$ and $B^O$, irrespective of the local operation of the other agent. The above proof easily generalises to arbitrary input states $\rho^{C^O_CC^O_TA^OB^O}:=\ket{\psi_C}\bra{\psi_C}^{C^O_C}\otimes \ket{\psi_T}\bra{\psi_T}^{C^O_T}\otimes \rho^{A^OB^O}$, where the input state $\rho^{A^OB^O}$ on $A^O$ and $B^O$ may be an entangled state, $\tr_{D^I_CD^I_TA^I}\big[\mathcal{W}_{QS}(\ket{\Psi}\bra{\Psi})\big]$  depends only on the marginal of the initial state over $A^O$ which is unaffected by local operations on $B^O$.

\bigskip

This establishes that the network $\mathfrak{N}_{\mathcal{W}_{QS},U,V}$ gives rise to a directed cycle of signalling relations $\{A^O\}\longrightarrow \{B^I\} \longrightarrow \{B^O\} \longrightarrow \{A^I\} \longrightarrow \{A^O\}$ for any non-trivial $\mathcal{U}^A$ and $\mathcal{V}^B$ (where $\{A^I\}\longrightarrow \{A^O\}$ and $\{B^I\}\longrightarrow \{B^O\}$ come from the non-triviality of these operations). Then implies (by relativistic causality) that the spacetime regions must satisfy $\mathcal{P}^{A^O}\xrightarrow[]{R}\mathcal{P}^{B^I}\xrightarrow[]{R}\mathcal{P}^{B^O}\xrightarrow[]{R}\mathcal{P}^{A^I}\xrightarrow[]{R}\mathcal{P}^{A^O}$, which violates assumption 3. This completes the proof. 

\end{proof}

\end{document}